\documentclass{hcij}

\usepackage{tikz,amsmath, amssymb,bm,color}
\usetikzlibrary{shapes,arrows}
\usetikzlibrary{calc}
\usetikzlibrary{shapes,decorations,arrows,decorations.markings,shapes.geometric,positioning,chains,shadows,arrows.meta,shapes.misc}
\usepackage{pgfplots}
\usetikzlibrary{pgfplots.groupplots}
\usetikzlibrary{positioning}

\definecolor{KITblue}{RGB}{70,100,170}
\definecolor{KITred}{RGB}{162,34,35}
\definecolor{KITgreen}{RGB}{0,150,130}
\definecolor{KITmaygreen}{RGB}{140,182,60}
\definecolor{KITyellow}{RGB}{252,229,0}
\definecolor{KITorange}{RGB}{223,155,27}
\definecolor{KITbrown}{RGB}{167,130,46}
\definecolor{KITyellow}{RGB}{252,229,0}
\definecolor{KITlila}{RGB}{163,16,124}
\definecolor{KITcyan}{RGB}{34,161,224}

\usepackage[T1]{fontenc}

\begin{document}

\articletitle{Human-machine Symbiosis: A Multivariate Perspective for Physically Coupled Human-machine Systems}

\authorlist{Jairo Inga\textsuperscript{1$\dagger$},  Miriam Ruess\textsuperscript{2$\dagger$}, Jan Heinrich Robens\textsuperscript{3}, Thomas Nelius\textsuperscript{3}, Sean Kille\textsuperscript{1}, Philipp Dahlinger\textsuperscript{4}, Roland Thomaschke\textsuperscript{2}, Gerhard Neumann\textsuperscript{4}, Sven Matthiesen\textsuperscript{3}, Sören Hohmann\textsuperscript{1}, and Andrea Kiesel\textsuperscript{2}}

\affiliationlist{\textsuperscript{1}Institute of Control Systems (IRS), Karlsruhe Institute of Technology (KIT), Germany\\ \textsuperscript{2}Cognition, Action, and Sustainability Unit, Department of Psychology, University of Freiburg, Germany, \\ \textsuperscript{3}Institute of Product Engineering (IPEK), Karlsruhe Institute of Technology (KIT), Germany\\ \textsuperscript{4}Autonomous Learning Robots, Institute of Anthropomatics and Robotics (IAR), Karlsruhe Institute of Technology (KIT), Germany
}


\titleacknowledgments{$\dagger$ Jairo Inga and Miriam Ruess are co-first authors.}
\titlefunding{This research was funded by the Federal Ministry of Education and Research (BMBF) and the Baden-Württemberg Ministry of Science as part of the Excellence Strategy of the German Federal and State Governments. This work was also funded by the Deutsche Forschungsgemeinschaft (DFG, German Research Foundation).}

\newpage
\begin{center}
	\textbf{\Large ABSTRACT}
	\par
\end{center}
{The notion of symbiosis has been increasingly mentioned in research on physically coupled human-machine systems. Yet, a uniform specification on which aspects constitute human-machine symbiosis is missing. By combining the expertise of different disciplines, we elaborate on a multivariate perspective of symbiosis as the highest form of physically coupled human-machine systems. Four dimensions are considered: Task, interaction, performance, and experience. First, human and machine work together to accomplish a common task conceptualized on both a decision and an action level (task dimension). Second, each partner possesses an internal representation of own as well as the other partner’s intentions and influence on the environment. This alignment, which is the core of the interaction, constitutes the symbiotic understanding between both partners, being the basis of a joint, highly coordinated and effective action (interaction dimension). Third, the symbiotic interaction leads to synergetic effects regarding the intention recognition and complementary strengths of the partners, resulting in a higher overall performance (performance dimension). Fourth, symbiotic systems specifically change the user’s experiences, like flow, acceptance, sense of agency, and embodiment (experience dimension). This multivariate perspective is flexible and generic and is also applicable in diverse human-machine scenarios, helping to bridge barriers between different disciplines.}
\medskip{}

\newpage

\thickhrrule

\tableofcontents

\thickhrrule

\articlebodystart

\section{Introduction}
Automatic and intelligent machines have become ever-present in today’s society whereby machines increasingly interact with humans. It was already in 1960 that Licklider \cite{Licklider.1960} had the visionary idea to use the biological analogy of a symbiosis when referring to human-machine interaction. With this idea of human-machine symbiosis, he intended to describe the close union and living together of humans and highly intelligent cybernetical machines in a manner that both benefit from it. He had the idea of humans and machines operating together to solve problems dynamically and interactively in real time. 
The vision of such a symbiosis becomes more and more present in contemporary research on human-machine interaction (e.g., \cite{Abbink.2018,Grigsby.2018,Jarrahi.2018,Wang.2019}). Yet, the concept itself, its drivers, boundary conditions, and the applied scenarios of human-machine symbiosis are still very diffuse. Importantly, a highly interdisciplinary effort is needed \cite{Rahwan.2019} in order to optimize the complexity of human-machine interactions in the future. Consequently, an approach taking into account the perspectives of both counterparts of the symbiosis as well as a multivariate construct thereof is needed. Here, we propose such a multivariate perspective on human-machine symbiosis based on an interdisciplinary approach. We focus on physically coupled human-machine systems and take into account multiple dimensions. The result is a generic approach which can be applied flexibly to various scenarios of physical human-machine interaction.

\subsection{Development of Human-machine Interaction}\label{subsec:intro_dev_HMS}

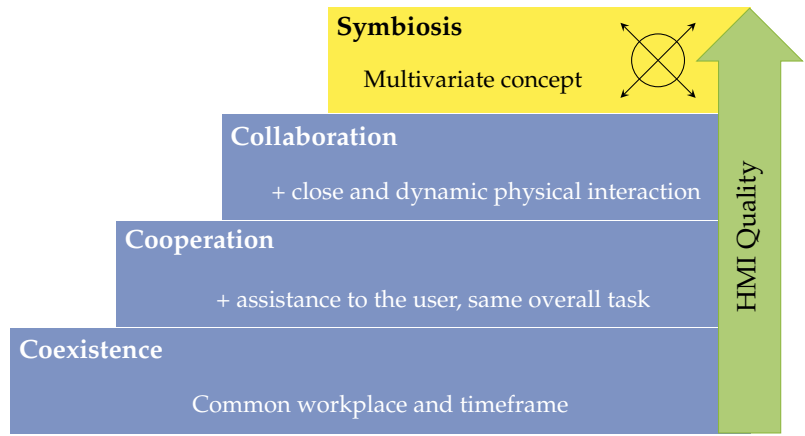
\begin{figure}[t]
	\centering
	\newlength{\mywidth}
\setlength{\mywidth}{4em}

\newlength{\myheight}
\setlength{\myheight}{4em}
    
\tikzset{
}

\tikzstyle{rect1} = [rectangle, fill = KITblue!70, 
minimum width = 7\mywidth, minimum height = \myheight]
\tikzstyle{rect2} = [rectangle, fill = KITblue!70,
minimum width = 6\mywidth, minimum height = \myheight]
\tikzstyle{rect3} = [rectangle, fill = KITblue!70, 
minimum width = 5\mywidth, minimum height = \myheight]
\tikzstyle{rect4} = [rectangle, fill = KITyellow!70, 
minimum width = 4\mywidth, minimum height = \myheight]

\begin{tikzpicture}
\node[rect1,align=right](coexistence){};
\node[anchor = west, yshift = 1.2em] at (coexistence.west){{\color{white}\textbf{Coexistence}}};
\node[anchor = west, yshift = -1em,anchor=center] at (coexistence.center){\small{\color{white}Common workplace and timeframe}};

\node[rect2, above = 0em of coexistence.north east, anchor = south east](cooperation){};
\node[anchor = west, yshift = 1.2em] at (cooperation.west){{\color{white}\textbf{Cooperation}}};
\node[anchor = west, yshift = -1em,anchor=center] at (cooperation.center){\small{\color{white}+~assistance to the user, same overall task}};

\node[rect3, above = 0em of cooperation.north east, anchor = south east,align=right](collaboration){};
\node[anchor = west, yshift = 1.2em] at (collaboration.west){{\color{white}\textbf{Collaboration}}};
\node[anchor = west, yshift = -1em,anchor=center] at (collaboration.center){\small{\color{white}+~close and dynamic physical interaction}};

\node[rect4, above = 0em of collaboration.north east, anchor = south east](symbiosis){};
\node[anchor = west, yshift = 1.2em] at (symbiosis.west){{\color{black}\textbf{Symbiosis}}};
\node[anchor = west, xshift=-2.5em, yshift = -0.8em,anchor=center,align = left] at (symbiosis.center){\small{Multivariate} {concept}};

\node[single arrow, draw=KITmaygreen, fill=KITmaygreen!70, rotate = 90,
      minimum width = \mywidth, single arrow head extend=6pt,
      minimum height=4.02*\myheight, anchor = north, above right = 6.29em and 1.14em of coexistence.south east)]{HMI Quality};

\node[circle, draw, xshift = 4.5em, minimum size = 2em] at (symbiosis.center)(dimCircle){};
\draw[-stealth] (dimCircle.center) -- ++(-0.5,-0.5);
\draw[-stealth] (dimCircle.center) -- ++(+0.5,-0.5);
\draw[-stealth] (dimCircle.center) -- ++(+0.5,+0.5);
\draw[-stealth] (dimCircle.center) -- ++(-0.5,+0.5);

\end{tikzpicture}
	\caption{Symbiosis as a multivariate concept and as the highest form of human-machine interaction (HMI) in physically coupled human-machine systems.}
	\label{fig:HMS_levels_up_to_symbiosis}
\end{figure}

Initially, robots and automation systems were developed for industrial environments to perform repetitive tasks on their own and out of human reach. Today these machines increasingly interact with humans. In the following, we first describe different levels of the development of human-machine interaction as suggested by \cite{Schmidtler.2015} and \cite{Matheson.2019} in order to state the preliminary stages of symbiotic systems (see Figure \ref{fig:HMS_levels_up_to_symbiosis}). We note that the terminologies in human-machine interaction (e.g., coexistence, cooperation, or collaboration) are neither defined precisely nor uniformly and may, thus, be used differently by several authors. The evolution of this interaction started, in the simplest form, with a human-machine coexistence in the sense that machines just share a common workplace and timeframe with the user \cite{Schmidtler.2015}. Yet, humans and machines do not necessarily work conjointly since the individual tasks are different and independent of each other.

If, in addition to a shared workspace and timeframe, the machine provides some kind of assistance to the user, a system results which can be termed human-machine cooperation. Importantly, in this system, the human and the machine share a common goal. Over the last years, a considerable body of literature has steadily grown around the study and design of such assistance systems which are based on the communication of human and machine, in particular via gesture and speech (see e.g. \cite{Green.2007,Ende.2011} \cite[p. 1837 f.]{Haddadin.2016}). For instance, many systems enhance human perception, like a handheld drilling machine proposed by \cite{Schoop.2016} which displays the angle of the drilling hole and the actual depth.


Current technological trends enable an increasing contact between humans and machines in a physical manner. Humans and machines act conjointly and communicate with each other, not only via gesture and speech but mainly via the haptic channel. This current stage of human-machine interaction can be termed human-machine collaboration. The terms collaboration and cooperation are sometimes used synonymously \cite{Mortl.2012,Flemisch.2016,Abbink.2018}, while others propose to distinguish both concepts \cite{Butepage.2017,Jarrasse.2014,Matheson.2019,Silverman.1992}. Here, we describe human-machine collaboration as the next developmental stage in the evolution of human-machine interaction, and this corresponds to the description of a higher form of cooperation. In addition to the haptic interaction, we adopt the differentiation that collaboration includes a dynamic distribution of roles \cite{Jarrasse.2014}. This distribution of roles, as also mentioned by \cite{Flemisch.2016,Kucukyilmaz.2013,Mortl.2012,Music.2017,Oguz.2010,Wang.2019} is accomplished through negotiation based on communication \cite{Oguz.2010}, intention interpretation on the basis of shared representations \cite{Butepage.2017,Flemisch.2016,Oguz.2010} and/or interaction history \cite{Butepage.2017,Jarrasse.2014}. 

If the physical interaction is tight and ongoing, human behavior depends on the behavior of the machine and the other way round. Moreover, a mutual interference of their expectations may exist. A prominent example are exoskeletons, robotic-like devices for use in home care environments or in hospitals, where both the goals and the individual physical effort applied for their achievement are entangled and have to be jointly determined for a seamless and effective interaction. Nonetheless, similar examples arise in other domains: Collaborative workbenches are designed in Industry 4.0 settings \cite{Unhelkar.2018}, where the physical coupling is established through the workpiece and the expectations with respect to the next movement and its execution interfere. Even though human-machine collaborative systems are no longer a fiction but have a realistic chance of being realized on a mechanical, sensorial, and power supply level in the future, the first realizations of these machines are often found to be barely intuitive and to have a lack of usability, leading to an overall system’s performance which is determined by the deficiencies of the human and the machine \cite{Chen.2016}. From human-human physical interaction, we know from daily experience and literature reports that efficiency can be increased and individual effort can be reduced if the interaction is adequate \cite{Curioni.2019}. Therefore, addressing the interaction between humans and machines is a fundamental challenge towards the breakthrough of physically coupled human-machine systems in applications, as mentioned, for example, in \cite{Gupta.2020} for rehabilitation robotics. 

The role each partner adopts with respect to the overall task is also a crucial aspect in physical human-machine interaction since it correlates with performance \cite{Kucukyilmaz.2011,Mortl.2012,Reed.2006}. Previously, a leader-follower (e.g., teleoperation) approach was widely followed, in which the human is always in the lead and has the authority (see e.g. \cite{Clarke.2007,Passenberg.2010,Tzafestas.2008}). Obviously, such an approach has deficiencies, especially in cases where the machine has more information than the human has. The naive engineer’s reflex is a complete replacement of the human by the machine. However, the absence of the human does not expel the deficiencies of the machine. Consequently, the last decade has seen a considerable increase in scientific articles which assert that an optimal human-machine system should be more flexible by operating without superior control by either side \cite{Butepage.2017,Flemisch.2016,Gerber.2020,Jarrasse.2014,Li.2015,Rothfuss.2020}. This implies the possibility of a context-sensitive shift of the desired level of automation, that is, the extent to which the machine acts autonomously (cf., definition of level of automation in \cite{Endsley.1999} or \cite{Sheridan.1978}. This concept is similar to the H-Mode(s) based on the Horse Metaphor by \cite{Flemisch.2014} who introduce a scale of assistance ranging from 100\% human control up to 100\% autonomy/machine control. In between, several levels are possible which correspond to having a tight rein, loose rein, or a secured rein on a horse. The importance of flexible switching of roles and the level of automation is supported by studies indicating its impact on performance \cite{Chiou.2021}, user-specific efficiency \cite{Kucukyilmaz.2013}, but also the avoidance of human agency loss due to an inadequate selection of the level of automation \cite{Berberian.2012}. 

Despite this symmetric relationship with respect to the overall task, the human should still maintain the ability of switching the system completely off, that is, the human has a superordinate (societal) position as advocated, for example, by \cite{Griffith.2006}. This was stated as a response to previous usage of the term symbiosis, which we review in the following.




\subsection{Previous ideas related to a symbiotic human-machine interaction}\label{subsec:intro_prev_HMS}

In biology, the concept of symbiosis has a long history and has been used in large variety with a lot of confusion about its meaning \cite{Martin.2012}. The terminology in its simplest form means just living together. Referring to the coevolution of species interacting closely together, de Bary (1879) applied the term symbiosis in biology in order to describe the living together of two organisms of different species \cite{Bary.1879}. In this respect, it may delineate the intimate physical association between two species which lasts for a considerable span within their lifetimes \cite[p. 235]{Levin.2012}. Of course, human-machine symbiosis only partly mimics the biological symbiosis concept. The close physical association between human and machine is stressed – which might be restricted to short time periods during which human and machine cooperate.

There exist ideas to describe interactions between humans and machines as symbiotic (e.g., \cite{Griffith.2006,Lesh.2004,Parker.1988,Tzafestas.2006}). In these works, the characteristics of symbiosis are said to include a capitalization on the individual strengths of humans and machines in order to assist the human in satisfying individual goals, that is, automation should be human-centered. In other studies, the term symbiosis describes scenarios characterized by a potentially permanent and close physical interaction between a human and a (passive or active) machine. Examples are power-extending exoskeletons or a robotic kinematic chain supporting human movement for rehabilitation purposes such as muscle strengthening \cite{Pervez.2008}, haptic feedback-based co-adaptation of human and machine during physical interaction, with the possibility of arbitrary role-switching \cite{Wang.2016}, or industrial human-robot collaborative assembly contexts in the sense of a continuous mutual engagement via a multimodal communication (for instance, via voice commands and gesture instructions; \cite{Wang.2019}). Some effort has been made to transfer the idea of symbiosis to concrete interaction scenarios, for example, in decision making problems \cite{Jarrahi.2018,Grigsby.2018}, or in manufacturing \cite{Ferreira.2014}. In addition, the idea of machines establishing a direct communication with the human brain to achieve symbiosis has also been mentioned \cite{Schalk.2008,vanErp.2010}. However, to the best of the authors’ knowledge, a clear presentation and discussion of the main aspects of a symbiotic interaction in physically coupled human-machine systems does not exist to date.

Other literature proposes frameworks for the description of human-machine cooperation or collaboration models \cite{Abbink.2018,Flemisch.2016,PacauxLemoine.2015,Rothfuss.2020}. However, these works attempt to describe human-machine interaction by a one-dimensional model, therefore missing a clear description of the various dimensions influencing the interaction. In addition, the term symbiosis is used in \cite{Abbink.2018} to describe relationships between humans and machines with shared information, demonstration, and action, and in \cite{Flemisch.2019} as a similar term to “cooperation”. To summarize, the literature does not exhibit a clear consensus on the meaning of symbiosis, its main drivers, characteristics, and its effects.

\subsection{Human and machine shared control - a joint framework in terms of feedback control systems}\label{subsec:intro_sharedcontrol}

\begin{figure}[t]
	\centering

\usetikzlibrary{calc}
\begin{tikzpicture}[thick,>=latex',scale=0.85, every node/.style={scale=0.85}]
  	\input{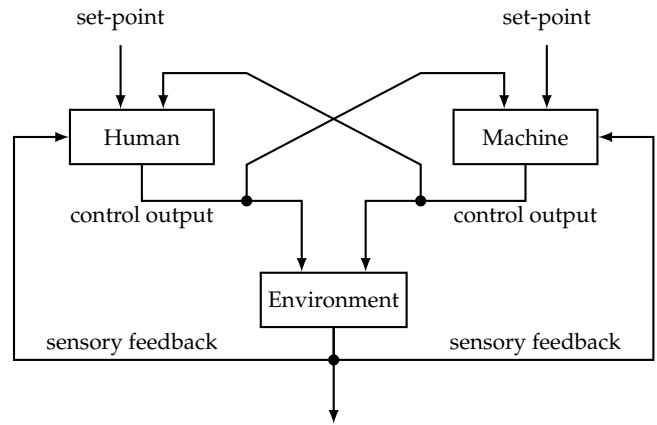}
	\coordinate[](bplayer);
	\node [block] at ($(bplayer)$)(system){Environment} ;
	\coordinate(sysoutsplit) at ($(system.south)+(0,-\nodedist/4)$);
	\draw[fill] (sysoutsplit) circle (2pt);
	\coordinate[label=above:sensory feedback]() at ($(system.east)+(\nodedist,-\nodedist/2)$);
	\coordinate[label=above:sensory feedback]() at ($(system.west)+(-\nodedist,-\nodedist/2)$);
	\coordinate(sysout) at ($(system.south)+(0,-\nodedist*3/4)$);
	\coordinate(sysinu1) at ($(system.north)+(-\nodedist/4,0)$);
	\coordinate(sysinu1ex) at ($(system.north)+(-\nodedist/4,1/2*\height)$);
	\coordinate(sysinu2) at ($(system.north)+(\nodedist/4,0)$);
	\coordinate(sysinu2ex) at ($(system.north)+(\nodedist/4,1/2*\height)$);
	\node [block] at ($(bplayer)+(-1.5*\nodedist,3*\height)$)(p1){Human};
	\coordinate(splitp1) at ($(p1.east)+(\nodedist/4,-\nodedist/2)$);
	\draw[fill] (splitp1) circle (2pt);
	\coordinate[label=above:control output]() at ($(p1.south)+(0,-0.56*\nodedist)$);
	\coordinate(u1r) at ($(splitp1)+(0,4/6*\height)$);
	\coordinate(sysinp1) at ($(p1.west)+(0,1/6*\height)$);
	\coordinate(sysinp1ex) at ($(sysinp1)+(-\nodedist/2,0)$);
	\coordinate(u2inp1) at ($(p1.north)+(1/6*\nodedist,0)$);
	\coordinate(setinp1) at ($(p1.north)+(-1/6*\nodedist,0)$);
	\coordinate(u2inp1ex) at ($(splitp1)+(0,\nodedist)$);
	\coordinate(sysr1) at ($(p1.west)+(-\height,0)$);
	\node [block] at ($(bplayer)+(1.5*\nodedist,3*\height)$)(p2){Machine};
	\coordinate(splitp2) at ($(p2.west)+(-\nodedist/4,-\nodedist/2)$);
	\draw[fill] (splitp2) circle (2pt);
	\coordinate[label=above:control output]() at ($(p2.south)+(0,-0.56*\nodedist)$);
	\coordinate(u2r) at ($(splitp2)+(0,4/6*\height)$);
	\coordinate(sysinp2) at ($(p2.west)+(0,-1/6*\height)$);
	\coordinate(sysinp2ex) at ($(sysinp2)+(-\nodedist/2,0)$);
	\coordinate(u1inp2) at ($(p2.north)+(-1/6*\nodedist,0)$);
	\coordinate(setinp2) at ($(p2.north)+(1/6*\nodedist,0)$);
	\coordinate(u1inp2ex) at ($(splitp2)+(0,\nodedist)$);
	\coordinate(sysr2) at ($(p2.east)+(\height,0)$);
	\draw[-latex](system)--(sysoutsplit)--++(2.5*\nodedist,0) |- (p2.east);
	\draw[-latex](system)--(sysoutsplit)--++(-2.5*\nodedist,0) |- (p1.west);
 	\draw[-latex](sysoutsplit)--(sysout);

 	\draw[-latex](p1)|-(splitp1)-|(sysinu1ex)--(sysinu1ex)--(sysinu1);
 	\draw[-latex](p2)|-(splitp2)-|(sysinu2ex)--(sysinu2ex)--(sysinu2);
 	
 	\draw[-latex](splitp2)--(u2r)--(u2inp1ex)-|(u2inp1);
 	\draw[-latex](splitp1)--(u1r)--(u1inp2ex)-|(u1inp2);
 	
 	\draw[-latex](setinp1)+(0,0.5*\nodedist) -- node[above, pos = -0.1]{set-point} (setinp1);
 	\draw[-latex](setinp2)+(0,0.5*\nodedist) -- node[above, pos = -0.1]{set-point} (setinp2);
\end{tikzpicture}
	\caption{Human-machine feedback control system as a representation for shared control, where the human and the machine simultaneously influence the environment and receive haptic feedback from the counterpart’s control, and sensory feedback from the environment (based on Abbink et al., 2012; Flad et al., 2014; Inga et al., 2018; Mörtl et al., 2012).}
	\label{fig:HMS_shared_control}
\end{figure}

Due to the significance of haptic interaction in physically coupled human-machine systems, an adequate framework is needed to describe the control system which emerges from the human-machine interaction. Human-machine systems including continuous physical interaction and haptic feedback are called shared control systems \cite{Abbink.2012,Flemisch.2016}. They are characterized by humans and machines “interacting congruently in a perception-action cycle to perform a dynamic task that either the human or the robot could execute individually under ideal circumstances” \cite[p. 511]{Abbink.2018}. Figure \ref{fig:HMS_shared_control} shows a schematic representation of a general human-machine shared control system.

The human and the machine can be seen as controllers within a feedback control system, simultaneously influencing the environment based on continuously obtained sensory information \cite{Flad.2014}. This information may include direct haptic feedback from the partner’s control output. Both feedforward and feedback elements can be utilized by each of the two partners in order to select actions to achieve a desired goal, for example, in the form of a position set-point, which in the literature is usually assumed to be given and be even identical for both partners (e.g., \cite{Honing.2014,Mortl.2012,Inga.2018}). Different and in particular non-compatible goals lead to disagreements in the control output and generate conflicts. Hence, the use of human behavior models becomes essential \cite{Boink.2014,Mars.2017,Inga.2018}. Shared control systems have been mostly considered as a framework for joint action at a haptic level in various applications including driving assistance systems \cite{Flad.2017,Mars.2017}, or teleoperation in robotics \cite{Mortl.2012,Smisek.2017}. Shared control is deemed an important basis for human-machine cooperation and collaboration \cite{Flemisch.2016}. However, it only provides a basic framework for the perception-action cycle in human-machine haptic interaction which is not sufficient for a full description of all aspects in human-machine interaction, e.g. the influence of the interferences of actions and decision \cite{PacauxLemoine.2015}. For example, besides the coupling depicted in Figure \ref{fig:HMS_shared_control}, the overall system behavior is affected by an upper level of interaction, usually not included in shared control approaches. This upper level defines the goals of the shared control system on the lower haptic level, which are set-points in the feedback control system of Figure \ref{fig:HMS_shared_control}. In addition, the relationship between shared control and upper interaction levels is discussed in some works (e.g., \cite{Abbink.2018,Flemisch.2016}), but the connection between the levels is mostly reduced to the reception of feedback from the environment and the transmission of goals from upper levels to lower levels. Given the entanglement of the human and machine decisions and the strong dynamic interaction between controllers with potentially incomplete information, it is conceivable that a highly performant human-machine feedback control system demands a kind of alignment between internal models or expectations.

\subsection{From shared control to human-machine symbiosis}

Shared control and its control-theoretical representation lay the foundation for an analysis of the interplay between humans and machines during a cooperation or collaboration. However, shared control systems are focused on the perception-action cycle at a haptic interaction level and thus are only one possible aspect of human-machine interaction, including symbiosis. Moreover, as reviewed in Section \ref{subsec:intro_prev_HMS}, previous usage of the term symbiosis is mostly not tailored for physically coupled human-machine systems. We propose human-machine symbiosis as a term implying not only the close, potentially continuous physical interaction in human-machine systems, but also a higher form of collaboration. This includes 
\begin{itemize}
	\item a symmetric relationship with respect to the overall task,
	\item the possibility of smoothly switching roles and the level of automation across various task abstraction levels,
	\item synergetic effects of each partner’s strengths which include the seamless fusion of information and mutual understanding towards a highly coordinated action resulting in a higher overall performance which is more than the sum of its parts,
	\item and an interaction form with the highest acceptance and the best human-sided experience.
\end{itemize}

These properties, their mechanisms, and their drivers are strongly correlated with each other. Due to the involvement of both human and machine partners, the evaluation and design of physically coupled human-machine systems has to include engineering, information technology and psychological disciplines. In light of the need of an interdisciplinary analysis, we aim for a multivariate perspective on symbiosis as the highest form of interaction in physically coupled human-machine systems, as depicted in Figure 1.

\section{Idea of a multivariate perspective}

Describing human-machine symbiosis demands an approach that, on the one hand, allows for an adequate framework in which automation design can be conducted, and on the other hand, complies with an optimal involvement of the users’ capacities and internal models of the interaction. However, as discussed in Section 1, previous models of human-machine interaction do not sufficiently describe the various mechanisms and properties of physically coupled human-machine systems. Thus, we propose a novel multivariate approach for the investigation of the features of human-machine symbiosis including the following dimensions: Task, interaction, performance, and experience. We note that the first three dimensions refer to the human-machine system while the dimension of experience is restricted to the human part of the symbiotic system.

\subsection{Task Dimension}\label{subsec:mult_task_dim}

The task dimension in symbiotic human-machine systems defines the scope of the interaction within the different degrees of abstraction of an overall complex task. An overall complex task can represent, for example, “assembly of product X”, “drive from A to B”. In stand-alone scenarios, that is, either the human or machine without the counterpart, modeling the behavior (human) or stating a framework (machine) for the completion of such an overall complex task involves a subdivision into several hierarchical levels. For example, three-level architectures in \cite{Saridis.1983,Volpe.2001} have been proposed for machine or automation design. Similarly, such a subdivision is also present in several human behavior models \cite{Donges.1999,Michon.1985}, which were mainly proposed towards the development of human-behavior models to be included in driving assistance systems. The levels are defined in an application-specific way, that is, navigation level (route selection), guidance level (maneuver selection, e.g., overtaking), and stabilization level (vehicle control) to achieve a desired position or trajectory \cite{Donges.1999}.

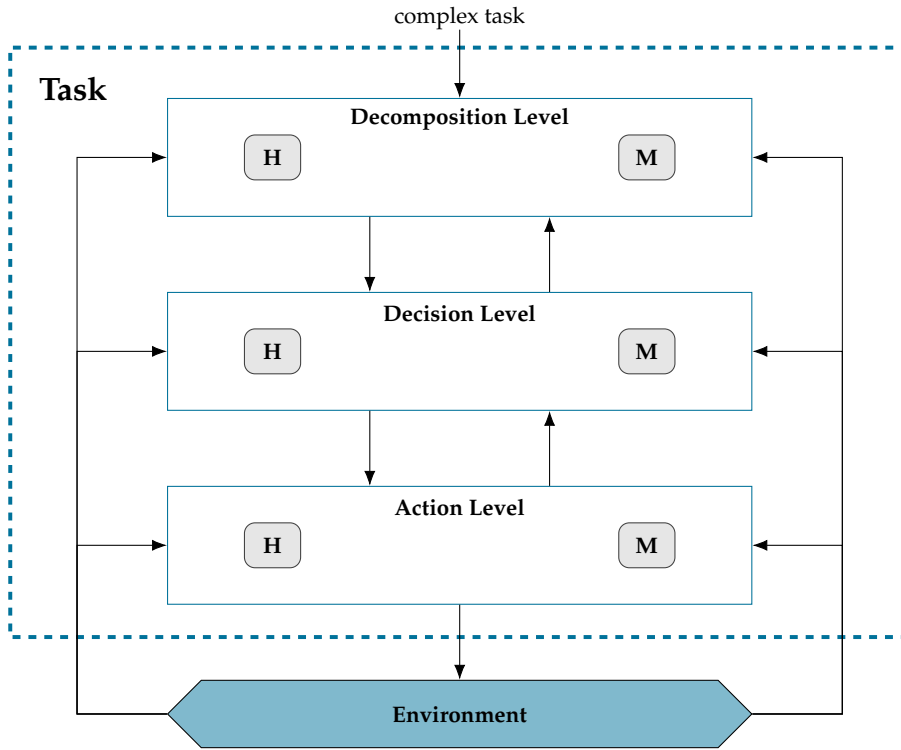
\begin{figure}[t]
	\centering
\setlength{\mywidth}{26em}

\setlength{\myheight}{3em}

\newlength{\arrowdist}
\setlength{\arrowdist}{4em}

\definecolor{environBlue}{RGB}{0,114,154}
    
\tikzset{
}

\tikzstyle{arrow} = [-{Latex[length=2mm]}]
\tikzstyle{level} = [rectangle, fill = white, draw = environBlue, draw,
minimum width = \mywidth, minimum height = 1.75*\myheight]
\tikzstyle{environ} = [signal, signal to = west and east, inner xsep=0mm, fill = environBlue!50, draw,
minimum width = \mywidth, minimum height = \myheight]
\tikzstyle{hmrep} = [rectangle,  fill = gray!20, draw = gray!50!black!, draw, rounded corners,
minimum width = 2.5em, minimum height = 2em]
\tikzstyle{task} = [rectangle, draw = environBlue, line width = 1.5, dashed,
minimum width = 40em, minimum height = 8.75*\myheight]

\pgfdeclarelayer{background}
\pgfsetlayers{background,main}

\begin{tikzpicture}[scale=0.85, every node/.style={scale=0.85}]

\node at (0,0)(ursprung){};


\node[level] at (ursprung)(decision){};
\coordinate(dec_NW) at ($(decision.north)+(-\arrowdist,0)$);
\coordinate(dec_NE) at ($(decision.north)+(\arrowdist,0)$);
\coordinate(dec_SW) at ($(decision.south)+(-\arrowdist,0)$);
\coordinate(dec_SE) at ($(decision.south)+(\arrowdist,0)$);
\node[hmrep, left = 6em of decision.center](H_dec){{\color{black}\textbf{H}}};
\node[hmrep, right = 6em of decision.center](M_dec){{\color{black}\textbf{M}}};

\node[anchor = south, yshift = 1em] at (decision.center){\bfseries{Decision Level}};

\node[level, above = of decision](decomposition){};
\coordinate(deco_left) at ($(decomposition.south)+(-\arrowdist,0)$);
\coordinate(deco_right) at ($(decomposition.south)+(\arrowdist,0)$);
\node[anchor = south, yshift = 0.9em] at (decomposition.center){\bfseries{Decomposition Level}};
\node[hmrep, left = 6em of decomposition.center](H_dec){{\color{black}\textbf{H}}};
\node[hmrep, right = 6em of decomposition.center](M_dec){{\color{black}\textbf{M}}};

\node[level, below = of decision](action){};
\coordinate(act_NW) at ($(action.north)+(-\arrowdist,0)$);
\coordinate(act_NE) at ($(action.north)+(\arrowdist,0)$);
\coordinate(act_SW) at ($(action.south)+(-\arrowdist,0)$);
\coordinate(act_SE) at ($(action.south)+(\arrowdist,0)$);
\node[anchor = south, yshift = 1em] at (action.center){\bfseries{Action Level}};
\node[hmrep, left = 6em of action.center](H_dec){{\color{black}\textbf{H}}};
\node[hmrep, right = 6em of action.center](M_dec){{\color{black}\textbf{M}}};

\node[environ, below = of action](environment){\bfseries{\color{black}Environment}};
\coordinate(env_NW) at ($(environment.north)+(-\arrowdist,0)$);
\coordinate(env_NE) at ($(environment.north)+(\arrowdist,0)$);

\node[task, below = 7em of decomposition.center, anchor= center](task){};
\node[anchor = north west, yshift = -1em, xshift = 1em] at (task.north west){\bfseries\Large{Task}};


\draw[arrow](decomposition)++(0,2) -- node [above, yshift=1.2em] {complex task}(decomposition);

\draw[arrow](deco_left) -- (dec_NW);
\draw[arrow] (dec_NE) -- (deco_right);

\draw[arrow](dec_SW) -- (act_NW);
\draw[arrow] (act_NE) -- (dec_SE);

\draw[arrow] (action) -- (environment);

\draw[arrow] (environment.east) --++(\arrowdist,0) |- (action.east);
\draw[arrow] (environment.east) --++(\arrowdist,0) |- (decision.east);
\draw[arrow] (environment.east) --++(\arrowdist,0) |- (decomposition.east);

\draw[arrow] (environment.west) --++(-\arrowdist,0) |- (action.west);
\draw[arrow] (environment.west) --++(-\arrowdist,0) |- (decision.west);
\draw[arrow] (environment.west) --++(-\arrowdist,0) |- (decomposition.west);


\end{tikzpicture}
	\caption{Task dimension: Layer model for the three task levels (i.e., action, decision, decomposition) of the human-machine symbiosis. The gray blocks represent the control mechanisms of the human (H) and the machine (M) at each level.}
	\label{fig:HMS_task_levels}
\end{figure}

Recent literature proposes human-machine interaction models \cite{Abbink.2018,Flemisch.2016,Rothfuss.2020} based on similar hierarchical levels. \cite{Flemisch.2016} define three levels, that is, strategic, tactical, and operational levels, and explain the analogy to the aforementioned navigation, guidance, and stabilization levels proposed for driving models. \cite{Abbink.2018} propose the same layers as \cite{Flemisch.2016}, yet, adding a fourth execution level below the operational level. They suggest that the operational level defines a goal in terms of a desired system control action (e.g., necessary acceleration to reach a desired velocity in a car) and that the execution level defines lower-level control actions (e.g., muscle activation or neural signals). \cite{Flemisch.2019} add an upper fourth level of cooperation and metacommunication, which is a level transversal to the previous three. This fourth level defines the needed abilities of both partners to cooperate with each other on all levels. \cite{Flemisch.2019} emphasize that this “communication about cooperation” is comparable to the extensions of shared control proposed in \cite{PacauxLemoine.2015} in the sense that it defines the modus of the cooperation. \cite{Rothfuss.2020} propose a model with four levels including decomposition, decision, trajectory, and action level. The first two levels can be associated with the strategic and guidance levels, respectively. The trajectory level gives the desired environmental state values - for example, sequence of set-points in Figure \ref{fig:HMS_shared_control} - to the action level. Finally, the action level selects the necessary control values to influence the environment to achieve the desired state values and the chosen “maneuver”. This action level is therefore comparable to the operational level in the three-level model of \cite{Flemisch.2016}. In this paper, building upon the mentioned literature, we define the task dimension of human-machine symbiosis in physically coupled systems using a layer model consisting of the decomposition level, decision level, and action level, depicted in Figure \ref{fig:HMS_task_levels}.

The model defines an interplay between each of the levels, similar to the previously proposed human-machine interaction models in the literature \cite{Abbink.2018,Flemisch.2016,Rothfuss.2020}. We consider the corresponding input of the action level to result from the decision level, for example, a subtask in the form of a desired position of a jointly controlled object. The input of the decision level, that is, the set of possible maneuvers/subtasks to complete the complex task, is passed from a third level, for example, the so-called decomposition level \cite{Rothfuss.2020}. While currently far from reality, it is theoretically possible to determine a higher-level goal based on the individual plans of both the human and the machine in a symbiotic way, see, for example, the original vision of symbiosis of \cite{Licklider.1960} or the concept of “joint cognitive system” of \cite{Silverman.1992}. However, this would imply a long-term communication, negotiation, and conflict arbitration process \cite{Flemisch.2016}. In this paper, we focus on symbiosis for physically interacting human-machine systems, and thus consider the output of the strategic level to be fixed. Despite our focus on the lower task levels, an extension of human-machine symbiosis towards higher, more cognitive levels - in the spirit of the first work of \cite{Licklider.1960} - is conceivable.
We also disregard the trajectory level included in the model by \cite{Rothfuss.2020} in favor of a higher compliance with current theories of human motion behavior, which support the idea that decision-making and sensorimotor control are entangled. This also explains the necessary direct feedback from the action level to the decision level. The literature conjectures that sensorimotor control is based on optimality principles \cite{Todorov.2004} and the movements are goal-directed, complying with the theory of planned behavior of \cite{Ajzen.1985}. The theories based on optimality principles are currently replacing previous approaches of open-loop trajectory planning (e.g., \cite{Flash.1985,Uno.1989}) which is realized by a lower-level control mechanism afterwards. We follow the approach by \cite{Flemisch.2016,Rothfuss.2020} and consider a single action level which also encompasses the interaction with the machine, yet, still allowing for the integration of current theories of human sensorimotor control. Indeed, the neuroscientific community seeks a cohesive framework \cite{Scott.2004}. Thus, recent literature suggests that the aforementioned optimality principles, and in particular optimal feedback control theory, are strong candidates to describe various aspects of motor behavior including goal-directed actions and task-related fast reactions with time constants in the order of magnitude of seconds and milliseconds, respectively \cite{Gallivan.2018,Nashed.2012,Pruszynski.2012}.
In this respect, sensory feedback from the environment, feedback from the action level to the decision level, and the definition and specification of the interaction mechanisms within each level are crucial for a description of symbiotic human-machine systems. The next section specifies the interplay between the human and the machine within and across the decision and action levels.

\subsection{Interaction Dimension}\label{subsec:mult_inter_dim}

While the task dimension defines the level of interaction in terms of an overall task, the mechanisms of interaction at each of these levels need to be defined such that a symbiotic human-machine system arises. The interaction dimension defines the main characteristics of the interplay between the partners across the levels of the task dimension. We conjecture the following features of a symbiotic interaction

\begin{enumerate}
	\item  a highly skilled and coordinated joint action: Symbiotic action is characterized by a low cognitive effort for task completion. Therefore, in the well-known subdivision by human behavior in knowledge-, rule-, and skill-based (KRS)  behavior of \cite{Rasmussen.1983}, symbiosis implies skill-based behavior on both the action and decision levels, while permitting rule-based behavior on the decision level.
	\item the possibility of a coordinated shift of the desired level of automation. In particular, a hybrid approach optimally combining context-sensitive, machine-induced shifts based on task-related triggers and human-induced shifts is characteristic to symbiotic systems.
	\item congruent individual representations of the interaction, including each partner’s own intention and the expectation about the intentions of the other partner, as well as the effect of the execution of the decisions and actions, including
	\item feedback from each partner and from the environment within a perception and shared-action cycle.
\end{enumerate}

In the following, we propose an interaction framework for symbiotic human-machine systems which captures the proposed features. The representation of the interaction can be interpreted in terms of own intention and expectation about the intentions of the other partner.

\subsubsection{Interaction on the Decision Level}
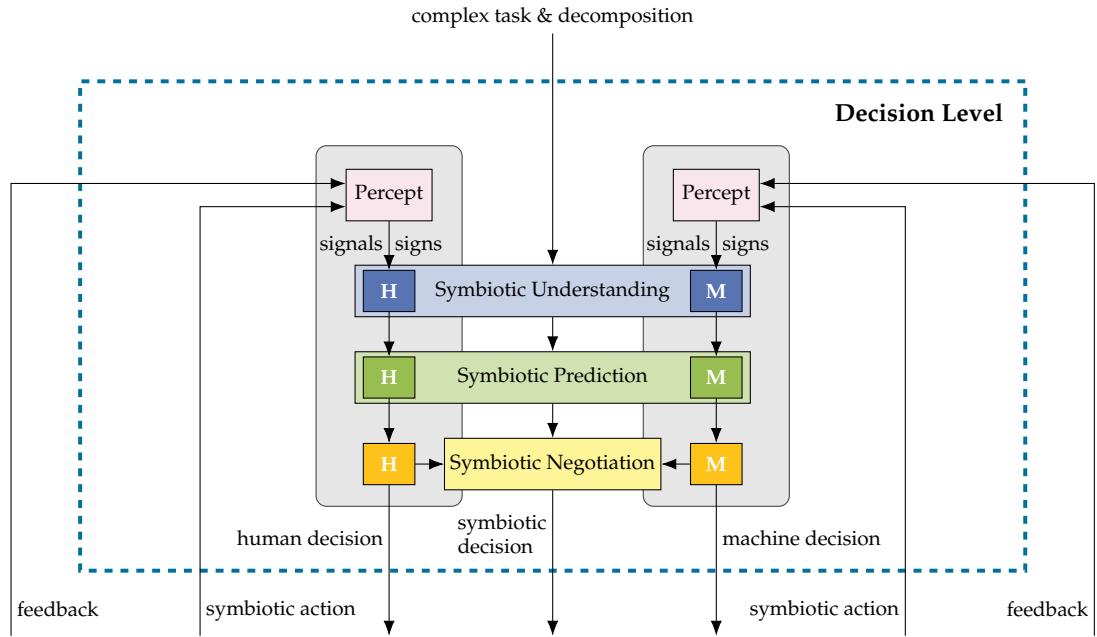
\begin{figure}[t]
	\centering
\setlength{\mywidth}{23em}

\setlength{\myheight}{3em}
\definecolor{environBlue}{RGB}{0,114,154}
    
\tikzset{
}

\tikzstyle{symbblock} = [rectangle, fill = KITblue!30, draw,
minimum width = \mywidth, minimum height = \myheight]
\tikzstyle{symbblock2} = [rectangle,fill = KITmaygreen!40, draw,
minimum width = \mywidth, minimum height = \myheight]
\tikzstyle{symbblock3} = [rectangle,fill = KITyellow!40, draw,
minimum width = 0.5*\mywidth, minimum height = \myheight]
\tikzstyle{hmrep} = [rectangle, fill = KITblue!90,draw,
minimum width = 3em, minimum height = 0.8\myheight]
\tikzstyle{perc} = [rectangle, fill = purple!10, draw ,
minimum width = 5em, minimum height = \myheight]
\tikzstyle{partner} = [rectangle, fill = gray!20, draw = gray!50!black!,
minimum width = 8.5em, minimum height = 7*\myheight, rounded corners]
\tikzstyle{arrow} = [-{Latex[length=2mm]}]
\tikzstyle{level} = [rectangle, draw = environBlue, line width = 1.5, dashed,
minimum width = 55em, minimum height = 9.5*\myheight]

\pgfdeclarelayer{background}
\pgfsetlayers{background,main}

\begin{tikzpicture}[font=\large, scale=0.65, transform shape]

\node at (0,0)(ursprung){};


\node[symbblock] at (ursprung)(understanding){Symbiotic Understanding};
\node[hmrep, left = 8em of understanding.center](H_under){{\color{white}\textbf{H}}};
\node[hmrep, right = 8em of understanding.center](M_under){{\color{white}\textbf{M}}};

\node[symbblock2, below =  2em of understanding](prediction){Symbiotic Prediction};
\node[hmrep, fill = KITmaygreen!90, left = 8em of prediction.center](H_pred){{\color{white}\textbf{H}}};
\node[hmrep, fill = KITmaygreen!90, right = 8em of prediction.center](M_pred){{\color{white}\textbf{M}}};

\node[symbblock3, below =  2em of prediction](decision){Symbiotic Negotiation};
\node[hmrep, fill = KITyellow!60!orange!90!, left = 8em of decision.center](H_neg){{\color{white}\textbf{H}}};
\node[hmrep, fill =  KITyellow!60!orange!90!, right = 8em of decision.center](M_neg){{\color{white}\textbf{M}}};

\node[perc, above = of H_under](H_perc){Percept};
\node[perc, above = of M_under](M_perc){Percept};

\node[level, above = 3em of prediction.center, anchor= center](decisionlevel){};
\node[anchor = north east, yshift = -1em, xshift = -1em] at (decisionlevel.north east){\bfseries\Large{Decision Level}};

\begin{pgfonlayer}{background}
\node[partner,above = 3em of H_pred.center, anchor = center]{};
\node[partner,above = 3em of M_pred.center, anchor = center]{};
\end{pgfonlayer}

\draw[arrow](H_perc) -- node [right]{signs} node [left]{signals}  (H_under);
\draw[arrow](H_under)--(H_pred);
\draw[arrow](H_pred) -- (H_neg);
\draw[arrow](H_neg) -- (decision);
\draw[arrow](H_neg) -- node[left, align = left, pos =0.35]{human decision}++(0,-10em);

\draw[arrow](M_perc) -- node [right]{signs} node [left]{signals} (M_under);
\draw[arrow](M_under)--(M_pred);
\draw[arrow](M_pred) -- (M_neg);
\draw[arrow](M_neg) -- (decision);
\draw[arrow](M_neg) -- node[right, pos =0.35]{machine decision}++(0,-10em);
\draw[arrow](understanding)+(0,15em) -- node[above, pos = 0]{complex task \& decomposition} (understanding);
\draw[arrow](understanding) -- (prediction);
\draw[arrow](prediction) -- (decision);
\draw[arrow](decision) -- node[left, align = left, pos =0.3]{symbiotic \\ decision}++(0,-10em);

\draw[arrow](H_neg)+(-11em,-10em) |- node[right, pos = 0.03]{symbiotic action} (H_perc.195);
\draw[arrow](M_neg)+(11em,-10em) |- node[left, pos = 0.03]{symbiotic action} (M_perc.345);
\draw[arrow](H_neg)+(-22em,-10em) |- node[right, pos = 0.03]{feedback} (H_perc.165);
\draw[arrow](M_neg)+(22em,-10em) |- node[left, pos = 0.03]{feedback} (M_perc.15);

\end{tikzpicture}
	\caption{Interaction dimension on the decision level for human-machine symbiosis: Symbiotic understanding allows the prediction of the overall system behavior and thus a symbiotic decision emerges which consists of each partner’s own decision. Symbiotic understanding is based on the complex task and its decomposition, as well as on the perception of environmental feedback as continuous signals and signs in the form of skilled behavior activation cues (cf. (Rasmussen, 1983)). Additionally, feedback from the environment and of the symbiotic action permits the consideration of the behavior at the lower action level in the chosen decisions. Human (H) and machine (M) blocks denote individual representations with respect to the corresponding elements.}
	\label{fig:HMS_interaction_decision}
\end{figure}

 Figure \ref{fig:HMS_interaction_decision} shows the interaction on the decision level. Both the human and the machine (gray rounded rectangles) obtain sensory feedback from the environment that is processed in the perception block. This feedback includes information from the environment in the form of states, for example, positions, velocities, orientations of the jointly manipulated object or systems. However, it also includes performance metrics (cf., Section \ref{subsec:mult_perf_dim}). This feedback is processed by the human and the machine (percept block) and is interpreted in the form of signs or cues, as well as continuous and task-relevant signals. These are swiftly processed to activate single or sequences of skill-based behaviors of the human-machine system. The heart of the interaction dimension is the symbiotic understanding between the human and the machine. The blue blocks depict the individual representation each of them has considering the other partner, including their own expectations on the interaction as well as a model which describes the interference of the expectations and potential actions. Symbiotic understanding implies that these representations and thus the human and machine conceptions of each other and of the unity they form, while not necessarily being identical, are congruent to each other and compatible with the common higher-level goal, allowing the capitalization on individual strengths and the goal-oriented fusion of information available to each partner. These individual conceptions must include the recursive connections and coupled control loops with potentially both feedforward and feedback components (cf., Section \ref{subsec:intro_sharedcontrol}). In addition, symbiotic understanding on the decision level permits the shift of the desired level of automation (cf., Section \ref{subsec:intro_dev_HMS}), that is, the extent to which the machine acts autonomously. Symbiotic understanding also enables a flexible initiation and realization of various levels of automation, thus allowing for an adequate individual prediction of the overall human-machine system behavior (green H/M blocks), after which the negotiation of the chosen decision takes place (yellow H/M blocks, see e.g. \cite{Rothfuss.2021} for first realizations of human-machine negotiation on the decision level). These decisions are passed down to the lower action level as individual, yet mutually congruent goals.

\subsubsection{Interaction on the Action Level}

\begin{figure}[t]
	\centering
\setlength{\mywidth}{23em}

\setlength{\myheight}{3em}

\definecolor{environBlue}{RGB}{0,114,154}
    
\tikzset{
}

\tikzstyle{symbblock} = [rectangle, fill = KITblue!30, draw,
minimum width = \mywidth, minimum height = \myheight]
\tikzstyle{symbblock2} = [rectangle,fill = KITmaygreen!40, draw,
minimum width = \mywidth, minimum height = \myheight]
\tikzstyle{symbblock3} = [rectangle,fill = KITyellow!40, draw,
minimum width = \mywidth, minimum height = 2.5*\myheight]
\tikzstyle{hmrep} = [rectangle, fill = KITblue!90,draw,
minimum width = 3em, minimum height = 0.8\myheight]
\tikzstyle{perc} = [rectangle, fill = purple!10, draw ,
minimum width = 5em, minimum height = \myheight]
\tikzstyle{partner} = [rectangle, fill = gray!20, draw = gray!50!black!,
minimum width = 8.5em, minimum height = 7*\myheight, rounded corners]
\tikzstyle{arrow} = [-{Latex[length=2mm]}]
\tikzstyle{level} = [rectangle, draw = environBlue, line width = 1.5, dashed,
minimum width = 50em, minimum height = 10*\myheight]

\tikzstyle{CC} = [rectangle, fill = environBlue!50, draw,
minimum width = 15em, minimum height = 0.6*\myheight]

\tikzstyle{environ} = [signal, signal to = west and east, inner xsep=0mm, fill = environBlue!50, draw,
minimum width = 55em, minimum height = \myheight]

\pgfdeclarelayer{background}
\pgfsetlayers{background,main}

\begin{tikzpicture}[font=\large,scale=0.65, transform shape]

\node at (0,0)(ursprung){};


\node[symbblock] at (ursprung)(understanding){Symbiotic Understanding};
\node[hmrep, left = 8em of understanding.center](H_under){{\color{white}\textbf{H}}};
\node[hmrep, right = 8em of understanding.center](M_under){{\color{white}\textbf{M}}};

\node[symbblock2, below =  2em of understanding](prediction){Symbiotic Prediction};
\node[hmrep, fill = KITmaygreen!90, left = 8em of prediction.center](H_pred){{\color{white}\textbf{H}}};
\node[hmrep, fill = KITmaygreen!90, right = 8em of prediction.center](M_pred){{\color{white}\textbf{M}}};

\node[symbblock3, below =  2em of prediction](decision){};
\node[anchor = north, yshift = -1em] at (decision.north)(label_exec){Symbiotic Execution};
\node[hmrep, fill = KITyellow!60!orange!90!, left = 8em of label_exec.center](H_neg){{\color{white}\textbf{H}}};
\node[hmrep, fill =  KITyellow!60!orange!90!, right = 8em of label_exec.center](M_neg){{\color{white}\textbf{M}}};

\node[perc, above = of H_under](H_perc){Percept};
\node[perc, above = of M_under](M_perc){Percept};

\node[level, above = 0.5em of prediction.center, anchor= center](decisionlevel){};
\node[anchor = north east, yshift = -1em, xshift = -1em] at (decisionlevel.north east){\bfseries\Large{Action Level}};

\begin{pgfonlayer}{background}
\node[partner,above = 3em of H_pred.center, anchor = center]{};
\node[partner,above = 3em of M_pred.center, anchor = center]{};
\end{pgfonlayer}

\node[CC, below = 1.5em of decision.center](coupling){{\color{black}Common coupling}};

\node[environ, below = 6em of coupling](environment){\textbf{\color{black}Environment}};

\draw[arrow](H_perc) --  node [left]{signals}  (H_under);
\draw[arrow](H_under)--(H_pred);
\draw[arrow](H_pred) -- (H_neg);
\draw[arrow](H_neg) -- ++(0,-2em) -| (coupling.160);
\draw[fill] (H_neg)+(0,-2em) circle (2pt);
\draw[arrow](H_neg) -- ++(0,-2em)--++(-6em,0) |- (H_under.-160);
\draw[arrow](H_under)+(-6em,13em) |- node[above, pos = 0]{human decision} (H_under.160); 

\draw[arrow](M_perc) -- node [left]{signals} (M_under);
\draw[arrow](M_under)--(M_pred);
\draw[arrow](M_pred) -- (M_neg);
\draw[arrow](M_neg) -- ++(0,-2em) -| (coupling.20);
\draw[fill] (M_neg)+(0,-2em) circle (2pt);
\draw[arrow](M_neg) -- ++(0,-2em)--++(6em,0) |- (M_under.-20);

\draw[arrow](M_under)+(6em,13em) |- node[above, pos = 0]{machine decision}(M_under.20);
\draw[arrow](understanding)+(0,13em) -- node[above, pos = 0]{symbiotic decision} (understanding);
\draw[arrow](understanding) -- (prediction);
\draw[arrow](prediction) -- (decision);
\draw[arrow](decision) -- node[right, pos =0.5]{symbiotic action}(environment);

\draw[arrow](coupling)--++(0,-3em) -- ++ (-22em,0) -- ++(0, 34em);
\draw[fill] (coupling)+(0,-3em) circle (2pt);
\draw[arrow](coupling)--++(0,-3em) -- ++ (22em,0) -- ++(0, 34em);

\draw[arrow](environment.west)-|node[right, pos = 0.7]{feedback}++(-3em, 39.5em);
\draw[arrow](environment.east)-|node[left,pos = 0.7]{feedback}++(3em, 39.5em);

\coordinate(H_perc_in1) at ($(H_perc.west)+(0,-0.7em)$);
\draw[arrow](H_perc)+(-12.5em,-0.7em) -- (H_perc_in1);
\draw[fill] (H_perc)+(-12.5em,-0.7em) circle (2pt);

\coordinate(H_perc_in2) at ($(H_perc.west)+(0,0.7em)$);
\draw[arrow](H_perc)+(-21em,0.7em) -- (H_perc_in2);
\draw[fill] (H_perc)+(-21em,0.7em) circle (2pt);

\coordinate(M_perc_in1) at ($(M_perc.east)+(0,-0.7em)$);
\draw[arrow](M_perc)+(12.5em,-0.7em) -- (M_perc_in1);
\draw[fill] (M_perc)+(12.5em,-0.7em) circle (2pt);

\coordinate(M_perc_in2) at ($(M_perc.east)+(0,0.7em)$);
\draw[arrow](M_perc)+(21em,0.7em) -- (M_perc_in2);
\draw[fill] (M_perc)+(21em,0.7em) circle (2pt);


\end{tikzpicture}
	\caption{Interaction dimension at the action level for human-machine symbiosis: Symbiotic understanding is the foundation of an overall system behavior prediction and thus of a highly skilled symbiotic action execution (joint action). Symbiotic understanding is based on the perception of environmental feedback as continuous signals, the individual decisions and the symbiotic decision from the upper decision level. Human (H) and machine (M) blocks denote individual representations with respect to the corresponding elements.}
	\label{fig:HMS_interaction_action}
\end{figure}
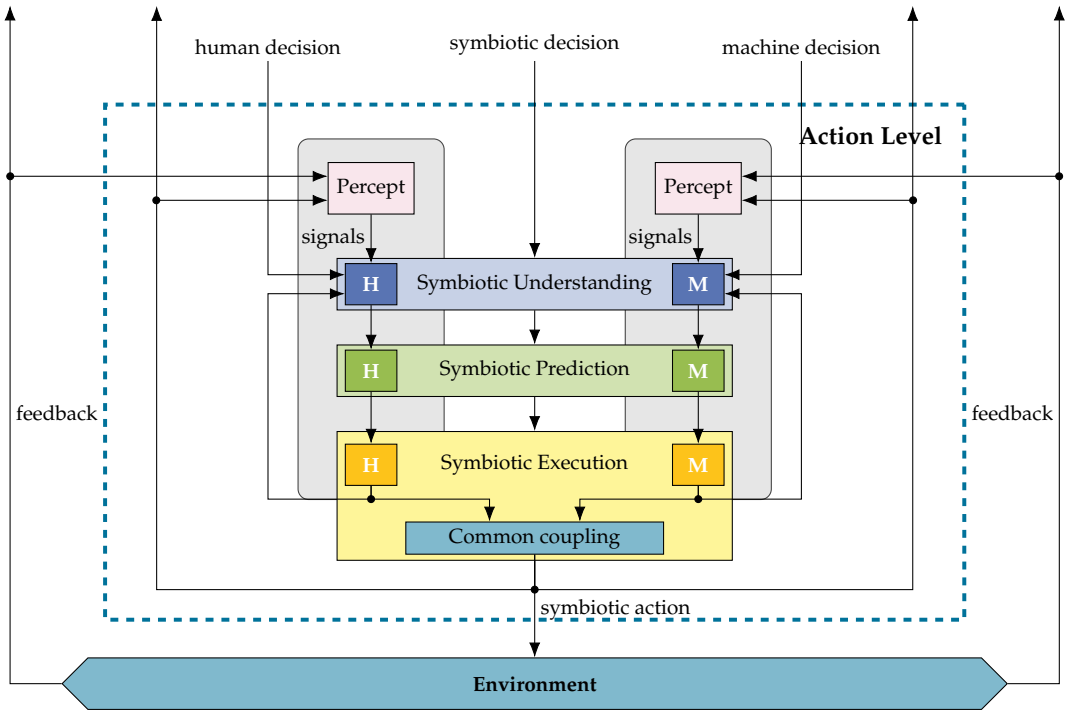

Figure \ref{fig:HMS_interaction_action} depicts the interaction dimension on the action level. On this level, we similarly have individual perception and representations or models of each partner as on the decision level. However, the action level includes the common coupling due to the simultaneous action execution at this lower haptic level. The output of the yellow H-block is the internal representation of the optimal human action, which is combined with perceived feedback from the environment and the information from the decision level for a symbiotic understanding at this level. An analogous procedure takes place on the machine side. Symbiotic understanding at this action level also includes an efficient intention recognition based on the haptic feedback, which also allows for the perception and emergence of a desired level of automation. We note that the level of automation at the action level has to be described with a continuous-valued approach accordingly to the continuous interaction \cite{Braun.2019}, contrary to its counterpart at the decision level where discrete-valued shifts (as defined by \cite{Endsley.1999} or \cite{Sheridan.1978}) are also conceivable (see, e.g., \cite{Chiou.2021}).
We hypothesize that symbiotic understanding, prediction, and execution can be merged into a consistent symbiotic interaction model which allows the continuous description of the overall symbiotic human-machine system behavior at a particular task level. The symbiotic interaction, which yields highly skilled behavior, results in a better performance compared to individual actions by both partners when solving a task. This performance dimension will be elaborated in the next section. 

\subsection{Performance Dimension}\label{subsec:mult_perf_dim}

Symbiosis in human-machine systems constitutes the most effective interaction form which allows for the emergence of synergies. Therefore, we conjecture that the performance of the overall system goes beyond the possible performance each partner could deliver on his own. It also allows for various kinds of interaction, including shared control, where the human or the robot could execute the task individually \cite{Abbink.2018}. The evaluation variables of the performance of a system depend on the specific use case and can therefore vary considerably (e.g., \cite{Gray.1993,Khademian.2011,Uhl.2021}). Generally speaking, however, there are three different aspects for evaluating performance: Efficiency as the ratio between work and time, the number of application errors, as well as the quality of the work result \cite{Nielsen.1994}. In order to enable an increase in performance, the technical system must be adapted to the application (e.g., \cite{Hoelz.2021}) and especially to the user \cite{Matthiesen.2018,Vedder.2005}. Performance is also an aspect of usability \cite{Germann.2019}, which according to \cite{.2018} is the extent to which a technical system can be used effectively, efficiently, and satisfactorily by a defined user in its relevant application. Therefore, a high usability is a necessary condition for symbiotic human-machine interaction to emerge and systematic approaches for its analysis are needed, e.g. the Usability Study Evaluation Process of \cite{Germann.2021}.

\begin{figure}[t]
	\centering
\setlength{\mywidth}{23em}

\setlength{\myheight}{3em}

\definecolor{environBlue}{RGB}{0,114,154}
\definecolor{interGreen}{RGB}{229,232,197}
    
\tikzset{
}

\tikzstyle{symbblock} = [rectangle, fill = interGreen, draw,
minimum width = \mywidth, minimum height = \myheight]
\tikzstyle{perform} = [rectangle,fill = gray!10, draw,
minimum width = 0.8*\mywidth, minimum height = 4*\myheight]
\tikzstyle{usability} = [rectangle, rounded corners, fill = gray!10, draw,
minimum width = 0.6*\mywidth, minimum height = 5*\myheight]
\tikzstyle{aspects} = [rectangle,fill = white!40, draw,
minimum width = 0.7*\mywidth, minimum height = 0.8*\myheight]
\tikzstyle{Uaspects} = [rectangle,fill = white!40, draw,
minimum width = 0.5*\mywidth, minimum height = 0.8*\myheight]
\tikzstyle{hmrep} = [rectangle, fill = KITgreen!80!black!80,draw,
minimum width = 3em, minimum height = 0.8\myheight]
\tikzstyle{perc} = [rectangle, fill = purple!10, draw ,
minimum width = 5em, minimum height = \myheight]
\tikzstyle{partner} = [rectangle, fill = gray!20, draw = gray!50!black!,
minimum width = 8em, minimum height = 2*\myheight, rounded corners]
\tikzstyle{arrow} = [-{Latex[length=2mm]}]
\tikzstyle{level} = [rectangle, draw = environBlue, line width = 1.5, dashed,
minimum width = 55em, minimum height = 10*\myheight]

\tikzstyle{CC} = [rectangle, fill = environBlue, draw,
minimum width = 15em, minimum height = 0.6*\myheight]

\tikzstyle{environ} = [chamfered rectangle,chamfered rectangle xsep=5cm, chamfered rectangle angle=20, inner xsep = 10em,
inner ysep=8em, fill = environBlue!50, draw, minimum height = 4*\myheight]

\pgfdeclarelayer{background}
\pgfsetlayers{background,main}

\begin{tikzpicture}[font=\large,scale=0.65, transform shape]

\node at (0,0)(ursprung){};


\node[symbblock] at (ursprung)(understanding){Symbiotic Interaction};
\node[hmrep, left = 8em of understanding.center](H_under){{\color{white}\textbf{H}}};
\node[hmrep, right = 8em of understanding.center](M_under){{\color{white}\textbf{M}}};

\node[level, below = 8em of understanding.center, anchor= center](task){};
\node[anchor = north west, yshift = -1em, xshift = 1em] at (task.north west){\bfseries\Large{Task}};

\begin{pgfonlayer}{background}
\node[partner] at (H_under.center){};
\node[partner] at (M_under.center){};
\end{pgfonlayer}

\node[environ, below = 6em of understanding](environment){};
\node[anchor = north, yshift = -1em] at (environment.north){\textbf{\Large\color{black}Environment}};

\node[perform, below = 4em of environment.north](performance){};
\node[anchor = north, yshift = -0.5em, xshift= 4.5em] at (performance.north west){\textbf{Performance}};
\node[aspects, above = 1em of performance.center, anchor = south](quality){Quality of the work result};
\node[aspects, below = 0.5em of quality](effic){Efficiency};
\node[aspects, below = 0.5em of effic](errors){Number of application errors};

\node[usability, above left = 3em and 22em of performance.center,anchor = center](usability){};
\node[anchor = north, yshift = -0.5em, xshift= 6em] at (usability.north west){\textbf{Usability aspects}};
\node[Uaspects, above = 2.3em of usability.center, anchor = south](perform){\textbf{Performance}};
\node[Uaspects, below = 0.5em of perform](ergo){Ergonomics};
\node[Uaspects, below = 0.5em of ergo](vibr){Vibration comfort};
\node[Uaspects, below = 0.5em of vibr](etc){...};

\draw[arrow](understanding)+(0,8em) -- node[above, pos = 0]{complex task} (understanding);
\draw[arrow](understanding) -- node[left, pos =0.5]{symbiotic action}(environment);

\draw[arrow](performance) -- ++(-12em,0) |- (perform.east);
\draw[arrow](usability) |- node[left, pos =0.3]{feedback}(H_under.west);

\coordinate(split) at ($(understanding)+(11em,-5em)$);
\draw[dashed, line width =1.0](understanding.south)+(4em,0)  |- (split);
\draw[arrow,dashed, line width =1.0](split)  |- node[above,pos = 0.7]{\textbf{+}} (quality.east);
\draw[arrow,dashed, line width =1.0](split)  |- node[above,pos = 0.7]{\textbf{+}} (effic.east);
\draw[arrow,dashed, line width =1.0](split)  |- node[above,pos = 0.7]{\textbf{-}} (errors.east);

\draw[arrow](environment.east)--+(4em,0)  |- node[right,pos = 0.42]{feedback} (M_under.east);

\end{tikzpicture}
	\caption{Influence model of the symbiotic interaction of human (H) and machine (M) on the quality of the performance, which is an aspect of usability. Symbiotic interaction leads to a high quality of the work result, high efficiency, and a low number of application errors.}
	\label{fig:HMS_performance_dimension}
\end{figure}
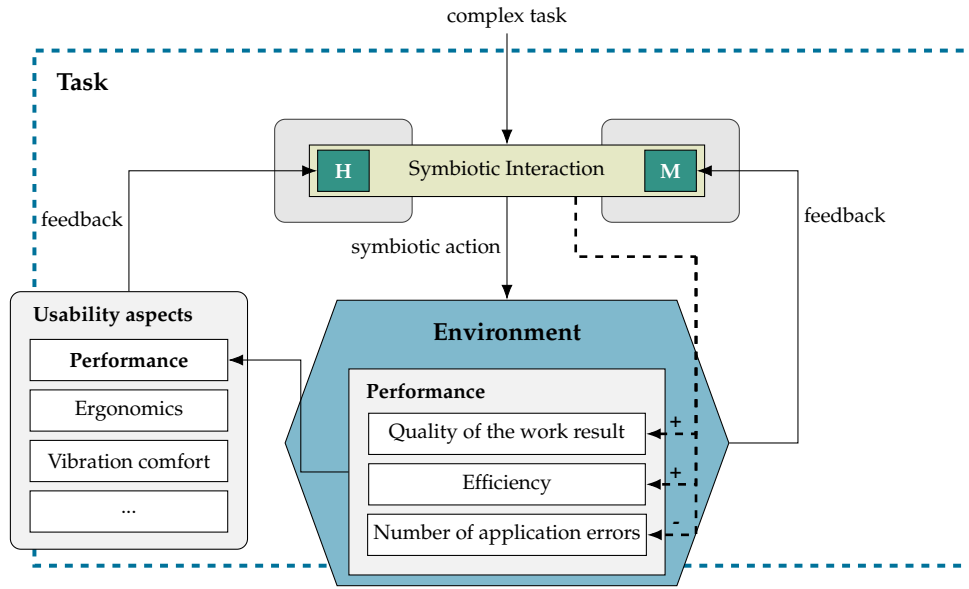

As described in the interaction dimension (cf., Section \ref{subsec:mult_inter_dim}), the symbiotic interaction between human and machine is characterized by mutual understanding which involves intention recognition as well as a highly coordinated joint execution of the task. The impact of this interaction on the performance (see Figure \ref{fig:HMS_performance_dimension}) is described in the following:
The two-sided understanding within a symbiotic human-machine interaction enables not only the human to recognize the intention of the machine for intuitive operation, but also the machine to recognize the intention of the human so that both partners can optimally and swiftly adapt to each other. This reduces the number of application errors that occur and simultaneously increases the efficiency of both partners. In a human-machine interaction, a particularly high increase in performance is achieved when the weakness of one partner can be compensated by the strength of the other \cite{Abbink.2018}. An example of this is a situation where either partner perceives task-relevant information which is unavailable to the other. Therefore, in a human-machine symbiosis, at each task level described in Section \ref{subsec:mult_task_dim}, the “stronger” partner takes the lead. This corresponds to the hybrid approach to switching the level of automation mentioned for the interaction dimension in Section \ref{subsec:mult_inter_dim}. We note that the symbiotic interaction also permits different levels of automation between task levels. For example, if it is favorable for the task, then the human can take the lead at the decision level, while the machine takes the lead at the action level. As soon as the subtask is completed, the machine may take the initiative to continue with another subtask, thus obtaining the lead at the decision level. Besides increasing efficiency, the individual strengths of the partners lead to an improved quality of the work result.

In addition to the high performance, long-term effects also are characteristic of symbiotic human-machine systems. Thus, the short-term increase in efficiency leads to energy savings in the medium term and due to the optimal interaction to a minimum risk of physical harm for the human being and less wear on the machine in the long term. Other aspects of usability are also necessary for symbiotic systems, such as ergonomics and vibration comfort. Indeed, for instance, bad ergonomics is reported in currently developed exoskeletons \cite{Fox.2020}. Therefore, it becomes apparent that the user's perception and experience play a decisive role in the consideration of usability in addition to performance \cite{Germann.2019}. The human-sided experience dimension will be described in more detail in the next section.

\subsection{Experience Dimension}

\begin{figure}[t]
	\centering
\setlength{\mywidth}{23em}

\setlength{\myheight}{3em}

\definecolor{environBlue}{RGB}{0,114,154}
\definecolor{interGreen}{RGB}{229,232,197}
    
\tikzset{
}

\tikzstyle{symbblock} = [rectangle, fill =interGreen, draw,
minimum width = \mywidth, minimum height = \myheight]
\tikzstyle{perform} = [rectangle,fill = KITmaygreen!40, draw,
minimum width = \mywidth, minimum height = \myheight]
\tikzstyle{aspects} = [rectangle,fill = KITyellow!40, draw,
minimum width = \mywidth, minimum height = 2.5*\myheight]
\tikzstyle{hmrep} = [rectangle, fill = KITgreen!80!black!80,draw,
minimum width = 3em, minimum height = 0.8\myheight]
\tikzstyle{perc} = [rectangle, fill = purple!10, draw ,
minimum width = 5em, minimum height = \myheight]
\tikzstyle{partner} = [rectangle, fill = gray!20, draw = gray!50!black!,
minimum width = 8em, minimum height = 2*\myheight, rounded corners]
\tikzstyle{partner_H} = [rectangle, fill = red!20, draw = red!50!black!,
minimum width = 9em, minimum height = 4.5*\myheight, rounded corners]

\tikzstyle{arrow} = [-{Latex[length=2mm]}]
\tikzstyle{level} = [rectangle, draw = environBlue, line width = 1.5, dashed,
minimum width = 55em, minimum height = 10*\myheight]

\tikzstyle{CC} = [rectangle, fill = environBlue!50, draw,
minimum width = 15em, minimum height = 0.6*\myheight]

\tikzstyle{environ} = [signal, signal to = west and east, inner xsep=0mm, fill = environBlue!50, draw,
minimum width = 40em, minimum height = \myheight]

\tikzstyle{cloud1} = [cloud,cloud puffs = 15, cloud puff arc = 110, line width = 0.7,
draw,minimum width = 16em, minimum height = 6em, fill = red!10, draw=red,opacity = 0.5]
\tikzstyle{cloud2} = [cloud,cloud puffs = 19, cloud puff arc = 80, line width = 0.7,
draw,minimum width = 13em, minimum height = 3em, fill = red!10, draw=red,opacity = 0.5]
\tikzstyle{cloud3} = [cloud,cloud puffs = 10, cloud puff arc = 130, line width = 0.7,
draw,minimum width = 5em, minimum height = 4em, fill = red!10, draw=red,opacity = 0.5]
\tikzstyle{cloud4} = [cloud,cloud puffs = 23, cloud puff arc = 90,line width = 0.7,
draw,minimum width = 16em, minimum height = 5em, fill = red!10,draw=red, opacity = 0.4]

\tikzstyle{kringel1} = [circle,draw = red!70, line width = 0.7, dashed,inner sep = 0.36em]
\tikzstyle{kringel2} = [circle,draw = red!70, line width = 0.7,dashed,inner sep = 0.25em]
\tikzstyle{kringel3} = [circle,draw = red!70, line width = 0.7, dashed,inner sep = 0.15em]
\tikzstyle{kringel4} = [circle,draw = red!70, line width = 0.7, dashed,inner sep = 0.13em]

\pgfdeclarelayer{background}
\pgfdeclarelayer{foreground}
\pgfsetlayers{background,main,foreground}

\begin{tikzpicture}[font=\large,scale=0.65, transform shape]

\node at (0,0)(ursprung){};


\node[symbblock] at (ursprung)(understanding){};
\node[hmrep, left = 8em of understanding.center](H_under){{\color{white}\textbf{H}}};
\begin{pgfonlayer}{foreground}
\node[hmrep, right = 8em of understanding.center](M_under){{\color{white}\textbf{M}}};
\end{pgfonlayer}

\node[level, below = 2.5em of understanding.center, anchor= center](task){};
\node[anchor = north west, yshift = -1em, xshift = 1em] at (task.north west)(task_caption){\bfseries\Large{Task}};

\begin{pgfonlayer}{background}
\node[partner_H] at (H_under.center)(human){};
\node[partner] at (M_under.center)(machine){};
\end{pgfonlayer}
\node[anchor = center, yshift = -1.5em, font = \normalsize] at (human.north)(flow){\bfseries \color{KITred} Flow experience};
\node[below = 1em of flow,font = \normalsize](accep){\bfseries \color{KITred} Acceptance};
\node[below = 4em of accep,font = \normalsize](embod){ \bfseries \color{KITred} Embodiment};
\node[below = 1em of embod,font = \normalsize](SOA){\bfseries \color{KITred} Sense of agency};

\node[CC, below = 11em of understanding.south](coupling){{\color{black}Common coupling}};

\node[environ, below = 4em of coupling](environment){\textbf{\color{black}Environment}};

\draw[arrow](understanding)+(0,14em) -- node[above, pos = 0]{complex task} (understanding);

\draw[arrow, dashed, KITred](M_under)+(0,-9em) -| node[above, pos = 0.25]{\normalsize{control of machine effects}}($(H_under)+(4em,-11em)$);


\begin{pgfonlayer}{background}
\node[cloud1, right = 5em of task_caption.center,anchor = center](wolk1){};
\node[anchor = south east, right = 5em of wolk1.west,font=\normalsize, align=left,KITred]{Optimal fit of \\ challenge \& capacities};
\node[kringel1, below right = 3em and 2em of wolk1.center](kr1_1){};
\node[kringel2, below right = 0.5em and 1.2em of kr1_1.center](kr1_2){};
\node[kringel3, below right = 0.5em and 1.2em of kr1_2.center](kr1_3){};

\node[cloud4, below right = 9em and 2em of understanding.center,anchor = center](wolk4){};
\node[kringel2, above left = 2.2em and 3.6em of wolk4.center](kr4_1){};
\node[kringel3, above left = 0.2em and 1.5em of kr4_1.center](kr4_2){};
\node[kringel4, above left = 0.2em and 1.5em of kr4_2.center](kr4_3){};

\end{pgfonlayer}

\node[cloud2] at (understanding.center)(wolk2){};
\node[kringel2, above left = 1.75em and 1.5em of wolk2.center](kr2_1){};
\node[kringel3, above left = 0.15em and 1.75em of kr2_1.center](kr2_2){};
\node[kringel4, above left = 0.15em and 1.75em of kr2_2.center](kr2_3){};

\node[cloud3] at (M_under)(wolk3){};
\node[kringel1, below left = 1.5em and 3em of wolk3.center](kr3_1){};
\node[kringel2, below left = 0.1em and 3.6em of kr3_1.center](kr3_2){};
\node[kringel3, below left = 0.1em and 3.6em of kr3_2.center](kr3_3){};
\node[kringel4, below left = 0.1em and 3.6em of kr3_3.center](kr3_4){};

\node[] at (ursprung){Symbiotic Interaction};

\draw[arrow](H_under) -- ++(0,-11em) -| (coupling.160);
\draw[arrow](M_under) -- ++(0,-11em) -| (coupling.20);

\draw[arrow](coupling) -- node[right, pos =0.5]{symbiotic action}(environment);

\draw[arrow](environment.east)--+(3em,0)  |- node[left,pos = 0.42]{feedback} (M_under.east);
\draw[arrow](environment.west)--+(-3em,0)  |- node[right,pos = 0.42]{feedback} (H_under.west);

\end{tikzpicture}
	\caption{Experience dimension for the symbiotic interaction of human (H) and machine (M): A symbiotic system may change the user’s experience resulting (1) in some flow experience due to the fact that the task challenges fit optimally to the user’s capacities, (2) in a high acceptance for the machine use in terms of the symbiotic interaction, (3) in an embodiment of (parts of) the machine, and (4) in a change of the sense of agency as the user may perceive control not just for own, but also for effects elicited by the machine.}
	\label{fig:HMS_experience_dimension}
\end{figure}
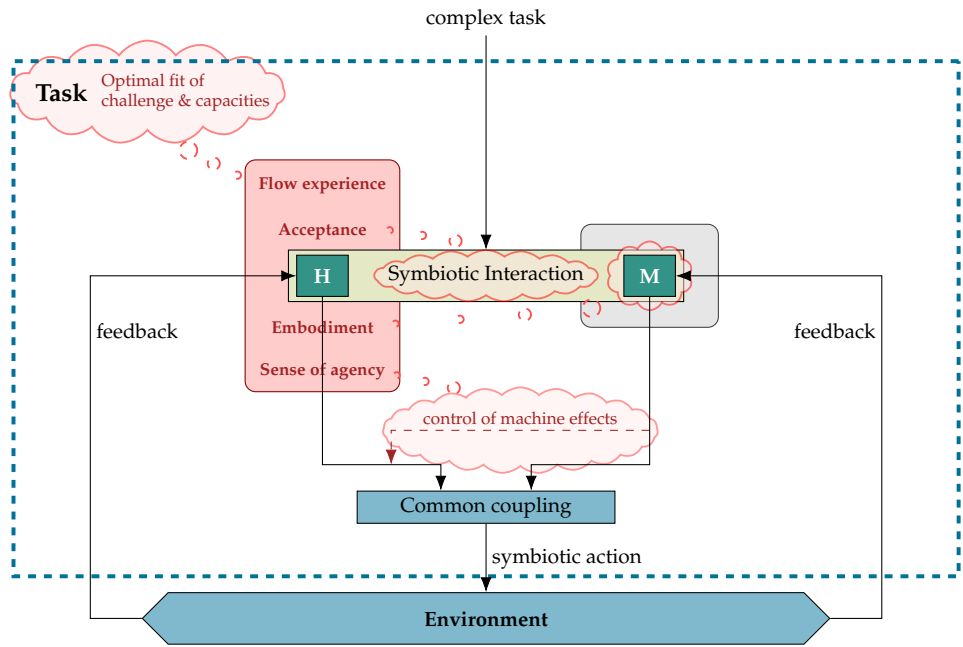

Optimization in the sense of a better overall performance has been a common focus in the development of human-machine systems (cf., Section \ref{subsec:mult_perf_dim}). In the multivariate perspective for symbiotic systems, the first three dimensions, that is task, interaction, and performance dimension, refer to the whole human-machine system. The fourth dimension is somewhat special, as the experience dimension is restricted to the human partner of the interaction system. Here, we refer to four constructs of human experience that may be of special interest: Flow experience, acceptance, sense of agency, and embodiment (see Figure \ref{fig:HMS_experience_dimension}).

\subsubsection{Flow Experience}

Highest quality of the human’s experience is expected in the case of a high balance between perceived challenge and perceived skill. This highest overall quality of subjective experience is referred to as a flow state in which the person feels cognitively efficient, motivated, happy, and is acting with full involvement (e.g., \cite{Csikszentmihalyi.1989,Moneta.1996}, for a meta-analysis, see \cite{Fong.2015}). Flow states have been observed to have a greater influence on the quality of our experiences irrespective of whether we are working or leisuring \cite{Csikszentmihalyi.1989}, to predict performance \cite{Engeser.2008}, and they may optimize energy expenditure \cite{Peifer.2020}. Consequently, when thinking about optimizing the interaction of human and machine, the experience of flow while conjointly performing a challenging task is the final goal, from the humans’ point of view.

In computer-mediated environments, a model has been proposed separating person (trait and state aspects) from artifact (e.g., computer, tool, toy, or software aspects) and from task components (PAT), as well as interactions of these components (e.g., person-task interaction in terms of the challenge-skill balance, an immediate feedback, perceived control, or clear goals) as antecedents of flow \cite{Finneran.2003}. Interestingly, these different PAT components have some main similarities to our multivariate perspective on human-machine interaction. For example, the main importance of the challenge-skill balance of the person-task interaction directly links to our described task dimension (cf., Section \ref{subsec:mult_task_dim}) in terms of an optimal balance between task affordances and the capacities and skills perceived by the user. Thus, if the task challenges fit optimally to the perceived capacities of the user, we would expect the user to experience some flow. Moreover, the capacities of the user may be perceived higher in the case of an optimal coupling with the machine. 

Flow can be operationalized using unidimensional or multidimensional methods \cite{Hoffman.2009}. Unidimensional methods ask participants to report flow experience using only up to three items, referring to a brief narrative description of flow experience which had been presented to the participants beforehand \cite{Hoffman.2009}. Multidimensional measures employ validated and commonly used scales relating to several constituent constructs related to flow (e.g., Flow Kurzskala FKS, \cite{Rheinberg.2003}; Dispositional Flow Scale 2 DFS-2, Flow State Scale 2 FSS-2, and their short versions, \cite{Jackson.2008}; short and core flow scale, \cite{Martin.2008}; WOrk-reLated Flow inventory WOLF, \cite{Bakker.2008}). Moreover, an experience sampling methodology can be employed, asking participants repeatedly to indicate their current flow state while being engaged in a certain activity \cite{Fullagar.2009}.

Apart from the human’s experience itself (i.e., in terms of flow experience), the optimization of the interaction of the human and the machine is of main interest \cite{Berberian.2019}. A highly adaptive partnership would be characterized by mutual understanding and trust, leading to an effortless interaction where a new joint agent identity, a kind of "we" might emerge \cite{Jenkins.2021}. In terms of the human’s experience, this relates to the acceptance of human-machine systems (for a meta-analysis, see \cite{King.2006}), to sense of agency (e.g., \cite{Ruess.2017,Ruess.2018}, as well as embodiment (for a review, see \cite{Schettler.2019}). 

\subsubsection{Acceptance}

An important basis for a successful human-machine interaction is the acceptance of the technology. In the Technology Acceptance Model (TAM; for a meta-analysis, see \cite{King.2006}; and the latest version TAM3, \cite{Venkatesh.2008}), it is proposed that behavioral intention influences behavior. The behavioral intention itself is predicted by the perceived ease of use, the perceived usefulness, attitudes, subjective norms and perceived behavioral control (e.g., \cite{King.2006}). Thus, acceptance may relate to the user’s experience of the symbiotic interaction with the machine, as mentioned in the interaction dimension (cf., Section \ref{subsec:mult_inter_dim}): High acceptance of the machine should result in interacting with the machine in a symbiotic manner. The other way round, if the interaction with the machine is symbiotic, high acceptance ratings would be expected. Acceptance has been operationalized employing scales that refer to existing technologies (e.g., \cite{Chintalapati.2017,vanderLaan.1997,Vantrepotte.2021}) or to future technologies described in scenario texts \cite{ReinaresLara.2018,Toft.2014}. Additionally, Cognitive-Affective Mapping, a recently developed method that bridges the gap between quantitative and qualitative research traditions \cite{Reuter.2021,Thagard.2010}, may be employed in terms of anticipatory acceptance prediction of emerging technologies.

\subsubsection{Sense of Agency}

The feeling of controlling and causing changes in the environment is called sense of agency (e.g., \cite{Haggard.2012}) and suggested to be an important cue in optimizing human-machine interaction \cite{Berberian.2019}. It implies that the human senses agency for environmental changes caused by the symbiotic human-machine system. Thus, this sense of agency relates to the environment and to the performance dimension (cf., Section \ref{subsec:mult_perf_dim}). If the interaction is symbiotic, a high sense of agency, also for positive performance effects caused by the machine, would be expected. Sense of agency is said to influence cooperativeness, team performance, and fluency \cite{Dagioglou.2021}. Consequently, a high sense of agency should also increase the performance itself. 

Sense of agency can be operationalized both explicitly as well as implicitly. Explicit measures use one single item asking on a Likert-scale for the amount of control (e.g., \cite{Wen.2019}) or employ some validated and standardized scales (e.g., Sense of Agency Scale SoAS, \cite{Tapal.2017}; Sense of Agency Scale for Heavy Machine Operation SoAS-HMO, \cite{Raima.2020}; or Sense of Agency Rating Scale SOARS, \cite{Polito.2013}). Implicit measures assess sense of agency in terms of intentional binding, a biased time perception in action contexts \cite{Haggard.2012,Humphreys.2009,Ruess.2020}, or sensory attenuation, a biased intensity perception in action contexts (e.g., \cite{Hughes.2013}). Implicit and explicit measures may relate to different processes of sense of agency (e.g., \cite{Dewey.2014,Imaizumi.2019}). Thus, for optimizing human-machine interaction on both action and decision levels of a task and also towards an intuitive manner, implicit and explicit sense of agency should be assessed. 

\subsubsection{Embodiment}

Finally, symbiotic systems may result in a changed experience of the own body. This refers to the embodiment, that is, the feeling that an object is integrated into the body scheme (for a review, see \cite{Schettler.2019}). For example, tool-use changed the estimated forearm length of the participants \cite{Sposito.2012}, some embodiment has been observed for robotic hands \cite{Alimardani.2013}, or, in a more anecdotal way, car drivers often report to even sense pain in case they witness a car crash \cite{Sheller.2004}. Symbiosis in physically coupled human-machine systems is expected to change the embodiment by incorporating at least parts of the machine into the own body scheme. 

Embodiment can be operationalized either explicitly, employing questionnaires (e.g., \cite{Hoffmann.2018}, or implicitly, in terms of the rubber hand illusion \cite{Botvinick.1998}. In this illusion, participants see stimulations exerted to a rubber hand and perceive this stimulation as if it was exerted to their own hand \cite{Botvinick.1998}. 

Currently, the relation of all four variables of the experience dimension to each other (i.e., flow experience, acceptance, sense of agency, and embodiment) is unclear. Elaborating on the interplay of these variables will increase our understanding and thus the accuracy to predict human experience in human-machine interactions and to develop future interaction scenarios in terms of symbiotic systems.

\subsection{Limitations and Future Perspectives}

We see the multivariate perspective to provide a holistic, far-end, and long-term optimization of physically coupled human-machine systems as it takes into account various disciplines integrated in the four dimensions. Positive effects may be achieved both in multiple  dimensions and in the long-run by asking for the optimal trade-off between different dimensions instead of singularly focused optimization in one dimension. For example, our multivariate perspective asks for a trade-off between closely coupled physical interaction and interacting at all levels of a common task, or for a trade-off between performative optimization and improving experience variables at the same time. Thereby, some negative side-effects of the development of human-machine interaction may be prevented, presumably being especially important in the long term. Other examples may be energy efficiency, wearout of the machine, or dependency and loss of capacity of the human, but also ethical or sustainability aspects. So far, negative side-effects or long-term negative effects of human-machine interaction were mainly evaluated and addressed separately from the development of human-machine interaction. Our multivariate perspective also allows to consider negative (side-)effects and to find an optimal balance of different challenges in the development of human-machine interaction. 

We note that the multivariate perspective is a framework which, in the future, can be specified in more detail: For example, the task dimension may be amplified separating not just the action and decision level, but an operational or the decomposition level. Similarly, the interaction dimension in terms of the feedback control loops and a recursive intention recognition with real-time updating of human and machine representations may become more sophisticated. For the performance dimension, so far, we refer to two main advantages of symbiotic systems (i.e., intention recognition and complementary strengths) which need to be specified and complemented. For the experience dimension, additionally to the suggested variables (i.e., flow experience, acceptance, sense of agency, and embodiment), variables like trust or ethical aspects, may be addressed in the future.

\section{Conclusion}

In the field of physically coupled human-machine systems, the idea of a “true collaboration” or “symbiosis” has been recently mentioned in the literature. Here, we propose a multivariate perspective as necessary aspects to consider in the future development of physically coupled human-machine systems, which consists of four dimensions: Task, interaction, performance, and experience dimension. First, human and machine form a unity as they (physically) work together and complement each other to accomplish a common task. The interaction between both partners as well as the utilization of their individual strengths may occur on different levels of a common task, on which each of the partners may take the lead depending on the individual strengths (task dimension). Both partners recognize the intention of the other in the sense that they are able to predict each other’s actions; they continuously adapt to each other and complement their individual strengths. However, the human and machine partners are considered equals with respect to the task and each of them may take the initiative (interaction dimension). The resulting symbiotic state is characterized by a higher overall performance (performance dimension). In addition, the users experience some kind of flow, higher acceptance of the technology, some sense of agency for effects elicited by the machine, as well as embodiment for parts of the machine devices (experience dimension). This multivariate perspective is  generic, flexible and applicable as a shared, interdisciplinary terminology towards an optimal interaction between humans and machines in various application scenarios.

\articlebodyend


\begingroup
\raggedright
\titleformat*{\section}{\bfseries\Large\centering\MakeUppercase}
\bibliographystyle{apacite}
\bibliography{HMS_bibliography}

\begin{thebibliography}{}

\bibitem [\protect \citeauthoryear {%
Abbink%
\ \protect \BOthers {.}}{%
Abbink%
\ \protect \BOthers {.}}{%
{\protect \APACyear {2018}}%
}]{%
Abbink.2018}
\APACinsertmetastar {%
Abbink.2018}%
\begin{APACrefauthors}%
Abbink, D\BPBI A.%
, Carlson, T.%
, Mulder, M.%
, de Winter, J\BPBI C\BPBI F.%
, Aminravan, F.%
, Gibo, T\BPBI L.%
\BCBL {}\ \BBA {} Boer, E\BPBI R.%
\end{APACrefauthors}%
\unskip\
\newblock
\APACrefYearMonthDay{2018}{}{}.
\newblock
{\BBOQ}\APACrefatitle {{A topology of shared control systems---finding common
  ground in diversity}} {{A topology of shared control systems---finding common
  ground in diversity}}.{\BBCQ}
\newblock
\APACjournalVolNumPages{{IEEE Transactions on Human-Machine
  Systems}}{48}{5}{509--525}.
\newblock
\begin{APACrefDOI} \doi{10.1109/THMS.2018.2791570} \end{APACrefDOI}
\PrintBackRefs{\CurrentBib}

\bibitem [\protect \citeauthoryear {%
Abbink%
, Mulder%
\BCBL {}\ \BBA {} Boer%
}{%
Abbink%
\ \protect \BOthers {.}}{%
{\protect \APACyear {2012}}%
}]{%
Abbink.2012}
\APACinsertmetastar {%
Abbink.2012}%
\begin{APACrefauthors}%
Abbink, D\BPBI A.%
, Mulder, M.%
\BCBL {}\ \BBA {} Boer, E\BPBI R.%
\end{APACrefauthors}%
\unskip\
\newblock
\APACrefYearMonthDay{2012}{}{}.
\newblock
{\BBOQ}\APACrefatitle {{Haptic shared control: smoothly shifting control
  authority?}} {{Haptic shared control: smoothly shifting control
  authority?}}{\BBCQ}
\newblock
\APACjournalVolNumPages{{Cognition, Technology {\&} Work}}{14}{1}{19--28}.
\newblock
\begin{APACrefDOI} \doi{10.1007/s10111-011-0192-5} \end{APACrefDOI}
\PrintBackRefs{\CurrentBib}

\bibitem [\protect \citeauthoryear {%
Ajzen%
}{%
Ajzen%
}{%
{\protect \APACyear {1985}}%
}]{%
Ajzen.1985}
\APACinsertmetastar {%
Ajzen.1985}%
\begin{APACrefauthors}%
Ajzen, I.%
\end{APACrefauthors}%
\unskip\
\newblock
\APACrefYearMonthDay{1985}{}{}.
\newblock
{\BBOQ}\APACrefatitle {{From intentions to actions: A theory of planned
  behavior}} {{From intentions to actions: A theory of planned
  behavior}}.{\BBCQ}
\newblock
\BIn{} \APACrefbtitle {{Action Control}} {{Action Control}}\ (\BPGS\ 11--39).
\newblock
\APACaddressPublisher{}{{Springer, Berlin, Heidelberg}}.
\newblock
\begin{APACrefDOI} \doi{10.1007/978-3-642-69746-3_2} \end{APACrefDOI}
\PrintBackRefs{\CurrentBib}

\bibitem [\protect \citeauthoryear {%
Alimardani%
, Nishio%
\BCBL {}\ \BBA {} Ishiguro%
}{%
Alimardani%
\ \protect \BOthers {.}}{%
{\protect \APACyear {2013}}%
}]{%
Alimardani.2013}
\APACinsertmetastar {%
Alimardani.2013}%
\begin{APACrefauthors}%
Alimardani, M.%
, Nishio, S.%
\BCBL {}\ \BBA {} Ishiguro, H.%
\end{APACrefauthors}%
\unskip\
\newblock
\APACrefYearMonthDay{2013}{}{}.
\newblock
{\BBOQ}\APACrefatitle {{Humanlike robot hands controlled by brain activity
  arouse illusion of ownership in operators}} {{Humanlike robot hands
  controlled by brain activity arouse illusion of ownership in
  operators}}.{\BBCQ}
\newblock
\APACjournalVolNumPages{{Scientific Reports}}{3}{}{2396}.
\newblock
\begin{APACrefDOI} \doi{10.1038/srep02396} \end{APACrefDOI}
\PrintBackRefs{\CurrentBib}

\bibitem [\protect \citeauthoryear {%
Bakker%
}{%
Bakker%
}{%
{\protect \APACyear {2008}}%
}]{%
Bakker.2008}
\APACinsertmetastar {%
Bakker.2008}%
\begin{APACrefauthors}%
Bakker, A\BPBI B.%
\end{APACrefauthors}%
\unskip\
\newblock
\APACrefYearMonthDay{2008}{}{}.
\newblock
{\BBOQ}\APACrefatitle {{The work-related flow inventory: Construction and
  initial validation of the WOLF}} {{The work-related flow inventory:
  Construction and initial validation of the WOLF}}.{\BBCQ}
\newblock
\APACjournalVolNumPages{{Journal of Vocational Behavior}}{72}{3}{400--414}.
\newblock
\begin{APACrefDOI} \doi{10.1016/j.jvb.2007.11.007} \end{APACrefDOI}
\PrintBackRefs{\CurrentBib}

\bibitem [\protect \citeauthoryear {%
Berberian%
}{%
Berberian%
}{%
{\protect \APACyear {2019}}%
}]{%
Berberian.2019}
\APACinsertmetastar {%
Berberian.2019}%
\begin{APACrefauthors}%
Berberian, B.%
\end{APACrefauthors}%
\unskip\
\newblock
\APACrefYearMonthDay{2019}{}{}.
\newblock
{\BBOQ}\APACrefatitle {{Man-machine teaming: a problem of agency}}
  {{Man-machine teaming: a problem of agency}}.{\BBCQ}
\newblock
\APACjournalVolNumPages{{IFAC-PapersOnLine}}{51}{34}{118--123}.
\newblock
\begin{APACrefDOI} \doi{10.1016/j.ifacol.2019.01.049} \end{APACrefDOI}
\PrintBackRefs{\CurrentBib}

\bibitem [\protect \citeauthoryear {%
Berberian%
, Sarrazin%
, {Le Blaye}%
\BCBL {}\ \BBA {} Haggard%
}{%
Berberian%
\ \protect \BOthers {.}}{%
{\protect \APACyear {2012}}%
}]{%
Berberian.2012}
\APACinsertmetastar {%
Berberian.2012}%
\begin{APACrefauthors}%
Berberian, B.%
, Sarrazin, J\BHBI C.%
, {Le Blaye}, P.%
\BCBL {}\ \BBA {} Haggard, P.%
\end{APACrefauthors}%
\unskip\
\newblock
\APACrefYearMonthDay{2012}{}{}.
\newblock
{\BBOQ}\APACrefatitle {{Automation technology and sense of control: a window on
  human agency}} {{Automation technology and sense of control: a window on
  human agency}}.{\BBCQ}
\newblock
\APACjournalVolNumPages{{PLOS ONE}}{7}{3}{e34075}.
\newblock
\begin{APACrefDOI} \doi{10.1371/journal.pone.0034075} \end{APACrefDOI}
\PrintBackRefs{\CurrentBib}

\bibitem [\protect \citeauthoryear {%
Boink%
, {van Paassen}%
, Mulder%
\BCBL {}\ \BBA {} Abbink%
}{%
Boink%
\ \protect \BOthers {.}}{%
{\protect \APACyear {2014}}%
}]{%
Boink.2014}
\APACinsertmetastar {%
Boink.2014}%
\begin{APACrefauthors}%
Boink, R.%
, {van Paassen}, M\BPBI M.%
, Mulder, M.%
\BCBL {}\ \BBA {} Abbink, D\BPBI A.%
\end{APACrefauthors}%
\unskip\
\newblock
\APACrefYearMonthDay{2014}{}{}.
\newblock
{\BBOQ}\APACrefatitle {{Understanding and reducing conflicts between driver and
  haptic shared control}} {{Understanding and reducing conflicts between driver
  and haptic shared control}}.{\BBCQ}
\newblock
\BIn{} \APACrefbtitle {{2014 IEEE International Conference on Systems, Man, and
  Cybernetics (SMC)}} {{2014 IEEE International Conference on Systems, Man, and
  Cybernetics (SMC)}}\ (\BPGS\ 1510--1515).
\newblock
\APACaddressPublisher{}{IEEE}.
\newblock
\begin{APACrefDOI} \doi{10.1109/smc.2014.6974130} \end{APACrefDOI}
\PrintBackRefs{\CurrentBib}

\bibitem [\protect \citeauthoryear {%
Botvinick%
\ \BBA {} Cohen%
}{%
Botvinick%
\ \BBA {} Cohen%
}{%
{\protect \APACyear {1998}}%
}]{%
Botvinick.1998}
\APACinsertmetastar {%
Botvinick.1998}%
\begin{APACrefauthors}%
Botvinick, M.%
\BCBT {}\ \BBA {} Cohen, J.%
\end{APACrefauthors}%
\unskip\
\newblock
\APACrefYearMonthDay{1998}{}{}.
\newblock
{\BBOQ}\APACrefatitle {{Rubber hands 'feel' touch that eyes see}} {{Rubber
  hands 'feel' touch that eyes see}}.{\BBCQ}
\newblock
\APACjournalVolNumPages{{Nature}}{391}{6669}{756}.
\newblock
\begin{APACrefDOI} \doi{10.1038/35784} \end{APACrefDOI}
\PrintBackRefs{\CurrentBib}

\bibitem [\protect \citeauthoryear {%
Braun%
, Flad%
\BCBL {}\ \BBA {} Hohmann%
}{%
Braun%
\ \protect \BOthers {.}}{%
{\protect \APACyear {2019}}%
}]{%
Braun.2019}
\APACinsertmetastar {%
Braun.2019}%
\begin{APACrefauthors}%
Braun, C\BPBI A.%
, Flad, M.%
\BCBL {}\ \BBA {} Hohmann, S.%
\end{APACrefauthors}%
\unskip\
\newblock
\APACrefYearMonthDay{2019}{}{}.
\newblock
{\BBOQ}\APACrefatitle {{A continuous and quantitative metric for the levels of
  automation}} {{A continuous and quantitative metric for the levels of
  automation}}.{\BBCQ}
\newblock
\APACjournalVolNumPages{{IFAC-PapersOnLine}}{52}{19}{37--42}.
\newblock
\begin{APACrefDOI} \doi{10.1016/j.ifacol.2019.12.081} \end{APACrefDOI}
\PrintBackRefs{\CurrentBib}

\bibitem [\protect \citeauthoryear {%
B{\"u}tepage%
\ \BBA {} Kragic%
}{%
B{\"u}tepage%
\ \BBA {} Kragic%
}{%
{\protect \APACyear {{\protect \bibnodate {}}}}%
}]{%
Butepage.2017}
\APACinsertmetastar {%
Butepage.2017}%
\begin{APACrefauthors}%
B{\"u}tepage, J.%
\BCBT {}\ \BBA {} Kragic, D.%
\end{APACrefauthors}%
\unskip\
\newblock
\APACrefYearMonthDay{{\protect \bibnodate {}}}{}{}.
\newblock
\APACrefbtitle {{Human-robot collaboration: From psychology to social
  robotics}.} {{Human-robot collaboration: From psychology to social
  robotics}.}
\newblock
\APACaddressPublisher{arXiv preprint arXiv:1705.10146}{}.
\PrintBackRefs{\CurrentBib}

\bibitem [\protect \citeauthoryear {%
Chen%
\ \protect \BOthers {.}}{%
Chen%
\ \protect \BOthers {.}}{%
{\protect \APACyear {2016}}%
}]{%
Chen.2016}
\APACinsertmetastar {%
Chen.2016}%
\begin{APACrefauthors}%
Chen, B.%
, Ma, H.%
, Qin, L\BHBI Y.%
, Gao, F.%
, Chan, K\BHBI M.%
, Law, S\BHBI W.%
\BDBL {}Liao, W\BHBI H.%
\end{APACrefauthors}%
\unskip\
\newblock
\APACrefYearMonthDay{2016}{}{}.
\newblock
{\BBOQ}\APACrefatitle {{Recent developments and challenges of lower extremity
  exoskeletons}} {{Recent developments and challenges of lower extremity
  exoskeletons}}.{\BBCQ}
\newblock
\APACjournalVolNumPages{{Journal of Orthopaedic Translation}}{5}{}{26--37}.
\newblock
\begin{APACrefDOI} \doi{10.1016/j.jot.2015.09.007} \end{APACrefDOI}
\PrintBackRefs{\CurrentBib}

\bibitem [\protect \citeauthoryear {%
Chintalapati%
\ \BBA {} Daruri%
}{%
Chintalapati%
\ \BBA {} Daruri%
}{%
{\protect \APACyear {2017}}%
}]{%
Chintalapati.2017}
\APACinsertmetastar {%
Chintalapati.2017}%
\begin{APACrefauthors}%
Chintalapati, N.%
\BCBT {}\ \BBA {} Daruri, V\BPBI S\BPBI K.%
\end{APACrefauthors}%
\unskip\
\newblock
\APACrefYearMonthDay{2017}{}{}.
\newblock
{\BBOQ}\APACrefatitle {{Examining the use of YouTube as a learning resource in
  higher education: Scale development and validation of TAM model}} {{Examining
  the use of YouTube as a learning resource in higher education: Scale
  development and validation of TAM model}}.{\BBCQ}
\newblock
\APACjournalVolNumPages{{Telematics and Informatics}}{34}{6}{853--860}.
\newblock
\begin{APACrefDOI} \doi{10.1016/j.tele.2016.08.008} \end{APACrefDOI}
\PrintBackRefs{\CurrentBib}

\bibitem [\protect \citeauthoryear {%
Chiou%
, Hawes%
\BCBL {}\ \BBA {} Stolkin%
}{%
Chiou%
\ \protect \BOthers {.}}{%
{\protect \APACyear {2021}}%
}]{%
Chiou.2021}
\APACinsertmetastar {%
Chiou.2021}%
\begin{APACrefauthors}%
Chiou, M.%
, Hawes, N.%
\BCBL {}\ \BBA {} Stolkin, R.%
\end{APACrefauthors}%
\unskip\
\newblock
\APACrefYearMonthDay{2021}{}{}.
\newblock
{\BBOQ}\APACrefatitle {{Mixed-initiative variable autonomy for remotely
  operated mobile robots}} {{Mixed-initiative variable autonomy for remotely
  operated mobile robots}}.{\BBCQ}
\newblock
\APACjournalVolNumPages{{ACM Transactions on Human-Robot Interaction
  (THRI)}}{10}{4}{1--34}.
\newblock
\begin{APACrefDOI} \doi{10.1145/3472206} \end{APACrefDOI}
\PrintBackRefs{\CurrentBib}

\bibitem [\protect \citeauthoryear {%
Clarke%
, Schillhuber%
, Zaeh%
\BCBL {}\ \BBA {} Ulbrich%
}{%
Clarke%
\ \protect \BOthers {.}}{%
{\protect \APACyear {2007}}%
}]{%
Clarke.2007}
\APACinsertmetastar {%
Clarke.2007}%
\begin{APACrefauthors}%
Clarke, S.%
, Schillhuber, G.%
, Zaeh, M\BPBI F.%
\BCBL {}\ \BBA {} Ulbrich, H.%
\end{APACrefauthors}%
\unskip\
\newblock
\APACrefYearMonthDay{2007}{}{}.
\newblock
{\BBOQ}\APACrefatitle {{Prediction-based methods for teleoperation across
  delayed networks}} {{Prediction-based methods for teleoperation across
  delayed networks}}.{\BBCQ}
\newblock
\APACjournalVolNumPages{{Multimedia Systems}}{13}{4}{253--261}.
\newblock
\begin{APACrefDOI} \doi{10.1007/s00530-007-0103-z} \end{APACrefDOI}
\PrintBackRefs{\CurrentBib}

\bibitem [\protect \citeauthoryear {%
Csikszentmihalyi%
\ \BBA {} LeFevre%
}{%
Csikszentmihalyi%
\ \BBA {} LeFevre%
}{%
{\protect \APACyear {1989}}%
}]{%
Csikszentmihalyi.1989}
\APACinsertmetastar {%
Csikszentmihalyi.1989}%
\begin{APACrefauthors}%
Csikszentmihalyi, M.%
\BCBT {}\ \BBA {} LeFevre, J.%
\end{APACrefauthors}%
\unskip\
\newblock
\APACrefYearMonthDay{1989}{}{}.
\newblock
{\BBOQ}\APACrefatitle {{Optimal experience in work and leisure}} {{Optimal
  experience in work and leisure}}.{\BBCQ}
\newblock
\APACjournalVolNumPages{{Journal of Personality and Social
  Psychology}}{56}{5}{815--822}.
\newblock
\begin{APACrefDOI} \doi{10.1037/0022-3514.56.5.815} \end{APACrefDOI}
\PrintBackRefs{\CurrentBib}

\bibitem [\protect \citeauthoryear {%
Curioni%
, Vesper%
, Knoblich%
\BCBL {}\ \BBA {} Sebanz%
}{%
Curioni%
\ \protect \BOthers {.}}{%
{\protect \APACyear {2019}}%
}]{%
Curioni.2019}
\APACinsertmetastar {%
Curioni.2019}%
\begin{APACrefauthors}%
Curioni, A.%
, Vesper, C.%
, Knoblich, G.%
\BCBL {}\ \BBA {} Sebanz, N.%
\end{APACrefauthors}%
\unskip\
\newblock
\APACrefYearMonthDay{2019}{}{}.
\newblock
{\BBOQ}\APACrefatitle {{Reciprocal information flow and role distribution
  support joint action coordination}} {{Reciprocal information flow and role
  distribution support joint action coordination}}.{\BBCQ}
\newblock
\APACjournalVolNumPages{{Cognition}}{187}{}{21--31}.
\newblock
\begin{APACrefDOI} \doi{10.1016/j.cognition.2019.02.006} \end{APACrefDOI}
\PrintBackRefs{\CurrentBib}

\bibitem [\protect \citeauthoryear {%
Dagioglou%
\ \BBA {} Karkaletsis%
}{%
Dagioglou%
\ \BBA {} Karkaletsis%
}{%
{\protect \APACyear {2021}}%
}]{%
Dagioglou.2021}
\APACinsertmetastar {%
Dagioglou.2021}%
\begin{APACrefauthors}%
Dagioglou, M.%
\BCBT {}\ \BBA {} Karkaletsis, V.%
\end{APACrefauthors}%
\unskip\
\newblock
\APACrefYearMonthDay{2021}{}{}.
\newblock
{\BBOQ}\APACrefatitle {{The sense of agency during human-agent collaboration}}
  {{The sense of agency during human-agent collaboration}}.{\BBCQ}
\newblock
\BIn{} \APACrefbtitle {{HRI 2021 Workshop: Robo-Identity: Artificial identity
  and multi-embodiment}.} {{HRI 2021 Workshop: Robo-Identity: Artificial
  identity and multi-embodiment}.}
\newblock
\APACaddressPublisher{Boulder (Virtual), USA}{}.
\PrintBackRefs{\CurrentBib}

\bibitem [\protect \citeauthoryear {%
De~Bary%
}{%
De~Bary%
}{%
{\protect \APACyear {1879}}%
}]{%
Bary.1879}
\APACinsertmetastar {%
Bary.1879}%
\begin{APACrefauthors}%
De~Bary, A.%
\end{APACrefauthors}%
\unskip\
\newblock
\APACrefYear{1879}.
\newblock
\APACrefbtitle {{Die Erscheinung der symbiose: Vortrag gehalten auf der
  versammlung deutscher naturforscher und aerzte zu cassel.}} {{Die Erscheinung
  der symbiose: Vortrag gehalten auf der versammlung deutscher naturforscher
  und aerzte zu cassel.}}
\newblock
\APACaddressPublisher{}{Tr{\"u}bner}.
\PrintBackRefs{\CurrentBib}

\bibitem [\protect \citeauthoryear {%
Dewey%
\ \BBA {} Knoblich%
}{%
Dewey%
\ \BBA {} Knoblich%
}{%
{\protect \APACyear {2014}}%
}]{%
Dewey.2014}
\APACinsertmetastar {%
Dewey.2014}%
\begin{APACrefauthors}%
Dewey, J\BPBI A.%
\BCBT {}\ \BBA {} Knoblich, G.%
\end{APACrefauthors}%
\unskip\
\newblock
\APACrefYearMonthDay{2014}{}{}.
\newblock
{\BBOQ}\APACrefatitle {{Do implicit and explicit measures of Sense of Agency
  measure the same thing?}} {{Do implicit and explicit measures of Sense of
  Agency measure the same thing?}}{\BBCQ}
\newblock
\APACjournalVolNumPages{{PloS one}}{9}{10}{e110118}.
\newblock
\begin{APACrefDOI} \doi{10.1371/journal.pone.0110118.g001} \end{APACrefDOI}
\PrintBackRefs{\CurrentBib}

\bibitem [\protect \citeauthoryear {%
\APACcitebtitle {{DIN EN ISO 9241-11:2018-11: Ergonomics of human-system
  interaction - Part 11: Usability: Definitions and concepts (ISO
  9241-11:2018); German version EN ISO 9241-11:2018}}}{%
\APACcitebtitle {{DIN EN ISO 9241-11:2018-11: Ergonomics of human-system
  interaction - Part 11: Usability: Definitions and concepts (ISO
  9241-11:2018); German version EN ISO 9241-11:2018}}}{%
{\protect \APACyear {2018}}%
}]{%
.2018}
\APACinsertmetastar {%
.2018}%
\APACrefbtitle {{DIN EN ISO 9241-11:2018-11: Ergonomics of human-system
  interaction - Part 11: Usability: Definitions and concepts (ISO
  9241-11:2018); German version EN ISO 9241-11:2018}.} {{DIN EN ISO
  9241-11:2018-11: Ergonomics of human-system interaction - Part 11: Usability:
  Definitions and concepts (ISO 9241-11:2018); German version EN ISO
  9241-11:2018}.}
\newblock
\APACrefYearMonthDay{2018}{}{}.
\newblock
\APACaddressPublisher{Berlin}{{Beuth Verlag GmbH}}.
\newblock
\begin{APACrefDOI} \doi{10.31030/2757945} \end{APACrefDOI}
\PrintBackRefs{\CurrentBib}

\bibitem [\protect \citeauthoryear {%
Donges%
}{%
Donges%
}{%
{\protect \APACyear {1999}}%
}]{%
Donges.1999}
\APACinsertmetastar {%
Donges.1999}%
\begin{APACrefauthors}%
Donges, E.%
\end{APACrefauthors}%
\unskip\
\newblock
\APACrefYearMonthDay{1999}{}{}.
\newblock
{\BBOQ}\APACrefatitle {{A conceptual framework for active safety in road
  traffic}} {{A conceptual framework for active safety in road
  traffic}}.{\BBCQ}
\newblock
\APACjournalVolNumPages{{Vehicle System Dynamics}}{2-3}{32}{113--128}.
\newblock
\begin{APACrefDOI} \doi{10.1076/vesd.32.2.113.2089} \end{APACrefDOI}
\PrintBackRefs{\CurrentBib}

\bibitem [\protect \citeauthoryear {%
Ende%
\ \protect \BOthers {.}}{%
Ende%
\ \protect \BOthers {.}}{%
{\protect \APACyear {2011}}%
}]{%
Ende.2011}
\APACinsertmetastar {%
Ende.2011}%
\begin{APACrefauthors}%
Ende, T.%
, Haddadin, S.%
, Parusel, S.%
, Wusthoff, T.%
, Hassenzahl, M.%
\BCBL {}\ \BBA {} Albu-Schaffer, A.%
\end{APACrefauthors}%
\unskip\
\newblock
\APACrefYearMonthDay{2011}{}{}.
\newblock
{\BBOQ}\APACrefatitle {{A human-centered approach to robot gesture based
  communication within collaborative working processes}} {{A human-centered
  approach to robot gesture based communication within collaborative working
  processes}}.{\BBCQ}
\newblock
\BIn{} \APACrefbtitle {{2011 IEEE/RSJ International Conference on Intelligent
  Robots and Systems}.} {{2011 IEEE/RSJ International Conference on Intelligent
  Robots and Systems}.}
\newblock
\APACaddressPublisher{}{IEEE}.
\newblock
\begin{APACrefDOI} \doi{10.1109/iros.2011.6094592} \end{APACrefDOI}
\PrintBackRefs{\CurrentBib}

\bibitem [\protect \citeauthoryear {%
Endsley%
\ \BBA {} Kaber%
}{%
Endsley%
\ \BBA {} Kaber%
}{%
{\protect \APACyear {1999}}%
}]{%
Endsley.1999}
\APACinsertmetastar {%
Endsley.1999}%
\begin{APACrefauthors}%
Endsley, M\BPBI R.%
\BCBT {}\ \BBA {} Kaber, D\BPBI B.%
\end{APACrefauthors}%
\unskip\
\newblock
\APACrefYearMonthDay{1999}{}{}.
\newblock
{\BBOQ}\APACrefatitle {{Level of automation effects on performance, situation
  awareness and workload in a dynamic control task}} {{Level of automation
  effects on performance, situation awareness and workload in a dynamic control
  task}}.{\BBCQ}
\newblock
\APACjournalVolNumPages{{Ergonomics}}{42}{3}{462--492}.
\newblock
\begin{APACrefDOI} \doi{10.1080/001401399185595} \end{APACrefDOI}
\PrintBackRefs{\CurrentBib}

\bibitem [\protect \citeauthoryear {%
Engeser%
\ \BBA {} Rheinberg%
}{%
Engeser%
\ \BBA {} Rheinberg%
}{%
{\protect \APACyear {2008}}%
}]{%
Engeser.2008}
\APACinsertmetastar {%
Engeser.2008}%
\begin{APACrefauthors}%
Engeser, S.%
\BCBT {}\ \BBA {} Rheinberg, F.%
\end{APACrefauthors}%
\unskip\
\newblock
\APACrefYearMonthDay{2008}{}{}.
\newblock
{\BBOQ}\APACrefatitle {{Flow, performance and moderators of challenge-skill
  balance}} {{Flow, performance and moderators of challenge-skill
  balance}}.{\BBCQ}
\newblock
\APACjournalVolNumPages{{Motivation and Emotion}}{32}{3}{158--172}.
\newblock
\begin{APACrefDOI} \doi{10.1007/s11031-008-9102-4} \end{APACrefDOI}
\PrintBackRefs{\CurrentBib}

\bibitem [\protect \citeauthoryear {%
Ferreira%
, Doltsinis%
\BCBL {}\ \BBA {} Lohse%
}{%
Ferreira%
\ \protect \BOthers {.}}{%
{\protect \APACyear {2014}}%
}]{%
Ferreira.2014}
\APACinsertmetastar {%
Ferreira.2014}%
\begin{APACrefauthors}%
Ferreira, P.%
, Doltsinis, S.%
\BCBL {}\ \BBA {} Lohse, N.%
\end{APACrefauthors}%
\unskip\
\newblock
\APACrefYearMonthDay{2014}{}{}.
\newblock
{\BBOQ}\APACrefatitle {{Symbiotic assembly systems -- A new paradigm}}
  {{Symbiotic assembly systems -- A new paradigm}}.{\BBCQ}
\newblock
\APACjournalVolNumPages{{Procedia CIRP}}{17}{}{26--31}.
\newblock
\begin{APACrefDOI} \doi{10.1016/j.procir.2014.01.066} \end{APACrefDOI}
\PrintBackRefs{\CurrentBib}

\bibitem [\protect \citeauthoryear {%
Finneran%
\ \BBA {} Zhang%
}{%
Finneran%
\ \BBA {} Zhang%
}{%
{\protect \APACyear {2003}}%
}]{%
Finneran.2003}
\APACinsertmetastar {%
Finneran.2003}%
\begin{APACrefauthors}%
Finneran, C\BPBI M.%
\BCBT {}\ \BBA {} Zhang, P.%
\end{APACrefauthors}%
\unskip\
\newblock
\APACrefYearMonthDay{2003}{}{}.
\newblock
{\BBOQ}\APACrefatitle {{A person--artefact--task (PAT) model of flow
  antecedents in computer-mediated environments}} {{A person--artefact--task
  (PAT) model of flow antecedents in computer-mediated environments}}.{\BBCQ}
\newblock
\APACjournalVolNumPages{{International Journal of Human-Computer
  Studies}}{59}{4}{475--496}.
\newblock
\begin{APACrefDOI} \doi{10.1016/S1071-5819(03)00112-5} \end{APACrefDOI}
\PrintBackRefs{\CurrentBib}

\bibitem [\protect \citeauthoryear {%
Flad%
, Frohlich%
\BCBL {}\ \BBA {} Hohmann%
}{%
Flad%
\ \protect \BOthers {.}}{%
{\protect \APACyear {2017}}%
}]{%
Flad.2017}
\APACinsertmetastar {%
Flad.2017}%
\begin{APACrefauthors}%
Flad, M.%
, Frohlich, L.%
\BCBL {}\ \BBA {} Hohmann, S.%
\end{APACrefauthors}%
\unskip\
\newblock
\APACrefYearMonthDay{2017}{}{}.
\newblock
{\BBOQ}\APACrefatitle {{Cooperative shared control driver assistance systems
  based on motion primitives and differential games}} {{Cooperative shared
  control driver assistance systems based on motion primitives and differential
  games}}.{\BBCQ}
\newblock
\APACjournalVolNumPages{{IEEE Transactions on Human-Machine
  Systems}}{47}{5}{711--722}.
\newblock
\begin{APACrefDOI} \doi{10.1109/thms.2017.2700435} \end{APACrefDOI}
\PrintBackRefs{\CurrentBib}

\bibitem [\protect \citeauthoryear {%
Flad%
, Otten%
, Schwab%
\BCBL {}\ \BBA {} Hohmann%
}{%
Flad%
\ \protect \BOthers {.}}{%
{\protect \APACyear {2014}}%
}]{%
Flad.2014}
\APACinsertmetastar {%
Flad.2014}%
\begin{APACrefauthors}%
Flad, M.%
, Otten, J.%
, Schwab, S.%
\BCBL {}\ \BBA {} Hohmann, S.%
\end{APACrefauthors}%
\unskip\
\newblock
\APACrefYearMonthDay{2014}{}{}.
\newblock
{\BBOQ}\APACrefatitle {{Necessary and sufficient conditions for the design of
  cooperative shared control}} {{Necessary and sufficient conditions for the
  design of cooperative shared control}}.{\BBCQ}
\newblock
\BIn{} \APACrefbtitle {{2014 IEEE International Conference on Systems, Man, and
  Cybernetics (SMC)}} {{2014 IEEE International Conference on Systems, Man, and
  Cybernetics (SMC)}}\ (\BPGS\ 1253--1259).
\newblock
\APACaddressPublisher{}{IEEE}.
\newblock
\begin{APACrefDOI} \doi{10.1109/smc.2014.6974086} \end{APACrefDOI}
\PrintBackRefs{\CurrentBib}

\bibitem [\protect \citeauthoryear {%
Flash%
\ \BBA {} Hogan%
}{%
Flash%
\ \BBA {} Hogan%
}{%
{\protect \APACyear {1985}}%
}]{%
Flash.1985}
\APACinsertmetastar {%
Flash.1985}%
\begin{APACrefauthors}%
Flash, T.%
\BCBT {}\ \BBA {} Hogan, N.%
\end{APACrefauthors}%
\unskip\
\newblock
\APACrefYearMonthDay{1985}{}{}.
\newblock
{\BBOQ}\APACrefatitle {{The coordination of arm movements: an experimentally
  confirmed mathematical model}} {{The coordination of arm movements: an
  experimentally confirmed mathematical model}}.{\BBCQ}
\newblock
\APACjournalVolNumPages{{Journal of Neuroscience}}{5}{7}{1688--1703}.
\newblock
\begin{APACrefDOI} \doi{10.1523/JNEUROSCI.05-07-01688.1985} \end{APACrefDOI}
\PrintBackRefs{\CurrentBib}

\bibitem [\protect \citeauthoryear {%
F.~Flemisch%
, Abbink%
, Itoh%
, Pacaux-Lemoine%
\BCBL {}\ \BBA {} We{\ss}el%
}{%
F.~Flemisch%
\ \protect \BOthers {.}}{%
{\protect \APACyear {2016}}%
}]{%
Flemisch.2016}
\APACinsertmetastar {%
Flemisch.2016}%
\begin{APACrefauthors}%
Flemisch, F.%
, Abbink, D.%
, Itoh, M.%
, Pacaux-Lemoine, M\BHBI P.%
\BCBL {}\ \BBA {} We{\ss}el, G.%
\end{APACrefauthors}%
\unskip\
\newblock
\APACrefYearMonthDay{2016}{}{}.
\newblock
{\BBOQ}\APACrefatitle {{Shared control is the sharp end of cooperation: Towards
  a common framework of joint action, shared control and human machine
  cooperation}} {{Shared control is the sharp end of cooperation: Towards a
  common framework of joint action, shared control and human machine
  cooperation}}.{\BBCQ}
\newblock
\APACjournalVolNumPages{{IFAC-PapersOnLine}}{49}{19}{72--77}.
\newblock
\begin{APACrefDOI} \doi{10.1016/j.ifacol.2016.10.464} \end{APACrefDOI}
\PrintBackRefs{\CurrentBib}

\bibitem [\protect \citeauthoryear {%
F.~Flemisch%
, Abbink%
, Itoh%
, Pacaux-Lemoine%
\BCBL {}\ \BBA {} We{\ss}el%
}{%
F.~Flemisch%
\ \protect \BOthers {.}}{%
{\protect \APACyear {2019}}%
}]{%
Flemisch.2019}
\APACinsertmetastar {%
Flemisch.2019}%
\begin{APACrefauthors}%
Flemisch, F.%
, Abbink, D\BPBI A.%
, Itoh, M.%
, Pacaux-Lemoine, M\BHBI P.%
\BCBL {}\ \BBA {} We{\ss}el, G.%
\end{APACrefauthors}%
\unskip\
\newblock
\APACrefYearMonthDay{2019}{}{}.
\newblock
{\BBOQ}\APACrefatitle {{Joining the blunt and the pointy end of the spear:
  towards a common framework of joint action, human--machine cooperation,
  cooperative guidance and control, shared, traded and supervisory control}}
  {{Joining the blunt and the pointy end of the spear: towards a common
  framework of joint action, human--machine cooperation, cooperative guidance
  and control, shared, traded and supervisory control}}.{\BBCQ}
\newblock
\APACjournalVolNumPages{{Cognition, Technology {\&} Work}}{21}{4}{555--568}.
\newblock
\begin{APACrefDOI} \doi{10.1007/s10111-019-00576-1} \end{APACrefDOI}
\PrintBackRefs{\CurrentBib}

\bibitem [\protect \citeauthoryear {%
F\BPBI O.~Flemisch%
, Bengler%
, Bubb%
, Winner%
\BCBL {}\ \BBA {} Bruder%
}{%
F\BPBI O.~Flemisch%
\ \protect \BOthers {.}}{%
{\protect \APACyear {2014}}%
}]{%
Flemisch.2014}
\APACinsertmetastar {%
Flemisch.2014}%
\begin{APACrefauthors}%
Flemisch, F\BPBI O.%
, Bengler, K.%
, Bubb, H.%
, Winner, H.%
\BCBL {}\ \BBA {} Bruder, R.%
\end{APACrefauthors}%
\unskip\
\newblock
\APACrefYearMonthDay{2014}{}{}.
\newblock
{\BBOQ}\APACrefatitle {{Towards cooperative guidance and control of highly
  automated vehicles: H-Mode and Conduct-by-Wire}} {{Towards cooperative
  guidance and control of highly automated vehicles: H-Mode and
  Conduct-by-Wire}}.{\BBCQ}
\newblock
\APACjournalVolNumPages{{Ergonomics}}{57}{3}{343--360}.
\newblock
\begin{APACrefDOI} \doi{10.1080/00140139.2013.869355} \end{APACrefDOI}
\PrintBackRefs{\CurrentBib}

\bibitem [\protect \citeauthoryear {%
Fong%
, Zaleski%
\BCBL {}\ \BBA {} Leach%
}{%
Fong%
\ \protect \BOthers {.}}{%
{\protect \APACyear {2015}}%
}]{%
Fong.2015}
\APACinsertmetastar {%
Fong.2015}%
\begin{APACrefauthors}%
Fong, C\BPBI J.%
, Zaleski, D\BPBI J.%
\BCBL {}\ \BBA {} Leach, J\BPBI K.%
\end{APACrefauthors}%
\unskip\
\newblock
\APACrefYearMonthDay{2015}{}{}.
\newblock
{\BBOQ}\APACrefatitle {{The challenge--skill balance and antecedents of flow: A
  meta-analytic investigation}} {{The challenge--skill balance and antecedents
  of flow: A meta-analytic investigation}}.{\BBCQ}
\newblock
\APACjournalVolNumPages{{The Journal of Positive Psychology}}{10}{5}{425--446}.
\newblock
\begin{APACrefDOI} \doi{10.1080/17439760.2014.967799} \end{APACrefDOI}
\PrintBackRefs{\CurrentBib}

\bibitem [\protect \citeauthoryear {%
Fox%
, Aranko%
, Heilala%
\BCBL {}\ \BBA {} Vahala%
}{%
Fox%
\ \protect \BOthers {.}}{%
{\protect \APACyear {2020}}%
}]{%
Fox.2020}
\APACinsertmetastar {%
Fox.2020}%
\begin{APACrefauthors}%
Fox, S.%
, Aranko, O.%
, Heilala, J.%
\BCBL {}\ \BBA {} Vahala, P.%
\end{APACrefauthors}%
\unskip\
\newblock
\APACrefYearMonthDay{2020}{}{}.
\newblock
{\BBOQ}\APACrefatitle {{Exoskeletons: Comprehensive, comparative and critical
  analyses of their potential to improve manufacturing performance}}
  {{Exoskeletons: Comprehensive, comparative and critical analyses of their
  potential to improve manufacturing performance}}.{\BBCQ}
\newblock
\APACjournalVolNumPages{{Journal of Manufacturing Technology
  Management}}{31}{6}{1261--1280}.
\newblock
\begin{APACrefDOI} \doi{10.1108/JMTM-01-2019-0023} \end{APACrefDOI}
\PrintBackRefs{\CurrentBib}

\bibitem [\protect \citeauthoryear {%
Fullagar%
\ \BBA {} Kelloway%
}{%
Fullagar%
\ \BBA {} Kelloway%
}{%
{\protect \APACyear {2009}}%
}]{%
Fullagar.2009}
\APACinsertmetastar {%
Fullagar.2009}%
\begin{APACrefauthors}%
Fullagar, C\BPBI J.%
\BCBT {}\ \BBA {} Kelloway, E\BPBI K.%
\end{APACrefauthors}%
\unskip\
\newblock
\APACrefYearMonthDay{2009}{}{}.
\newblock
{\BBOQ}\APACrefatitle {{Flow at work: An experience sampling approach}} {{Flow
  at work: An experience sampling approach}}.{\BBCQ}
\newblock
\APACjournalVolNumPages{{Journal of Occupational and Organizational
  Psychology}}{82}{3}{595--615}.
\newblock
\begin{APACrefDOI} \doi{10.1348/096317908X357903} \end{APACrefDOI}
\PrintBackRefs{\CurrentBib}

\bibitem [\protect \citeauthoryear {%
Gallivan%
, Chapman%
, Wolpert%
\BCBL {}\ \BBA {} Flanagan%
}{%
Gallivan%
\ \protect \BOthers {.}}{%
{\protect \APACyear {2018}}%
}]{%
Gallivan.2018}
\APACinsertmetastar {%
Gallivan.2018}%
\begin{APACrefauthors}%
Gallivan, J\BPBI P.%
, Chapman, C\BPBI S.%
, Wolpert, D\BPBI M.%
\BCBL {}\ \BBA {} Flanagan, J\BPBI R.%
\end{APACrefauthors}%
\unskip\
\newblock
\APACrefYearMonthDay{2018}{}{}.
\newblock
{\BBOQ}\APACrefatitle {{Decision-making in sensorimotor control}}
  {{Decision-making in sensorimotor control}}.{\BBCQ}
\newblock
\APACjournalVolNumPages{{Nature Reviews Neuroscience}}{19}{9}{519--534}.
\newblock
\begin{APACrefDOI} \doi{10.1038/s41583-018-0045-9} \end{APACrefDOI}
\PrintBackRefs{\CurrentBib}

\bibitem [\protect \citeauthoryear {%
Gerber%
, Derckx%
, D{\"o}ppner%
\BCBL {}\ \BBA {} Schoder%
}{%
Gerber%
\ \protect \BOthers {.}}{%
{\protect \APACyear {2020}}%
}]{%
Gerber.2020}
\APACinsertmetastar {%
Gerber.2020}%
\begin{APACrefauthors}%
Gerber, A.%
, Derckx, P.%
, D{\"o}ppner, D\BPBI A.%
\BCBL {}\ \BBA {} Schoder, D.%
\end{APACrefauthors}%
\unskip\
\newblock
\APACrefYearMonthDay{2020}{}{}.
\newblock
{\BBOQ}\APACrefatitle {{Conceptualization of the human-machine symbiosis -- A
  literature review}} {{Conceptualization of the human-machine symbiosis -- A
  literature review}}.{\BBCQ}
\newblock
\BIn{} \APACrefbtitle {{Proceedings of the 53rd Hawaii International Conference
  on System Sciences 2020}.} {{Proceedings of the 53rd Hawaii International
  Conference on System Sciences 2020}.}
\newblock
\begin{APACrefDOI} \doi{10.24251/HICSS.2020.036} \end{APACrefDOI}
\PrintBackRefs{\CurrentBib}

\bibitem [\protect \citeauthoryear {%
Germann%
, Helmstetter%
, Fotler%
\BCBL {}\ \BBA {} Matthiesen%
}{%
Germann%
\ \protect \BOthers {.}}{%
{\protect \APACyear {2021}}%
}]{%
Germann.2021}
\APACinsertmetastar {%
Germann.2021}%
\begin{APACrefauthors}%
Germann, R.%
, Helmstetter, S.%
, Fotler, D.%
\BCBL {}\ \BBA {} Matthiesen, S.%
\end{APACrefauthors}%
\unskip\
\newblock
\APACrefYearMonthDay{2021}{}{}.
\newblock
{\BBOQ}\APACrefatitle {{Perceived usability in user-centered design: Analysis
  of usability aspects for improving human-machine systems: (in press)}}
  {{Perceived usability in user-centered design: Analysis of usability aspects
  for improving human-machine systems: (in press)}}.{\BBCQ}
\newblock
\BIn{} \APACrefbtitle {{Human-Automation Interaction}.} {{Human-Automation
  Interaction}.}
\newblock
\APACaddressPublisher{}{{Springer Nature Switzerland AG (Ed.)}}.
\PrintBackRefs{\CurrentBib}

\bibitem [\protect \citeauthoryear {%
Germann%
, Jahnke%
\BCBL {}\ \BBA {} Matthiesen%
}{%
Germann%
\ \protect \BOthers {.}}{%
{\protect \APACyear {2019}}%
}]{%
Germann.2019}
\APACinsertmetastar {%
Germann.2019}%
\begin{APACrefauthors}%
Germann, R.%
, Jahnke, B.%
\BCBL {}\ \BBA {} Matthiesen, S.%
\end{APACrefauthors}%
\unskip\
\newblock
\APACrefYearMonthDay{2019}{}{}.
\newblock
{\BBOQ}\APACrefatitle {{Objective usability evaluation of drywall screwdriver
  under consideration of the user experience}} {{Objective usability evaluation
  of drywall screwdriver under consideration of the user experience}}.{\BBCQ}
\newblock
\APACjournalVolNumPages{{Applied Ergonomics}}{75}{}{170--177}.
\newblock
\begin{APACrefDOI} \doi{10.1016/j.apergo.2018.10.001} \end{APACrefDOI}
\PrintBackRefs{\CurrentBib}

\bibitem [\protect \citeauthoryear {%
Gray%
, John%
\BCBL {}\ \BBA {} Atwood%
}{%
Gray%
\ \protect \BOthers {.}}{%
{\protect \APACyear {1993}}%
}]{%
Gray.1993}
\APACinsertmetastar {%
Gray.1993}%
\begin{APACrefauthors}%
Gray, W.%
, John, B.%
\BCBL {}\ \BBA {} Atwood, M.%
\end{APACrefauthors}%
\unskip\
\newblock
\APACrefYearMonthDay{1993}{}{}.
\newblock
{\BBOQ}\APACrefatitle {{Project ernestine: Validating a GOMS analysis for
  predicting and explaining real-world task performance}} {{Project ernestine:
  Validating a GOMS analysis for predicting and explaining real-world task
  performance}}.{\BBCQ}
\newblock
\APACjournalVolNumPages{{Human-Computer Interaction}}{8}{3}{237--309}.
\newblock
\begin{APACrefDOI} \doi{10.1207/s15327051hci0803_3} \end{APACrefDOI}
\PrintBackRefs{\CurrentBib}

\bibitem [\protect \citeauthoryear {%
Green%
, Billinghurst%
, Chen%
\BCBL {}\ \BBA {} Chase%
}{%
Green%
\ \protect \BOthers {.}}{%
{\protect \APACyear {2007}}%
}]{%
Green.2007}
\APACinsertmetastar {%
Green.2007}%
\begin{APACrefauthors}%
Green, S.%
, Billinghurst, M.%
, Chen, X.%
\BCBL {}\ \BBA {} Chase, J\BPBI G.%
\end{APACrefauthors}%
\unskip\
\newblock
\APACrefYearMonthDay{2007}{}{}.
\newblock
{\BBOQ}\APACrefatitle {{Human robot collaboration: An augmented reality
  approach: A literature review and analysis}} {{Human robot collaboration: An
  augmented reality approach: A literature review and analysis}}.{\BBCQ}
\newblock
\APACjournalVolNumPages{{Proceedings of MESA 07}}{}{}{}.
\PrintBackRefs{\CurrentBib}

\bibitem [\protect \citeauthoryear {%
Griffith%
}{%
Griffith%
}{%
{\protect \APACyear {2006}}%
}]{%
Griffith.2006}
\APACinsertmetastar {%
Griffith.2006}%
\begin{APACrefauthors}%
Griffith, D.%
\end{APACrefauthors}%
\unskip\
\newblock
\APACrefYearMonthDay{2006}{}{}.
\newblock
{\BBOQ}\APACrefatitle {{Neo-symbiosis: a system design philosophy for diversity
  and enrichment}} {{Neo-symbiosis: a system design philosophy for diversity
  and enrichment}}.{\BBCQ}
\newblock
\APACjournalVolNumPages{{International Journal of Industrial
  Ergonomics}}{36}{12}{1075--1079}.
\newblock
\begin{APACrefDOI} \doi{10.1016/j.ergon.2006.09.008} \end{APACrefDOI}
\PrintBackRefs{\CurrentBib}

\bibitem [\protect \citeauthoryear {%
Grigsby%
}{%
Grigsby%
}{%
{\protect \APACyear {2018}}%
}]{%
Grigsby.2018}
\APACinsertmetastar {%
Grigsby.2018}%
\begin{APACrefauthors}%
Grigsby, S\BPBI S.%
\end{APACrefauthors}%
\unskip\
\newblock
\APACrefYearMonthDay{2018}{}{}.
\newblock
{\BBOQ}\APACrefatitle {{Artificial intelligence for advanced human-machine
  symbiosis}} {{Artificial intelligence for advanced human-machine
  symbiosis}}.{\BBCQ}
\newblock
\BIn{} \APACrefbtitle {{International Conference on Augmented Cognition}}
  {{International Conference on Augmented Cognition}}\ (\BPGS\ 255--266).
\newblock
\begin{APACrefDOI} \doi{10.1007/978-3-319-91470-1_22} \end{APACrefDOI}
\PrintBackRefs{\CurrentBib}

\bibitem [\protect \citeauthoryear {%
Gupta%
, Singh%
, Verma%
, Mondal%
\BCBL {}\ \BBA {} Gupta%
}{%
Gupta%
\ \protect \BOthers {.}}{%
{\protect \APACyear {2020}}%
}]{%
Gupta.2020}
\APACinsertmetastar {%
Gupta.2020}%
\begin{APACrefauthors}%
Gupta, A.%
, Singh, A.%
, Verma, V.%
, Mondal, A\BPBI K.%
\BCBL {}\ \BBA {} Gupta, M\BPBI K.%
\end{APACrefauthors}%
\unskip\
\newblock
\APACrefYearMonthDay{2020}{}{}.
\newblock
{\BBOQ}\APACrefatitle {{Developments and clinical evaluations of robotic
  exoskeleton technology for human upper-limb rehabilitation}} {{Developments
  and clinical evaluations of robotic exoskeleton technology for human
  upper-limb rehabilitation}}.{\BBCQ}
\newblock
\APACjournalVolNumPages{{Advanced Robotics}}{34}{15}{1023--1040}.
\newblock
\begin{APACrefDOI} \doi{10.1080/01691864.2020.1749926} \end{APACrefDOI}
\PrintBackRefs{\CurrentBib}

\bibitem [\protect \citeauthoryear {%
Haddadin%
\ \BBA {} Croft%
}{%
Haddadin%
\ \BBA {} Croft%
}{%
{\protect \APACyear {2016}}%
}]{%
Haddadin.2016}
\APACinsertmetastar {%
Haddadin.2016}%
\begin{APACrefauthors}%
Haddadin, S.%
\BCBT {}\ \BBA {} Croft, E.%
\end{APACrefauthors}%
\unskip\
\newblock
\APACrefYearMonthDay{2016}{}{}.
\newblock
{\BBOQ}\APACrefatitle {{Physical human--robot interaction}} {{Physical
  human--robot interaction}}.{\BBCQ}
\newblock
\BIn{} B.~Siciliano\ \BBA {} O.~Khatib\ (\BEDS), \APACrefbtitle {{Springer
  handbook of robotics}} {{Springer handbook of robotics}}\ (\BPGS\
  1835--1874).
\newblock
\APACaddressPublisher{Berlin and Heidelberg}{Springer}.
\newblock
\begin{APACrefDOI} \doi{10.1007/978-3-319-32552-1_69} \end{APACrefDOI}
\PrintBackRefs{\CurrentBib}

\bibitem [\protect \citeauthoryear {%
Haggard%
\ \BBA {} Chambon%
}{%
Haggard%
\ \BBA {} Chambon%
}{%
{\protect \APACyear {2012}}%
}]{%
Haggard.2012}
\APACinsertmetastar {%
Haggard.2012}%
\begin{APACrefauthors}%
Haggard, P.%
\BCBT {}\ \BBA {} Chambon, V.%
\end{APACrefauthors}%
\unskip\
\newblock
\APACrefYearMonthDay{2012}{}{}.
\newblock
{\BBOQ}\APACrefatitle {{Sense of agency}} {{Sense of agency}}.{\BBCQ}
\newblock
\APACjournalVolNumPages{{Current Biology}}{22}{10}{R390-R392}.
\newblock
\begin{APACrefDOI} \doi{10.1016/j.cub.2012.02.040} \end{APACrefDOI}
\PrintBackRefs{\CurrentBib}

\bibitem [\protect \citeauthoryear {%
Hoelz%
, Kleinhans%
\BCBL {}\ \BBA {} Matthiesen%
}{%
Hoelz%
\ \protect \BOthers {.}}{%
{\protect \APACyear {2021}}%
}]{%
Hoelz.2021}
\APACinsertmetastar {%
Hoelz.2021}%
\begin{APACrefauthors}%
Hoelz, K.%
, Kleinhans, L.%
\BCBL {}\ \BBA {} Matthiesen, S.%
\end{APACrefauthors}%
\unskip\
\newblock
\APACrefYearMonthDay{2021}{}{}.
\newblock
{\BBOQ}\APACrefatitle {{Wood screw design: influence of thread parameters on
  the withdrawal capacity}} {{Wood screw design: influence of thread parameters
  on the withdrawal capacity}}.{\BBCQ}
\newblock
\APACjournalVolNumPages{{European Journal of Wood and Wood
  Products}}{79}{4}{773--784}.
\newblock
\begin{APACrefDOI} \doi{10.1007/s00107-021-01668-4} \end{APACrefDOI}
\PrintBackRefs{\CurrentBib}

\bibitem [\protect \citeauthoryear {%
Hoffman%
\ \BBA {} Novak%
}{%
Hoffman%
\ \BBA {} Novak%
}{%
{\protect \APACyear {2009}}%
}]{%
Hoffman.2009}
\APACinsertmetastar {%
Hoffman.2009}%
\begin{APACrefauthors}%
Hoffman, D\BPBI L.%
\BCBT {}\ \BBA {} Novak, T\BPBI P.%
\end{APACrefauthors}%
\unskip\
\newblock
\APACrefYearMonthDay{2009}{}{}.
\newblock
{\BBOQ}\APACrefatitle {{Flow online: Lessons learned and future prospects}}
  {{Flow online: Lessons learned and future prospects}}.{\BBCQ}
\newblock
\APACjournalVolNumPages{{Journal of Interactive Marketing}}{23}{1}{23--34}.
\newblock
\begin{APACrefDOI} \doi{10.1016/j.intmar.2008.10.003} \end{APACrefDOI}
\PrintBackRefs{\CurrentBib}

\bibitem [\protect \citeauthoryear {%
Hoffmann%
, Bock%
\BCBL {}\ \BBA {} {Rosenthal v.d. P{\"u}tten, Astrid M.}%
}{%
Hoffmann%
\ \protect \BOthers {.}}{%
{\protect \APACyear {2018}}%
}]{%
Hoffmann.2018}
\APACinsertmetastar {%
Hoffmann.2018}%
\begin{APACrefauthors}%
Hoffmann, L.%
, Bock, N.%
\BCBL {}\ \BBA {} {Rosenthal v.d. P{\"u}tten, Astrid M.}%
\end{APACrefauthors}%
\unskip\
\newblock
\APACrefYearMonthDay{2018}{}{}.
\newblock
{\BBOQ}\APACrefatitle {{The peculiarities of robot embodiment (EmCorp-Scale)}}
  {{The peculiarities of robot embodiment (EmCorp-Scale)}}.{\BBCQ}
\newblock
\BIn{} \APACrefbtitle {{Proceedings of the 2018 ACM/IEEE International
  Conference on Human-Robot Interaction (HRI'18)}} {{Proceedings of the 2018
  ACM/IEEE International Conference on Human-Robot Interaction (HRI'18)}}\
  (\BPGS\ 370--378).
\newblock
\begin{APACrefDOI} \doi{10.1145/3171221.3171242} \end{APACrefDOI}
\PrintBackRefs{\CurrentBib}

\bibitem [\protect \citeauthoryear {%
Honing%
, Gibo%
, Kuiper%
\BCBL {}\ \BBA {} Abbink%
}{%
Honing%
\ \protect \BOthers {.}}{%
{\protect \APACyear {2014}}%
}]{%
Honing.2014}
\APACinsertmetastar {%
Honing.2014}%
\begin{APACrefauthors}%
Honing, V.%
, Gibo, T\BPBI L.%
, Kuiper, R\BPBI J.%
\BCBL {}\ \BBA {} Abbink, D\BPBI A.%
\end{APACrefauthors}%
\unskip\
\newblock
\APACrefYearMonthDay{2014}{}{}.
\newblock
{\BBOQ}\APACrefatitle {{Training with haptic shared control to learn a slow
  dynamic system}} {{Training with haptic shared control to learn a slow
  dynamic system}}.{\BBCQ}
\newblock
\BIn{} \APACrefbtitle {{2014 IEEE International Conference on Systems, Man, and
  Cybernetics (SMC)}.} {{2014 IEEE International Conference on Systems, Man,
  and Cybernetics (SMC)}.}
\newblock
\APACaddressPublisher{}{IEEE}.
\newblock
\begin{APACrefDOI} \doi{10.1109/smc.2014.6974408} \end{APACrefDOI}
\PrintBackRefs{\CurrentBib}

\bibitem [\protect \citeauthoryear {%
Hughes%
, Desantis%
\BCBL {}\ \BBA {} Waszak%
}{%
Hughes%
\ \protect \BOthers {.}}{%
{\protect \APACyear {2013}}%
}]{%
Hughes.2013}
\APACinsertmetastar {%
Hughes.2013}%
\begin{APACrefauthors}%
Hughes, G.%
, Desantis, A.%
\BCBL {}\ \BBA {} Waszak, F.%
\end{APACrefauthors}%
\unskip\
\newblock
\APACrefYearMonthDay{2013}{}{}.
\newblock
{\BBOQ}\APACrefatitle {{Mechanisms of intentional binding and sensory
  attenuation: the role of temporal prediction, temporal control, identity
  prediction, and motor prediction}} {{Mechanisms of intentional binding and
  sensory attenuation: the role of temporal prediction, temporal control,
  identity prediction, and motor prediction}}.{\BBCQ}
\newblock
\APACjournalVolNumPages{{Psychological Bulletin}}{139}{1}{133--151}.
\newblock
\begin{APACrefDOI} \doi{10.1037/a0028566} \end{APACrefDOI}
\PrintBackRefs{\CurrentBib}

\bibitem [\protect \citeauthoryear {%
Humphreys%
\ \BBA {} Buehner%
}{%
Humphreys%
\ \BBA {} Buehner%
}{%
{\protect \APACyear {2009}}%
}]{%
Humphreys.2009}
\APACinsertmetastar {%
Humphreys.2009}%
\begin{APACrefauthors}%
Humphreys, G\BPBI R.%
\BCBT {}\ \BBA {} Buehner, M\BPBI J.%
\end{APACrefauthors}%
\unskip\
\newblock
\APACrefYearMonthDay{2009}{}{}.
\newblock
{\BBOQ}\APACrefatitle {{Magnitude estimation reveals temporal binding at
  super-second intervals}} {{Magnitude estimation reveals temporal binding at
  super-second intervals}}.{\BBCQ}
\newblock
\APACjournalVolNumPages{{Journal of Experimental Psychology. Human Perception
  and Performance}}{35}{5}{1542--1549}.
\newblock
\begin{APACrefDOI} \doi{10.1037/a0014492} \end{APACrefDOI}
\PrintBackRefs{\CurrentBib}

\bibitem [\protect \citeauthoryear {%
Imaizumi%
\ \BBA {} Tanno%
}{%
Imaizumi%
\ \BBA {} Tanno%
}{%
{\protect \APACyear {2019}}%
}]{%
Imaizumi.2019}
\APACinsertmetastar {%
Imaizumi.2019}%
\begin{APACrefauthors}%
Imaizumi, S.%
\BCBT {}\ \BBA {} Tanno, Y.%
\end{APACrefauthors}%
\unskip\
\newblock
\APACrefYearMonthDay{2019}{}{}.
\newblock
{\BBOQ}\APACrefatitle {{Intentional binding coincides with explicit sense of
  agency}} {{Intentional binding coincides with explicit sense of
  agency}}.{\BBCQ}
\newblock
\APACjournalVolNumPages{{Consciousness and Cognition}}{67}{}{1--15}.
\newblock
\begin{APACrefDOI} \doi{10.1016/j.concog.2018.11.005} \end{APACrefDOI}
\PrintBackRefs{\CurrentBib}

\bibitem [\protect \citeauthoryear {%
Inga%
, Eitel%
, Flad%
\BCBL {}\ \BBA {} Hohmann%
}{%
Inga%
\ \protect \BOthers {.}}{%
{\protect \APACyear {2018}}%
}]{%
Inga.2018}
\APACinsertmetastar {%
Inga.2018}%
\begin{APACrefauthors}%
Inga, J.%
, Eitel, M.%
, Flad, M.%
\BCBL {}\ \BBA {} Hohmann, S.%
\end{APACrefauthors}%
\unskip\
\newblock
\APACrefYearMonthDay{2018}{}{}.
\newblock
{\BBOQ}\APACrefatitle {{Evaluating Human Behavior in Manual and Shared Control
  via Inverse Optimization}} {{Evaluating Human Behavior in Manual and Shared
  Control via Inverse Optimization}}.{\BBCQ}
\newblock
\BIn{} \APACrefbtitle {{2018 IEEE International Conference on Systems, Man, and
  Cybernetics (SMC)}.} {{2018 IEEE International Conference on Systems, Man,
  and Cybernetics (SMC)}.}
\newblock
\APACaddressPublisher{}{IEEE}.
\newblock
\begin{APACrefDOI} \doi{10.1109/smc.2018.00461} \end{APACrefDOI}
\PrintBackRefs{\CurrentBib}

\bibitem [\protect \citeauthoryear {%
Jackson%
, Martin%
\BCBL {}\ \BBA {} Eklund%
}{%
Jackson%
\ \protect \BOthers {.}}{%
{\protect \APACyear {2008}}%
}]{%
Jackson.2008}
\APACinsertmetastar {%
Jackson.2008}%
\begin{APACrefauthors}%
Jackson, S\BPBI A.%
, Martin, A\BPBI J.%
\BCBL {}\ \BBA {} Eklund, R\BPBI C.%
\end{APACrefauthors}%
\unskip\
\newblock
\APACrefYearMonthDay{2008}{}{}.
\newblock
{\BBOQ}\APACrefatitle {{Long and short measures of flow: Examining construct
  validity of the FSS-2, DFS-2, and new brief counterparts}} {{Long and short
  measures of flow: Examining construct validity of the FSS-2, DFS-2, and new
  brief counterparts}}.{\BBCQ}
\newblock
\APACjournalVolNumPages{{Journal of Sport and Exercise
  Psychology}}{}{30}{561--587}.
\PrintBackRefs{\CurrentBib}

\bibitem [\protect \citeauthoryear {%
Jarrahi%
}{%
Jarrahi%
}{%
{\protect \APACyear {2018}}%
}]{%
Jarrahi.2018}
\APACinsertmetastar {%
Jarrahi.2018}%
\begin{APACrefauthors}%
Jarrahi, M\BPBI H.%
\end{APACrefauthors}%
\unskip\
\newblock
\APACrefYearMonthDay{2018}{}{}.
\newblock
{\BBOQ}\APACrefatitle {{Artificial intelligence and the future of work:
  Human-AI symbiosis in organizational decision making}} {{Artificial
  intelligence and the future of work: Human-AI symbiosis in organizational
  decision making}}.{\BBCQ}
\newblock
\APACjournalVolNumPages{{Business Horizons}}{61}{4}{577--586}.
\newblock
\begin{APACrefDOI} \doi{10.1016/j.bushor.2018.03.007} \end{APACrefDOI}
\PrintBackRefs{\CurrentBib}

\bibitem [\protect \citeauthoryear {%
Jarrass{\'e}%
, Sanguineti%
\BCBL {}\ \BBA {} Burdet%
}{%
Jarrass{\'e}%
\ \protect \BOthers {.}}{%
{\protect \APACyear {2014}}%
}]{%
Jarrasse.2014}
\APACinsertmetastar {%
Jarrasse.2014}%
\begin{APACrefauthors}%
Jarrass{\'e}, N.%
, Sanguineti, V.%
\BCBL {}\ \BBA {} Burdet, E.%
\end{APACrefauthors}%
\unskip\
\newblock
\APACrefYearMonthDay{2014}{}{}.
\newblock
{\BBOQ}\APACrefatitle {{Slaves no longer: review on role assignment for
  human--robot joint motor action}} {{Slaves no longer: review on role
  assignment for human--robot joint motor action}}.{\BBCQ}
\newblock
\APACjournalVolNumPages{{Adaptive Behavior}}{22}{1}{70--82}.
\newblock
\begin{APACrefDOI} \doi{10.1177/1059712313481044} \end{APACrefDOI}
\PrintBackRefs{\CurrentBib}

\bibitem [\protect \citeauthoryear {%
Jenkins%
\ \protect \BOthers {.}}{%
Jenkins%
\ \protect \BOthers {.}}{%
{\protect \APACyear {2021}}%
}]{%
Jenkins.2021}
\APACinsertmetastar {%
Jenkins.2021}%
\begin{APACrefauthors}%
Jenkins, M.%
, Esemezie, O.%
, Lee, V.%
, Mensingh, M.%
, Nagales, K.%
\BCBL {}\ \BBA {} Obhi, S\BPBI S.%
\end{APACrefauthors}%
\unskip\
\newblock
\APACrefYearMonthDay{2021}{}{}.
\newblock
{\BBOQ}\APACrefatitle {{An investigation of {\textquotedbl}We{\textquotedbl}
  agency in co-operative joint actions}} {{An investigation of
  {\textquotedbl}We{\textquotedbl} agency in co-operative joint
  actions}}.{\BBCQ}
\newblock
\APACjournalVolNumPages{{Psychological Research}}{85}{8}{3167--3181}.
\newblock
\begin{APACrefDOI} \doi{10.1007/s00426-020-01462-6} \end{APACrefDOI}
\PrintBackRefs{\CurrentBib}

\bibitem [\protect \citeauthoryear {%
Khademian%
\ \BBA {} Hashtrudi-Zaad%
}{%
Khademian%
\ \BBA {} Hashtrudi-Zaad%
}{%
{\protect \APACyear {2011}}%
}]{%
Khademian.2011}
\APACinsertmetastar {%
Khademian.2011}%
\begin{APACrefauthors}%
Khademian, B.%
\BCBT {}\ \BBA {} Hashtrudi-Zaad, K.%
\end{APACrefauthors}%
\unskip\
\newblock
\APACrefYearMonthDay{2011}{}{}.
\newblock
{\BBOQ}\APACrefatitle {{Shared control architectures for haptic training:
  Performance and coupled stability analysis}} {{Shared control architectures
  for haptic training: Performance and coupled stability analysis}}.{\BBCQ}
\newblock
\APACjournalVolNumPages{{The International Journal of Robotics
  Research}}{30}{}{1627--1642}.
\newblock
\begin{APACrefDOI} \doi{10.1177/0278364910397559} \end{APACrefDOI}
\PrintBackRefs{\CurrentBib}

\bibitem [\protect \citeauthoryear {%
King%
\ \BBA {} He%
}{%
King%
\ \BBA {} He%
}{%
{\protect \APACyear {2006}}%
}]{%
King.2006}
\APACinsertmetastar {%
King.2006}%
\begin{APACrefauthors}%
King, W\BPBI R.%
\BCBT {}\ \BBA {} He, J.%
\end{APACrefauthors}%
\unskip\
\newblock
\APACrefYearMonthDay{2006}{}{}.
\newblock
{\BBOQ}\APACrefatitle {{A meta-analysis of the technology acceptance model}}
  {{A meta-analysis of the technology acceptance model}}.{\BBCQ}
\newblock
\APACjournalVolNumPages{{Information {\&} Management}}{43}{6}{740--755}.
\newblock
\begin{APACrefDOI} \doi{10.1016/j.im.2006.05.003} \end{APACrefDOI}
\PrintBackRefs{\CurrentBib}

\bibitem [\protect \citeauthoryear {%
Kucukyilmaz%
, Sezgin%
\BCBL {}\ \BBA {} Basdogan%
}{%
Kucukyilmaz%
\ \protect \BOthers {.}}{%
{\protect \APACyear {2011}}%
}]{%
Kucukyilmaz.2011}
\APACinsertmetastar {%
Kucukyilmaz.2011}%
\begin{APACrefauthors}%
Kucukyilmaz, A.%
, Sezgin, T\BPBI M.%
\BCBL {}\ \BBA {} Basdogan, C.%
\end{APACrefauthors}%
\unskip\
\newblock
\APACrefYearMonthDay{2011}{}{}.
\newblock
{\BBOQ}\APACrefatitle {{Conveying intentions through haptics in human-computer
  collaboration}} {{Conveying intentions through haptics in human-computer
  collaboration}}.{\BBCQ}
\newblock
\BIn{} \APACrefbtitle {{2011 IEEE World Haptics Conference}} {{2011 IEEE World
  Haptics Conference}}\ (\BPGS\ 421--426).
\newblock
\APACaddressPublisher{}{IEEE}.
\newblock
\begin{APACrefDOI} \doi{10.1109/WHC.2011.5945523} \end{APACrefDOI}
\PrintBackRefs{\CurrentBib}

\bibitem [\protect \citeauthoryear {%
Kucukyilmaz%
, Sezgin%
\BCBL {}\ \BBA {} Basdogan%
}{%
Kucukyilmaz%
\ \protect \BOthers {.}}{%
{\protect \APACyear {2013}}%
}]{%
Kucukyilmaz.2013}
\APACinsertmetastar {%
Kucukyilmaz.2013}%
\begin{APACrefauthors}%
Kucukyilmaz, A.%
, Sezgin, T\BPBI M.%
\BCBL {}\ \BBA {} Basdogan, C.%
\end{APACrefauthors}%
\unskip\
\newblock
\APACrefYearMonthDay{2013}{}{}.
\newblock
{\BBOQ}\APACrefatitle {{Intention recognition for dynamic role exchange in
  haptic collaboration}} {{Intention recognition for dynamic role exchange in
  haptic collaboration}}.{\BBCQ}
\newblock
\APACjournalVolNumPages{{IEEE Transactions on Haptics}}{6}{1}{58--68}.
\newblock
\begin{APACrefDOI} \doi{10.1109/TOH.2012.21} \end{APACrefDOI}
\PrintBackRefs{\CurrentBib}

\bibitem [\protect \citeauthoryear {%
Lesh%
, Marks%
, Rich%
\BCBL {}\ \BBA {} Sidner%
}{%
Lesh%
\ \protect \BOthers {.}}{%
{\protect \APACyear {2004}}%
}]{%
Lesh.2004}
\APACinsertmetastar {%
Lesh.2004}%
\begin{APACrefauthors}%
Lesh, N.%
, Marks, J.%
, Rich, C.%
\BCBL {}\ \BBA {} Sidner, C\BPBI L.%
\end{APACrefauthors}%
\unskip\
\newblock
\APACrefYearMonthDay{2004}{}{}.
\newblock
{\BBOQ}\APACrefatitle {{{\textquotedbl}Man-computer symbiosis{\textquotedbl}
  revisited: Achieving natural communication and collaboration with computers}}
  {{{\textquotedbl}Man-computer symbiosis{\textquotedbl} revisited: Achieving
  natural communication and collaboration with computers}}.{\BBCQ}
\newblock
\APACjournalVolNumPages{{IEICE Transactions on Information and
  Systems}}{E87-D}{6}{1290--1298}.
\PrintBackRefs{\CurrentBib}

\bibitem [\protect \citeauthoryear {%
Levin%
\ \protect \BOthers {.}}{%
Levin%
\ \protect \BOthers {.}}{%
{\protect \APACyear {2012}}%
}]{%
Levin.2012}
\APACinsertmetastar {%
Levin.2012}%
\begin{APACrefauthors}%
Levin, S\BPBI A.%
, Carpenter, S\BPBI R.%
, Godfray, H\BPBI C\BPBI J.%
, Kinzig, A\BPBI P.%
, Loreau, M.%
, Losos, J\BPBI B.%
\BDBL {}Wilcove, D\BPBI S.%
\end{APACrefauthors}%
\unskip\
\newblock
\APACrefYear{2012}.
\newblock
\APACrefbtitle {{The princeton guide to ecology}} {{The princeton guide to
  ecology}}.
\newblock
\APACaddressPublisher{Princeton}{{Princeton University Press}}.
\PrintBackRefs{\CurrentBib}

\bibitem [\protect \citeauthoryear {%
Li%
\ \protect \BOthers {.}}{%
Li%
\ \protect \BOthers {.}}{%
{\protect \APACyear {2015}}%
}]{%
Li.2015}
\APACinsertmetastar {%
Li.2015}%
\begin{APACrefauthors}%
Li, Y.%
, Tee, K\BPBI P.%
, Chan, W\BPBI L.%
, Yan, R.%
, Chua, Y.%
\BCBL {}\ \BBA {} Limbu, D\BPBI K.%
\end{APACrefauthors}%
\unskip\
\newblock
\APACrefYearMonthDay{2015}{}{}.
\newblock
{\BBOQ}\APACrefatitle {{Role adaptation of human and robot in collaborative
  tasks}} {{Role adaptation of human and robot in collaborative tasks}}.{\BBCQ}
\newblock
\BIn{} \APACrefbtitle {{2015 IEEE International Conference on Robotics and
  Automation (ICRA)}.} {{2015 IEEE International Conference on Robotics and
  Automation (ICRA)}.}
\newblock
\APACaddressPublisher{}{IEEE}.
\newblock
\begin{APACrefDOI} \doi{10.1109/icra.2015.7139983} \end{APACrefDOI}
\PrintBackRefs{\CurrentBib}

\bibitem [\protect \citeauthoryear {%
Licklider%
}{%
Licklider%
}{%
{\protect \APACyear {1960}}%
}]{%
Licklider.1960}
\APACinsertmetastar {%
Licklider.1960}%
\begin{APACrefauthors}%
Licklider, J\BPBI C\BPBI R.%
\end{APACrefauthors}%
\unskip\
\newblock
\APACrefYearMonthDay{1960}{}{}.
\newblock
{\BBOQ}\APACrefatitle {{Man-Computer Symbiosis}} {{Man-Computer
  Symbiosis}}.{\BBCQ}
\newblock
\APACjournalVolNumPages{{IRE Transactions on Human Factors in
  Electronics}}{HFE-1}{1}{4--11}.
\newblock
\begin{APACrefDOI} \doi{10.1109/THFE2.1960.4503259} \end{APACrefDOI}
\PrintBackRefs{\CurrentBib}

\bibitem [\protect \citeauthoryear {%
Mars%
\ \BBA {} Chevrel%
}{%
Mars%
\ \BBA {} Chevrel%
}{%
{\protect \APACyear {2017}}%
}]{%
Mars.2017}
\APACinsertmetastar {%
Mars.2017}%
\begin{APACrefauthors}%
Mars, F.%
\BCBT {}\ \BBA {} Chevrel, P.%
\end{APACrefauthors}%
\unskip\
\newblock
\APACrefYearMonthDay{2017}{}{}.
\newblock
{\BBOQ}\APACrefatitle {{Modelling human control of steering for the design of
  advanced driver assistance systems}} {{Modelling human control of steering
  for the design of advanced driver assistance systems}}.{\BBCQ}
\newblock
\APACjournalVolNumPages{{Annual Reviews in Control}}{44}{}{292--302}.
\newblock
\begin{APACrefDOI} \doi{10.1016/j.arcontrol.2017.09.011} \end{APACrefDOI}
\PrintBackRefs{\CurrentBib}

\bibitem [\protect \citeauthoryear {%
A\BPBI J.~Martin%
\ \BBA {} Jackson%
}{%
A\BPBI J.~Martin%
\ \BBA {} Jackson%
}{%
{\protect \APACyear {2008}}%
}]{%
Martin.2008}
\APACinsertmetastar {%
Martin.2008}%
\begin{APACrefauthors}%
Martin, A\BPBI J.%
\BCBT {}\ \BBA {} Jackson, S\BPBI A.%
\end{APACrefauthors}%
\unskip\
\newblock
\APACrefYearMonthDay{2008}{}{}.
\newblock
{\BBOQ}\APACrefatitle {{Brief approaches to assessing task absorption and
  enhanced subjective experience: Examining `short' and `core' flow in diverse
  performance domains}} {{Brief approaches to assessing task absorption and
  enhanced subjective experience: Examining `short' and `core' flow in diverse
  performance domains}}.{\BBCQ}
\newblock
\APACjournalVolNumPages{{Motivation and Emotion}}{32}{3}{141--157}.
\newblock
\begin{APACrefDOI} \doi{10.1007/s11031-008-9094-0} \end{APACrefDOI}
\PrintBackRefs{\CurrentBib}

\bibitem [\protect \citeauthoryear {%
B\BPBI D.~Martin%
\ \BBA {} Schwab%
}{%
B\BPBI D.~Martin%
\ \BBA {} Schwab%
}{%
{\protect \APACyear {2012}}%
}]{%
Martin.2012}
\APACinsertmetastar {%
Martin.2012}%
\begin{APACrefauthors}%
Martin, B\BPBI D.%
\BCBT {}\ \BBA {} Schwab, E.%
\end{APACrefauthors}%
\unskip\
\newblock
\APACrefYearMonthDay{2012}{}{}.
\newblock
{\BBOQ}\APACrefatitle {{Current usage of symbiosis and associated terminology}}
  {{Current usage of symbiosis and associated terminology}}.{\BBCQ}
\newblock
\APACjournalVolNumPages{{International Journal of Biology}}{5}{1}{32--45}.
\newblock
\begin{APACrefDOI} \doi{10.5539/ijb.v5n1p32} \end{APACrefDOI}
\PrintBackRefs{\CurrentBib}

\bibitem [\protect \citeauthoryear {%
Matheson%
, Minto%
, Zampieri%
, Faccio%
\BCBL {}\ \BBA {} Rosati%
}{%
Matheson%
\ \protect \BOthers {.}}{%
{\protect \APACyear {2019}}%
}]{%
Matheson.2019}
\APACinsertmetastar {%
Matheson.2019}%
\begin{APACrefauthors}%
Matheson, E.%
, Minto, R.%
, Zampieri, E\BPBI G\BPBI G.%
, Faccio, M.%
\BCBL {}\ \BBA {} Rosati, G.%
\end{APACrefauthors}%
\unskip\
\newblock
\APACrefYearMonthDay{2019}{}{}.
\newblock
{\BBOQ}\APACrefatitle {{Human--robot collaboration in manufacturing
  applications: A review}} {{Human--robot collaboration in manufacturing
  applications: A review}}.{\BBCQ}
\newblock
\APACjournalVolNumPages{{Robotics}}{8}{4}{100}.
\newblock
\begin{APACrefDOI} \doi{10.3390/robotics8040100} \end{APACrefDOI}
\PrintBackRefs{\CurrentBib}

\bibitem [\protect \citeauthoryear {%
Matthiesen%
\ \BBA {} Germann%
}{%
Matthiesen%
\ \BBA {} Germann%
}{%
{\protect \APACyear {2018}}%
}]{%
Matthiesen.2018}
\APACinsertmetastar {%
Matthiesen.2018}%
\begin{APACrefauthors}%
Matthiesen, S.%
\BCBT {}\ \BBA {} Germann, R.%
\end{APACrefauthors}%
\unskip\
\newblock
\APACrefYearMonthDay{2018}{}{}.
\newblock
{\BBOQ}\APACrefatitle {{Meaningful prediction parameters for evaluating the
  suitability of power tools for usage}} {{Meaningful prediction parameters for
  evaluating the suitability of power tools for usage}}.{\BBCQ}
\newblock
\APACjournalVolNumPages{{Procedia CIRP}}{70}{}{241--246}.
\newblock
\begin{APACrefDOI} \doi{10.1016/j.procir.2018.02.040} \end{APACrefDOI}
\PrintBackRefs{\CurrentBib}

\bibitem [\protect \citeauthoryear {%
Michon%
}{%
Michon%
}{%
{\protect \APACyear {1985}}%
}]{%
Michon.1985}
\APACinsertmetastar {%
Michon.1985}%
\begin{APACrefauthors}%
Michon, J\BPBI A.%
\end{APACrefauthors}%
\unskip\
\newblock
\APACrefYearMonthDay{1985}{}{}.
\newblock
{\BBOQ}\APACrefatitle {{A critical view of driver behavior models: What do we
  know, what should we do?}} {{A critical view of driver behavior models: What
  do we know, what should we do?}}{\BBCQ}
\newblock
\BIn{} \APACrefbtitle {{Human Behavior and Traffic Safety}} {{Human Behavior
  and Traffic Safety}}\ (\BPGS\ 485--524).
\newblock
\APACaddressPublisher{}{{Springer, Boston, MA}}.
\newblock
\begin{APACrefDOI} \doi{10.1007/978-1-4613-2173-6_19} \end{APACrefDOI}
\PrintBackRefs{\CurrentBib}

\bibitem [\protect \citeauthoryear {%
Moneta%
\ \BBA {} Csikszentmihalyi%
}{%
Moneta%
\ \BBA {} Csikszentmihalyi%
}{%
{\protect \APACyear {1996}}%
}]{%
Moneta.1996}
\APACinsertmetastar {%
Moneta.1996}%
\begin{APACrefauthors}%
Moneta, G\BPBI B.%
\BCBT {}\ \BBA {} Csikszentmihalyi, M.%
\end{APACrefauthors}%
\unskip\
\newblock
\APACrefYearMonthDay{1996}{}{}.
\newblock
{\BBOQ}\APACrefatitle {{The effect of perceived challenges and skills on the
  quality of subjective experience}} {{The effect of perceived challenges and
  skills on the quality of subjective experience}}.{\BBCQ}
\newblock
\APACjournalVolNumPages{{Journal of Personality}}{2}{64}{275--310}.
\newblock
\begin{APACrefDOI} \doi{10.1111/j.1467-6494.1996.tb00512.x} \end{APACrefDOI}
\PrintBackRefs{\CurrentBib}

\bibitem [\protect \citeauthoryear {%
M{\"o}rtl%
\ \protect \BOthers {.}}{%
M{\"o}rtl%
\ \protect \BOthers {.}}{%
{\protect \APACyear {2012}}%
}]{%
Mortl.2012}
\APACinsertmetastar {%
Mortl.2012}%
\begin{APACrefauthors}%
M{\"o}rtl, A.%
, Lawitzky, M.%
, Kucukyilmaz, A.%
, Sezgin, M.%
, Basdogan, C.%
\BCBL {}\ \BBA {} Hirche, S.%
\end{APACrefauthors}%
\unskip\
\newblock
\APACrefYearMonthDay{2012}{}{}.
\newblock
{\BBOQ}\APACrefatitle {{The role of roles: Physical cooperation between humans
  and robots}} {{The role of roles: Physical cooperation between humans and
  robots}}.{\BBCQ}
\newblock
\APACjournalVolNumPages{{The International Journal of Robotics
  Research}}{31}{13}{1656--1674}.
\newblock
\begin{APACrefDOI} \doi{10.1177/0278364912455366} \end{APACrefDOI}
\PrintBackRefs{\CurrentBib}

\bibitem [\protect \citeauthoryear {%
Musi{\'c}%
\ \BBA {} Hirche%
}{%
Musi{\'c}%
\ \BBA {} Hirche%
}{%
{\protect \APACyear {2017}}%
}]{%
Music.2017}
\APACinsertmetastar {%
Music.2017}%
\begin{APACrefauthors}%
Musi{\'c}, S.%
\BCBT {}\ \BBA {} Hirche, S.%
\end{APACrefauthors}%
\unskip\
\newblock
\APACrefYearMonthDay{2017}{}{}.
\newblock
{\BBOQ}\APACrefatitle {{Control sharing in human-robot team interaction}}
  {{Control sharing in human-robot team interaction}}.{\BBCQ}
\newblock
\APACjournalVolNumPages{{Annual Reviews in Control}}{44}{}{342--354}.
\newblock
\begin{APACrefDOI} \doi{10.1016/j.arcontrol.2017.09.017} \end{APACrefDOI}
\PrintBackRefs{\CurrentBib}

\bibitem [\protect \citeauthoryear {%
Nashed%
, Crevecoeur%
\BCBL {}\ \BBA {} Scott%
}{%
Nashed%
\ \protect \BOthers {.}}{%
{\protect \APACyear {2012}}%
}]{%
Nashed.2012}
\APACinsertmetastar {%
Nashed.2012}%
\begin{APACrefauthors}%
Nashed, J\BPBI Y.%
, Crevecoeur, F.%
\BCBL {}\ \BBA {} Scott, S\BPBI H.%
\end{APACrefauthors}%
\unskip\
\newblock
\APACrefYearMonthDay{2012}{}{}.
\newblock
{\BBOQ}\APACrefatitle {{Influence of the behavioral goal and environmental
  obstacles on rapid feedback responses}} {{Influence of the behavioral goal
  and environmental obstacles on rapid feedback responses}}.{\BBCQ}
\newblock
\APACjournalVolNumPages{{Journal of Neurophysiology}}{108}{4}{999--1009}.
\newblock
\begin{APACrefDOI} \doi{10.1152/jn.01089.2011} \end{APACrefDOI}
\PrintBackRefs{\CurrentBib}

\bibitem [\protect \citeauthoryear {%
Nielsen%
\ \BBA {} Levy%
}{%
Nielsen%
\ \BBA {} Levy%
}{%
{\protect \APACyear {1994}}%
}]{%
Nielsen.1994}
\APACinsertmetastar {%
Nielsen.1994}%
\begin{APACrefauthors}%
Nielsen, J.%
\BCBT {}\ \BBA {} Levy, J.%
\end{APACrefauthors}%
\unskip\
\newblock
\APACrefYearMonthDay{1994}{}{}.
\newblock
{\BBOQ}\APACrefatitle {{Measuring usability}} {{Measuring usability}}.{\BBCQ}
\newblock
\APACjournalVolNumPages{{Communications of the ACM}}{37}{4}{66--75}.
\newblock
\begin{APACrefDOI} \doi{10.1145/175276.175282} \end{APACrefDOI}
\PrintBackRefs{\CurrentBib}

\bibitem [\protect \citeauthoryear {%
Oguz%
, Kucukyilmaz%
, Sezgin%
\BCBL {}\ \BBA {} Basdogan%
}{%
Oguz%
\ \protect \BOthers {.}}{%
{\protect \APACyear {2010}}%
}]{%
Oguz.2010}
\APACinsertmetastar {%
Oguz.2010}%
\begin{APACrefauthors}%
Oguz, S\BPBI O.%
, Kucukyilmaz, A.%
, Sezgin, T\BPBI M.%
\BCBL {}\ \BBA {} Basdogan, C.%
\end{APACrefauthors}%
\unskip\
\newblock
\APACrefYearMonthDay{2010}{}{}.
\newblock
{\BBOQ}\APACrefatitle {{Haptic negotiation and role exchange for collaboration
  in virtual environments}} {{Haptic negotiation and role exchange for
  collaboration in virtual environments}}.{\BBCQ}
\newblock
\BIn{} \APACrefbtitle {{2010 IEEE Haptics Symposium}} {{2010 IEEE Haptics
  Symposium}}\ (\BPGS\ 371--378).
\newblock
\APACaddressPublisher{}{IEEE}.
\newblock
\begin{APACrefDOI} \doi{10.1109/HAPTIC.2010.5444628} \end{APACrefDOI}
\PrintBackRefs{\CurrentBib}

\bibitem [\protect \citeauthoryear {%
Pacaux-Lemoine%
\ \BBA {} Itoh%
}{%
Pacaux-Lemoine%
\ \BBA {} Itoh%
}{%
{\protect \APACyear {2015}}%
}]{%
PacauxLemoine.2015}
\APACinsertmetastar {%
PacauxLemoine.2015}%
\begin{APACrefauthors}%
Pacaux-Lemoine, M\BHBI P.%
\BCBT {}\ \BBA {} Itoh, M.%
\end{APACrefauthors}%
\unskip\
\newblock
\APACrefYearMonthDay{2015}{}{}.
\newblock
{\BBOQ}\APACrefatitle {{Towards vertical and horizontal extension of shared
  control concept}} {{Towards vertical and horizontal extension of shared
  control concept}}.{\BBCQ}
\newblock
\BIn{} \APACrefbtitle {{2015 IEEE International Conference on Robotics and
  Automation (ICRA)}} {{2015 IEEE International Conference on Robotics and
  Automation (ICRA)}}\ (\BPGS\ 3086--3091).
\newblock
\APACaddressPublisher{}{IEEE}.
\newblock
\begin{APACrefDOI} \doi{10.1109/smc.2015.536} \end{APACrefDOI}
\PrintBackRefs{\CurrentBib}

\bibitem [\protect \citeauthoryear {%
Parker%
\ \BBA {} Pin%
}{%
Parker%
\ \BBA {} Pin%
}{%
{\protect \APACyear {1988}}%
}]{%
Parker.1988}
\APACinsertmetastar {%
Parker.1988}%
\begin{APACrefauthors}%
Parker, L\BPBI E.%
\BCBT {}\ \BBA {} Pin, F\BPBI G.%
\end{APACrefauthors}%
\unskip\
\newblock
\APACrefYearMonthDay{1988}{}{}.
\newblock
{\BBOQ}\APACrefatitle {{Man-robot symbiosis: A framework for cooperative
  intelligence and control}} {{Man-robot symbiosis: A framework for cooperative
  intelligence and control}}.{\BBCQ}
\newblock
\BIn{} \APACrefbtitle {{Proceedings SPIE 1006, Space Station Automation IV}}
  {{Proceedings SPIE 1006, Space Station Automation IV}}\ (\BPGS\ 94--103).
\newblock
\begin{APACrefDOI} \doi{10.1117/12.949063} \end{APACrefDOI}
\PrintBackRefs{\CurrentBib}

\bibitem [\protect \citeauthoryear {%
Passenberg%
, Peer%
\BCBL {}\ \BBA {} BUSS%
}{%
Passenberg%
\ \protect \BOthers {.}}{%
{\protect \APACyear {2010}}%
}]{%
Passenberg.2010}
\APACinsertmetastar {%
Passenberg.2010}%
\begin{APACrefauthors}%
Passenberg, C.%
, Peer, A.%
\BCBL {}\ \BBA {} BUSS, M.%
\end{APACrefauthors}%
\unskip\
\newblock
\APACrefYearMonthDay{2010}{}{}.
\newblock
{\BBOQ}\APACrefatitle {{A survey of environment-, operator-, and task-adapted
  controllers for teleoperation systems}} {{A survey of environment-,
  operator-, and task-adapted controllers for teleoperation systems}}.{\BBCQ}
\newblock
\APACjournalVolNumPages{{Mechatronics}}{20}{7}{787--801}.
\newblock
\begin{APACrefDOI} \doi{10.1016/j.mechatronics.2010.04.005} \end{APACrefDOI}
\PrintBackRefs{\CurrentBib}

\bibitem [\protect \citeauthoryear {%
Peifer%
, Kluge%
, Rummel%
\BCBL {}\ \BBA {} Kolossa%
}{%
Peifer%
\ \protect \BOthers {.}}{%
{\protect \APACyear {2020}}%
}]{%
Peifer.2020}
\APACinsertmetastar {%
Peifer.2020}%
\begin{APACrefauthors}%
Peifer, C.%
, Kluge, A.%
, Rummel, N.%
\BCBL {}\ \BBA {} Kolossa, D.%
\end{APACrefauthors}%
\unskip\
\newblock
\APACrefYearMonthDay{2020}{}{}.
\newblock
{\BBOQ}\APACrefatitle {{Fostering flow experience in HCI to enhance and
  allocate human energy}} {{Fostering flow experience in HCI to enhance and
  allocate human energy}}.{\BBCQ}
\newblock
\BIn{} \APACrefbtitle {{International Conference on Human-Computer
  Interaction}} {{International Conference on Human-Computer Interaction}}\
  (\BPGS\ 204--220).
\PrintBackRefs{\CurrentBib}

\bibitem [\protect \citeauthoryear {%
Pervez%
\ \BBA {} Ryu%
}{%
Pervez%
\ \BBA {} Ryu%
}{%
{\protect \APACyear {2008}}%
}]{%
Pervez.2008}
\APACinsertmetastar {%
Pervez.2008}%
\begin{APACrefauthors}%
Pervez, A.%
\BCBT {}\ \BBA {} Ryu, J.%
\end{APACrefauthors}%
\unskip\
\newblock
\APACrefYearMonthDay{2008}{}{}.
\newblock
{\BBOQ}\APACrefatitle {{Safe physical human robot interaction-past, present and
  future}} {{Safe physical human robot interaction-past, present and
  future}}.{\BBCQ}
\newblock
\APACjournalVolNumPages{{Journal of Mechanical Science and
  Technology}}{22}{3}{469--483}.
\newblock
\begin{APACrefDOI} \doi{10.1007/s12206-007-1109-3} \end{APACrefDOI}
\PrintBackRefs{\CurrentBib}

\bibitem [\protect \citeauthoryear {%
Polito%
, Barnier%
\BCBL {}\ \BBA {} Woody%
}{%
Polito%
\ \protect \BOthers {.}}{%
{\protect \APACyear {2013}}%
}]{%
Polito.2013}
\APACinsertmetastar {%
Polito.2013}%
\begin{APACrefauthors}%
Polito, V.%
, Barnier, A\BPBI J.%
\BCBL {}\ \BBA {} Woody, E\BPBI Z.%
\end{APACrefauthors}%
\unskip\
\newblock
\APACrefYearMonthDay{2013}{}{}.
\newblock
{\BBOQ}\APACrefatitle {{Developing the Sense of Agency Rating Scale (SOARS): an
  empirical measure of agency disruption in hypnosis}} {{Developing the Sense
  of Agency Rating Scale (SOARS): an empirical measure of agency disruption in
  hypnosis}}.{\BBCQ}
\newblock
\APACjournalVolNumPages{{Consciousness and Cognition}}{22}{3}{684--696}.
\newblock
\begin{APACrefDOI} \doi{10.1016/j.concog.2013.04.003} \end{APACrefDOI}
\PrintBackRefs{\CurrentBib}

\bibitem [\protect \citeauthoryear {%
Pruszynski%
\ \BBA {} Scott%
}{%
Pruszynski%
\ \BBA {} Scott%
}{%
{\protect \APACyear {2012}}%
}]{%
Pruszynski.2012}
\APACinsertmetastar {%
Pruszynski.2012}%
\begin{APACrefauthors}%
Pruszynski, J\BPBI A.%
\BCBT {}\ \BBA {} Scott, S\BPBI H.%
\end{APACrefauthors}%
\unskip\
\newblock
\APACrefYearMonthDay{2012}{}{}.
\newblock
{\BBOQ}\APACrefatitle {{Optimal feedback control and the long-latency stretch
  response}} {{Optimal feedback control and the long-latency stretch
  response}}.{\BBCQ}
\newblock
\APACjournalVolNumPages{{Experimental Brain Research}}{218}{3}{341--359}.
\newblock
\begin{APACrefDOI} \doi{10.1007/s00221-012-3041-8} \end{APACrefDOI}
\PrintBackRefs{\CurrentBib}

\bibitem [\protect \citeauthoryear {%
Rahwan%
\ \protect \BOthers {.}}{%
Rahwan%
\ \protect \BOthers {.}}{%
{\protect \APACyear {2019}}%
}]{%
Rahwan.2019}
\APACinsertmetastar {%
Rahwan.2019}%
\begin{APACrefauthors}%
Rahwan, I.%
, Cebrian, M.%
, Obradovich, N.%
, Bongard, J.%
, Bonnefon, J\BHBI F.%
, Breazeal, C.%
\BDBL {}Wellman, M.%
\end{APACrefauthors}%
\unskip\
\newblock
\APACrefYearMonthDay{2019}{}{}.
\newblock
{\BBOQ}\APACrefatitle {{Machine behaviour}} {{Machine behaviour}}.{\BBCQ}
\newblock
\APACjournalVolNumPages{{Nature}}{568}{7753}{477--486}.
\newblock
\begin{APACrefDOI} \doi{10.1038/s41586-019-1138-y} \end{APACrefDOI}
\PrintBackRefs{\CurrentBib}

\bibitem [\protect \citeauthoryear {%
Raima%
, Ito%
, Saiki%
, Yamazaki%
\BCBL {}\ \BBA {} Kurita%
}{%
Raima%
\ \protect \BOthers {.}}{%
{\protect \APACyear {2020}}%
}]{%
Raima.2020}
\APACinsertmetastar {%
Raima.2020}%
\begin{APACrefauthors}%
Raima, C.%
, Ito, M.%
, Saiki, S.%
, Yamazaki, Y.%
\BCBL {}\ \BBA {} Kurita, Y.%
\end{APACrefauthors}%
\unskip\
\newblock
\APACrefYearMonthDay{2020}{}{}.
\newblock
{\BBOQ}\APACrefatitle {{Developing a sense of agency scale for heavy machinery
  operation}} {{Developing a sense of agency scale for heavy machinery
  operation}}.{\BBCQ}
\newblock
\BIn{} \APACrefbtitle {{2020 13th International Conference on Human System
  Interaction (HSI)}.} {{2020 13th International Conference on Human System
  Interaction (HSI)}.}
\newblock
\APACaddressPublisher{}{IEEE}.
\newblock
\begin{APACrefDOI} \doi{10.1109/hsi49210.2020.9142639} \end{APACrefDOI}
\PrintBackRefs{\CurrentBib}

\bibitem [\protect \citeauthoryear {%
Rasmussen%
}{%
Rasmussen%
}{%
{\protect \APACyear {1983}}%
}]{%
Rasmussen.1983}
\APACinsertmetastar {%
Rasmussen.1983}%
\begin{APACrefauthors}%
Rasmussen, J.%
\end{APACrefauthors}%
\unskip\
\newblock
\APACrefYearMonthDay{1983}{}{}.
\newblock
{\BBOQ}\APACrefatitle {{Skills, rules, and knowledge; signals, signs, and
  symbols, and other distinctions in human performance models}} {{Skills,
  rules, and knowledge; signals, signs, and symbols, and other distinctions in
  human performance models}}.{\BBCQ}
\newblock
\APACjournalVolNumPages{{IEEE Transactions on Systems, Man, and
  Cybernetics}}{SMC-13}{3}{257--266}.
\newblock
\begin{APACrefDOI} \doi{10.1109/tsmc.1983.6313160} \end{APACrefDOI}
\PrintBackRefs{\CurrentBib}

\bibitem [\protect \citeauthoryear {%
Reed%
\ \protect \BOthers {.}}{%
Reed%
\ \protect \BOthers {.}}{%
{\protect \APACyear {2006}}%
}]{%
Reed.2006}
\APACinsertmetastar {%
Reed.2006}%
\begin{APACrefauthors}%
Reed, K.%
, Peshkin, M.%
, Hartmann, M.%
, Patton, J.%
, Vishton, P.%
\BCBL {}\ \BBA {} Grabowecky, M.%
\end{APACrefauthors}%
\unskip\
\newblock
\APACrefYearMonthDay{2006}{}{}.
\newblock
{\BBOQ}\APACrefatitle {{Haptic cooperation between people, and between people
  and machines}} {{Haptic cooperation between people, and between people and
  machines}}.{\BBCQ}
\newblock
\BIn{} \APACrefbtitle {{2006 IEEE/RSJ International Conference on Intelligent
  Robots and Systems}} {{2006 IEEE/RSJ International Conference on Intelligent
  Robots and Systems}}\ (\BPGS\ 2109--2114).
\newblock
\APACaddressPublisher{}{IEEE}.
\newblock
\begin{APACrefDOI} \doi{10.1109/IROS.2006.282489} \end{APACrefDOI}
\PrintBackRefs{\CurrentBib}

\bibitem [\protect \citeauthoryear {%
Reinares-Lara%
, Olarte-Pascual%
\BCBL {}\ \BBA {} Pelegr{\'i}n-Borondo%
}{%
Reinares-Lara%
\ \protect \BOthers {.}}{%
{\protect \APACyear {2018}}%
}]{%
ReinaresLara.2018}
\APACinsertmetastar {%
ReinaresLara.2018}%
\begin{APACrefauthors}%
Reinares-Lara, E.%
, Olarte-Pascual, C.%
\BCBL {}\ \BBA {} Pelegr{\'i}n-Borondo, J.%
\end{APACrefauthors}%
\unskip\
\newblock
\APACrefYearMonthDay{2018}{}{}.
\newblock
{\BBOQ}\APACrefatitle {{Do you want to be a cyborg? The moderating effect of
  ethics on neural implant acceptance}} {{Do you want to be a cyborg? The
  moderating effect of ethics on neural implant acceptance}}.{\BBCQ}
\newblock
\APACjournalVolNumPages{{Computers in Human Behavior}}{85}{}{43--53}.
\newblock
\begin{APACrefDOI} \doi{10.1016/j.chb.2018.03.032} \end{APACrefDOI}
\PrintBackRefs{\CurrentBib}

\bibitem [\protect \citeauthoryear {%
Reuter%
, Mansell%
, Rhea%
\BCBL {}\ \BBA {} Kiesel%
}{%
Reuter%
\ \protect \BOthers {.}}{%
{\protect \APACyear {2021}}%
}]{%
Reuter.2021}
\APACinsertmetastar {%
Reuter.2021}%
\begin{APACrefauthors}%
Reuter, L.%
, Mansell, J.%
, Rhea, C.%
\BCBL {}\ \BBA {} Kiesel, A.%
\end{APACrefauthors}%
\unskip\
\newblock
\APACrefYearMonthDay{2021}{}{}.
\newblock
{\BBOQ}\APACrefatitle {{Direct Assessment of Individual Connotation and
  Experience: An Introduction to Cognitive-Affective Mapping.}} {{Direct
  Assessment of Individual Connotation and Experience: An Introduction to
  Cognitive-Affective Mapping.}}{\BBCQ}
\newblock
\APACjournalVolNumPages{{Politics and the Life Sciences}}{}{}{1--21}.
\PrintBackRefs{\CurrentBib}

\bibitem [\protect \citeauthoryear {%
Rheinberg%
\ \BBA {} Vollmeyer%
}{%
Rheinberg%
\ \BBA {} Vollmeyer%
}{%
{\protect \APACyear {2003}}%
}]{%
Rheinberg.2003}
\APACinsertmetastar {%
Rheinberg.2003}%
\begin{APACrefauthors}%
Rheinberg, F.%
\BCBT {}\ \BBA {} Vollmeyer, R.%
\end{APACrefauthors}%
\unskip\
\newblock
\APACrefYearMonthDay{2003}{}{}.
\newblock
{\BBOQ}\APACrefatitle {{Flow-Erleben in einem Computerspiel unter experimentell
  variierten Bedingungen}} {{Flow-Erleben in einem Computerspiel unter
  experimentell variierten Bedingungen}}.{\BBCQ}
\newblock
\APACjournalVolNumPages{{Zeitschrift f{\"u}r Psychologie}}{}{4}{161--170}.
\PrintBackRefs{\CurrentBib}

\bibitem [\protect \citeauthoryear {%
Rothfuß%
, Wörner%
, Inga%
\BCBL {}\ \BBA {} Hohmann%
}{%
Rothfuß%
\ \protect \BOthers {.}}{%
{\protect \APACyear {2020}}%
}]{%
Rothfuss.2020}
\APACinsertmetastar {%
Rothfuss.2020}%
\begin{APACrefauthors}%
Rothfuß, S.%
, Wörner, M.%
, Inga, J.%
\BCBL {}\ \BBA {} Hohmann, S.%
\end{APACrefauthors}%
\unskip\
\newblock
\APACrefYearMonthDay{2020}{}{}.
\newblock
{\BBOQ}\APACrefatitle {{A study on human-machine cooperation on decision
  level}} {{A study on human-machine cooperation on decision level}}.{\BBCQ}
\newblock
\BIn{} \APACrefbtitle {{2020 IEEE International Conference on Systems, Man, and
  Cybernetics (SMC)}} {{2020 IEEE International Conference on Systems, Man, and
  Cybernetics (SMC)}}\ (\BPGS\ 2291--2298).
\newblock
\APACaddressPublisher{}{IEEE}.
\newblock
\begin{APACrefDOI} \doi{10.1109/smc42975.2020.9282813} \end{APACrefDOI}
\PrintBackRefs{\CurrentBib}

\bibitem [\protect \citeauthoryear {%
Rothfuß%
, Wörner%
, Inga%
, Kiesel%
\BCBL {}\ \BBA {} Hohmann%
}{%
Rothfuß%
\ \protect \BOthers {.}}{%
{\protect \APACyear {2021}}%
}]{%
Rothfuss.2021}
\APACinsertmetastar {%
Rothfuss.2021}%
\begin{APACrefauthors}%
Rothfuß, S.%
, Wörner, M.%
, Inga, J.%
, Kiesel, A.%
\BCBL {}\ \BBA {} Hohmann, S.%
\end{APACrefauthors}%
\unskip\
\newblock
\APACrefYearMonthDay{2021}{}{}.
\newblock
{\BBOQ}\APACrefatitle {{Human-Machine Cooperative Decision Making Outperforms
  Individualism and Autonomy}} {{Human-Machine Cooperative Decision Making
  Outperforms Individualism and Autonomy}}.{\BBCQ}
\newblock

\newblock
\begin{APACrefDOI} \doi{10.36227/techrxiv.16780273.v1} \end{APACrefDOI}
\PrintBackRefs{\CurrentBib}

\bibitem [\protect \citeauthoryear {%
Ruess%
, Thomaschke%
\BCBL {}\ \BBA {} Kiesel%
}{%
Ruess%
\ \protect \BOthers {.}}{%
{\protect \APACyear {2017}}%
}]{%
Ruess.2017}
\APACinsertmetastar {%
Ruess.2017}%
\begin{APACrefauthors}%
Ruess, M.%
, Thomaschke, R.%
\BCBL {}\ \BBA {} Kiesel, A.%
\end{APACrefauthors}%
\unskip\
\newblock
\APACrefYearMonthDay{2017}{}{}.
\newblock
{\BBOQ}\APACrefatitle {{The time course of intentional binding}} {{The time
  course of intentional binding}}.{\BBCQ}
\newblock
\APACjournalVolNumPages{{Attention, Perception {\&}
  Psychophysics}}{79}{4}{1123--1131}.
\newblock
\begin{APACrefDOI} \doi{10.3758/s13414-017-1292-y} \end{APACrefDOI}
\PrintBackRefs{\CurrentBib}

\bibitem [\protect \citeauthoryear {%
Ruess%
, Thomaschke%
\BCBL {}\ \BBA {} Kiesel%
}{%
Ruess%
\ \protect \BOthers {.}}{%
{\protect \APACyear {2018}}%
}]{%
Ruess.2018}
\APACinsertmetastar {%
Ruess.2018}%
\begin{APACrefauthors}%
Ruess, M.%
, Thomaschke, R.%
\BCBL {}\ \BBA {} Kiesel, A.%
\end{APACrefauthors}%
\unskip\
\newblock
\APACrefYearMonthDay{2018}{}{}.
\newblock
{\BBOQ}\APACrefatitle {{Intentional binding of visual effects}} {{Intentional
  binding of visual effects}}.{\BBCQ}
\newblock
\APACjournalVolNumPages{{Attention, Perception {\&}
  Psychophysics}}{80}{3}{713--722}.
\newblock
\begin{APACrefDOI} \doi{10.3758/s13414-017-1479-2} \end{APACrefDOI}
\PrintBackRefs{\CurrentBib}

\bibitem [\protect \citeauthoryear {%
Ruess%
, Thomaschke%
\BCBL {}\ \BBA {} Kiesel%
}{%
Ruess%
\ \protect \BOthers {.}}{%
{\protect \APACyear {2020}}%
}]{%
Ruess.2020}
\APACinsertmetastar {%
Ruess.2020}%
\begin{APACrefauthors}%
Ruess, M.%
, Thomaschke, R.%
\BCBL {}\ \BBA {} Kiesel, A.%
\end{APACrefauthors}%
\unskip\
\newblock
\APACrefYearMonthDay{2020}{}{}.
\newblock
{\BBOQ}\APACrefatitle {{Acting and reacting: Is intentional binding due to
  sense of agency or to temporal expectancy?}} {{Acting and reacting: Is
  intentional binding due to sense of agency or to temporal
  expectancy?}}{\BBCQ}
\newblock
\APACjournalVolNumPages{{Journal of Experimental Psychology. Human Perception
  and Performance}}{46}{1}{1--9}.
\newblock
\begin{APACrefDOI} \doi{10.1037/xhp0000700} \end{APACrefDOI}
\PrintBackRefs{\CurrentBib}

\bibitem [\protect \citeauthoryear {%
Saridis%
}{%
Saridis%
}{%
{\protect \APACyear {1983}}%
}]{%
Saridis.1983}
\APACinsertmetastar {%
Saridis.1983}%
\begin{APACrefauthors}%
Saridis, G.%
\end{APACrefauthors}%
\unskip\
\newblock
\APACrefYearMonthDay{1983}{}{}.
\newblock
{\BBOQ}\APACrefatitle {{Intelligent robotic control}} {{Intelligent robotic
  control}}.{\BBCQ}
\newblock
\APACjournalVolNumPages{{IEEE Transactions on Automatic
  Control}}{28}{5}{547--557}.
\newblock
\begin{APACrefDOI} \doi{10.1109/tac.1983.1103278} \end{APACrefDOI}
\PrintBackRefs{\CurrentBib}

\bibitem [\protect \citeauthoryear {%
Schalk%
}{%
Schalk%
}{%
{\protect \APACyear {2008}}%
}]{%
Schalk.2008}
\APACinsertmetastar {%
Schalk.2008}%
\begin{APACrefauthors}%
Schalk, G.%
\end{APACrefauthors}%
\unskip\
\newblock
\APACrefYearMonthDay{2008}{}{}.
\newblock
{\BBOQ}\APACrefatitle {{Brain-computer symbiosis}} {{Brain-computer
  symbiosis}}.{\BBCQ}
\newblock
\APACjournalVolNumPages{{Journal of Neural Engineering}}{5}{1}{P1-P15}.
\newblock
\begin{APACrefDOI} \doi{10.1088/1741-2560/5/1/p01} \end{APACrefDOI}
\PrintBackRefs{\CurrentBib}

\bibitem [\protect \citeauthoryear {%
Schettler%
, Raja%
\BCBL {}\ \BBA {} Anderson%
}{%
Schettler%
\ \protect \BOthers {.}}{%
{\protect \APACyear {2019}}%
}]{%
Schettler.2019}
\APACinsertmetastar {%
Schettler.2019}%
\begin{APACrefauthors}%
Schettler, A.%
, Raja, V.%
\BCBL {}\ \BBA {} Anderson, M\BPBI L.%
\end{APACrefauthors}%
\unskip\
\newblock
\APACrefYearMonthDay{2019}{}{}.
\newblock
{\BBOQ}\APACrefatitle {{The Embodiment of Objects: Review, Analysis, and Future
  Directions}} {{The Embodiment of Objects: Review, Analysis, and Future
  Directions}}.{\BBCQ}
\newblock
\APACjournalVolNumPages{{Frontiers in Neuroscience}}{13}{}{1332}.
\newblock
\begin{APACrefDOI} \doi{10.3389/fnins.2019.01332} \end{APACrefDOI}
\PrintBackRefs{\CurrentBib}

\bibitem [\protect \citeauthoryear {%
Schmidtler%
, Knott%
, H{\"o}lzel%
\BCBL {}\ \BBA {} Bengler%
}{%
Schmidtler%
\ \protect \BOthers {.}}{%
{\protect \APACyear {2015}}%
}]{%
Schmidtler.2015}
\APACinsertmetastar {%
Schmidtler.2015}%
\begin{APACrefauthors}%
Schmidtler, J.%
, Knott, V.%
, H{\"o}lzel, C.%
\BCBL {}\ \BBA {} Bengler, K.%
\end{APACrefauthors}%
\unskip\
\newblock
\APACrefYearMonthDay{2015}{}{}.
\newblock
{\BBOQ}\APACrefatitle {{Human centered assistance applications for the working
  environment of the future}} {{Human centered assistance applications for the
  working environment of the future}}.{\BBCQ}
\newblock
\APACjournalVolNumPages{{Occupational Ergonomics}}{12}{3}{83--95}.
\newblock
\begin{APACrefDOI} \doi{10.3233/OER-150226} \end{APACrefDOI}
\PrintBackRefs{\CurrentBib}

\bibitem [\protect \citeauthoryear {%
Schoop%
\ \protect \BOthers {.}}{%
Schoop%
\ \protect \BOthers {.}}{%
{\protect \APACyear {2016}}%
}]{%
Schoop.2016}
\APACinsertmetastar {%
Schoop.2016}%
\begin{APACrefauthors}%
Schoop, E.%
, Nguyen, M.%
, Lim, D.%
, Savage, V.%
, Follmer, S.%
\BCBL {}\ \BBA {} Hartmann, B.%
\end{APACrefauthors}%
\unskip\
\newblock
\APACrefYearMonthDay{2016}{}{}.
\newblock
{\BBOQ}\APACrefatitle {{Drill Sergeant: Supporting physical construction
  projects through an ecosystem of augmented tools}} {{Drill Sergeant:
  Supporting physical construction projects through an ecosystem of augmented
  tools}}.{\BBCQ}
\newblock
\BIn{} \APACrefbtitle {{Proceedings of the 2016 CHI Conference Extended
  Abstracts on Human Factors in Computing Systems}.} {{Proceedings of the 2016
  CHI Conference Extended Abstracts on Human Factors in Computing Systems}.}
\PrintBackRefs{\CurrentBib}

\bibitem [\protect \citeauthoryear {%
Scott%
}{%
Scott%
}{%
{\protect \APACyear {2004}}%
}]{%
Scott.2004}
\APACinsertmetastar {%
Scott.2004}%
\begin{APACrefauthors}%
Scott, S\BPBI H.%
\end{APACrefauthors}%
\unskip\
\newblock
\APACrefYearMonthDay{2004}{}{}.
\newblock
{\BBOQ}\APACrefatitle {{Optimal feedback control and the neural basis of
  volitional motor control}} {{Optimal feedback control and the neural basis of
  volitional motor control}}.{\BBCQ}
\newblock
\APACjournalVolNumPages{{Nature Reviews Neuroscience}}{5}{7}{532--546}.
\newblock
\begin{APACrefDOI} \doi{10.1038/nrn1427} \end{APACrefDOI}
\PrintBackRefs{\CurrentBib}

\bibitem [\protect \citeauthoryear {%
Sheller%
}{%
Sheller%
}{%
{\protect \APACyear {2004}}%
}]{%
Sheller.2004}
\APACinsertmetastar {%
Sheller.2004}%
\begin{APACrefauthors}%
Sheller, M.%
\end{APACrefauthors}%
\unskip\
\newblock
\APACrefYearMonthDay{2004}{}{}.
\newblock
{\BBOQ}\APACrefatitle {{Automotive Emotions}} {{Automotive Emotions}}.{\BBCQ}
\newblock
\APACjournalVolNumPages{{Theory, Culture {\&} Society}}{21}{4-5}{221--242}.
\newblock
\begin{APACrefDOI} \doi{10.1177/0263276404046068} \end{APACrefDOI}
\PrintBackRefs{\CurrentBib}

\bibitem [\protect \citeauthoryear {%
Sheridan%
\ \BBA {} Verplank%
}{%
Sheridan%
\ \BBA {} Verplank%
}{%
{\protect \APACyear {1978}}%
}]{%
Sheridan.1978}
\APACinsertmetastar {%
Sheridan.1978}%
\begin{APACrefauthors}%
Sheridan, T\BPBI B.%
\BCBT {}\ \BBA {} Verplank, W\BPBI L.%
\end{APACrefauthors}%
\unskip\
\newblock
\APACrefYear{1978}.
\newblock
\APACrefbtitle {{Human and Computer Control of Undersea Teleoperators}} {{Human
  and Computer Control of Undersea Teleoperators}}.
\newblock
\APACaddressPublisher{Fort Belvoir, VA}{{Defense Technical Information
  Center}}.
\newblock
\begin{APACrefDOI} \doi{10.21236/ada057655} \end{APACrefDOI}
\PrintBackRefs{\CurrentBib}

\bibitem [\protect \citeauthoryear {%
Silverman%
}{%
Silverman%
}{%
{\protect \APACyear {1992}}%
}]{%
Silverman.1992}
\APACinsertmetastar {%
Silverman.1992}%
\begin{APACrefauthors}%
Silverman, B\BPBI G.%
\end{APACrefauthors}%
\unskip\
\newblock
\APACrefYearMonthDay{1992}{}{}.
\newblock
{\BBOQ}\APACrefatitle {{Human-Computer Collaboration}} {{Human-Computer
  Collaboration}}.{\BBCQ}
\newblock
\APACjournalVolNumPages{{Human-Computer Interaction}}{}{}{165--196}.
\PrintBackRefs{\CurrentBib}

\bibitem [\protect \citeauthoryear {%
Smisek%
, Sunil%
, {van Paassen}%
, Abbink%
\BCBL {}\ \BBA {} Mulder%
}{%
Smisek%
\ \protect \BOthers {.}}{%
{\protect \APACyear {2017}}%
}]{%
Smisek.2017}
\APACinsertmetastar {%
Smisek.2017}%
\begin{APACrefauthors}%
Smisek, J.%
, Sunil, E.%
, {van Paassen}, M\BPBI M.%
, Abbink, D\BPBI A.%
\BCBL {}\ \BBA {} Mulder, M.%
\end{APACrefauthors}%
\unskip\
\newblock
\APACrefYearMonthDay{2017}{}{}.
\newblock
{\BBOQ}\APACrefatitle {{Neuromuscular-system-based tuning of a haptic shared
  control interface for UAV teleoperation}} {{Neuromuscular-system-based tuning
  of a haptic shared control interface for UAV teleoperation}}.{\BBCQ}
\newblock
\APACjournalVolNumPages{{IEEE Transactions on Human-Machine
  Systems}}{47}{4}{449--461}.
\newblock
\begin{APACrefDOI} \doi{10.1109/thms.2016.2616280} \end{APACrefDOI}
\PrintBackRefs{\CurrentBib}

\bibitem [\protect \citeauthoryear {%
Sposito%
, Bolognini%
, Vallar%
\BCBL {}\ \BBA {} Maravita%
}{%
Sposito%
\ \protect \BOthers {.}}{%
{\protect \APACyear {2012}}%
}]{%
Sposito.2012}
\APACinsertmetastar {%
Sposito.2012}%
\begin{APACrefauthors}%
Sposito, A.%
, Bolognini, N.%
, Vallar, G.%
\BCBL {}\ \BBA {} Maravita, A.%
\end{APACrefauthors}%
\unskip\
\newblock
\APACrefYearMonthDay{2012}{}{}.
\newblock
{\BBOQ}\APACrefatitle {{Extension of perceived arm length following tool-use:
  clues to plasticity of body metrics}} {{Extension of perceived arm length
  following tool-use: clues to plasticity of body metrics}}.{\BBCQ}
\newblock
\APACjournalVolNumPages{{Neuropsychologia}}{50}{9}{2187--2194}.
\newblock
\begin{APACrefDOI} \doi{10.1016/j.neuropsychologia.2012.05.022}
  \end{APACrefDOI}
\PrintBackRefs{\CurrentBib}

\bibitem [\protect \citeauthoryear {%
Tapal%
, Oren%
, Dar%
\BCBL {}\ \BBA {} Eitam%
}{%
Tapal%
\ \protect \BOthers {.}}{%
{\protect \APACyear {2017}}%
}]{%
Tapal.2017}
\APACinsertmetastar {%
Tapal.2017}%
\begin{APACrefauthors}%
Tapal, A.%
, Oren, E.%
, Dar, R.%
\BCBL {}\ \BBA {} Eitam, B.%
\end{APACrefauthors}%
\unskip\
\newblock
\APACrefYearMonthDay{2017}{}{}.
\newblock
{\BBOQ}\APACrefatitle {{The Sense of Agency Scale: A measure of consciously
  perceived control over one's mind, body, and the immediate environment}}
  {{The Sense of Agency Scale: A measure of consciously perceived control over
  one's mind, body, and the immediate environment}}.{\BBCQ}
\newblock
\APACjournalVolNumPages{{Frontiers in Psychology}}{8}{}{1552}.
\newblock
\begin{APACrefDOI} \doi{10.3389/fpsyg.2017.01552} \end{APACrefDOI}
\PrintBackRefs{\CurrentBib}

\bibitem [\protect \citeauthoryear {%
Thagard%
}{%
Thagard%
}{%
{\protect \APACyear {2010}}%
}]{%
Thagard.2010}
\APACinsertmetastar {%
Thagard.2010}%
\begin{APACrefauthors}%
Thagard, P.%
\end{APACrefauthors}%
\unskip\
\newblock
\APACrefYearMonthDay{2010}{}{}.
\newblock
{\BBOQ}\APACrefatitle {{EMPATHICA: A computer support system with visual
  representations for Cognitive-Affective Mapping}} {{EMPATHICA: A computer
  support system with visual representations for Cognitive-Affective
  Mapping}}.{\BBCQ}
\newblock
\BIn{} \APACrefbtitle {{Workshops at the Twenty-Fourth AAAI Conference on
  Artificial Intelligence}.} {{Workshops at the Twenty-Fourth AAAI Conference
  on Artificial Intelligence}.}
\PrintBackRefs{\CurrentBib}

\bibitem [\protect \citeauthoryear {%
Todorov%
}{%
Todorov%
}{%
{\protect \APACyear {2004}}%
}]{%
Todorov.2004}
\APACinsertmetastar {%
Todorov.2004}%
\begin{APACrefauthors}%
Todorov, E.%
\end{APACrefauthors}%
\unskip\
\newblock
\APACrefYearMonthDay{2004}{}{}.
\newblock
{\BBOQ}\APACrefatitle {{Optimality principles in sensorimotor control}}
  {{Optimality principles in sensorimotor control}}.{\BBCQ}
\newblock
\APACjournalVolNumPages{{Nature Neuroscience}}{7}{9}{907--915}.
\newblock
\begin{APACrefDOI} \doi{10.1038/nn1309} \end{APACrefDOI}
\PrintBackRefs{\CurrentBib}

\bibitem [\protect \citeauthoryear {%
Toft%
, Schuitema%
\BCBL {}\ \BBA {} Th{\o}gersen%
}{%
Toft%
\ \protect \BOthers {.}}{%
{\protect \APACyear {2014}}%
}]{%
Toft.2014}
\APACinsertmetastar {%
Toft.2014}%
\begin{APACrefauthors}%
Toft, M\BPBI B.%
, Schuitema, G.%
\BCBL {}\ \BBA {} Th{\o}gersen, J.%
\end{APACrefauthors}%
\unskip\
\newblock
\APACrefYearMonthDay{2014}{}{}.
\newblock
{\BBOQ}\APACrefatitle {{Responsible technology acceptance: Model development
  and application to consumer acceptance of Smart Grid technology}}
  {{Responsible technology acceptance: Model development and application to
  consumer acceptance of Smart Grid technology}}.{\BBCQ}
\newblock
\APACjournalVolNumPages{{Applied Energy}}{134}{}{392--400}.
\newblock
\begin{APACrefDOI} \doi{10.1016/j.apenergy.2014.08.048} \end{APACrefDOI}
\PrintBackRefs{\CurrentBib}

\bibitem [\protect \citeauthoryear {%
C.~Tzafestas%
, Velanas%
\BCBL {}\ \BBA {} Fakiridis%
}{%
C.~Tzafestas%
\ \protect \BOthers {.}}{%
{\protect \APACyear {2008}}%
}]{%
Tzafestas.2008}
\APACinsertmetastar {%
Tzafestas.2008}%
\begin{APACrefauthors}%
Tzafestas, C.%
, Velanas, S.%
\BCBL {}\ \BBA {} Fakiridis, G.%
\end{APACrefauthors}%
\unskip\
\newblock
\APACrefYearMonthDay{2008}{}{}.
\newblock
{\BBOQ}\APACrefatitle {{Adaptive impedance control in haptic teleoperation to
  improve transparency under time-delay}} {{Adaptive impedance control in
  haptic teleoperation to improve transparency under time-delay}}.{\BBCQ}
\newblock
\BIn{} \APACrefbtitle {{2008 IEEE International Conference on Robotics and
  Automation}.} {{2008 IEEE International Conference on Robotics and
  Automation}.}
\newblock
\APACaddressPublisher{}{IEEE}.
\newblock
\begin{APACrefDOI} \doi{10.1109/robot.2008.4543211} \end{APACrefDOI}
\PrintBackRefs{\CurrentBib}

\bibitem [\protect \citeauthoryear {%
S\BPBI G.~Tzafestas%
}{%
S\BPBI G.~Tzafestas%
}{%
{\protect \APACyear {2006}}%
}]{%
Tzafestas.2006}
\APACinsertmetastar {%
Tzafestas.2006}%
\begin{APACrefauthors}%
Tzafestas, S\BPBI G.%
\end{APACrefauthors}%
\unskip\
\newblock
\APACrefYearMonthDay{2006}{}{}.
\newblock
{\BBOQ}\APACrefatitle {{Concerning human-automation symbiosis in the society
  and the nature}} {{Concerning human-automation symbiosis in the society and
  the nature}}.{\BBCQ}
\newblock
\APACjournalVolNumPages{{International Journal of Factory Automation, Robotics
  and Soft Computing}}{1}{3}{16--24}.
\PrintBackRefs{\CurrentBib}

\bibitem [\protect \citeauthoryear {%
Uhl%
, Lindenmann%
\BCBL {}\ \BBA {} Matthiesen%
}{%
Uhl%
\ \protect \BOthers {.}}{%
{\protect \APACyear {2021}}%
}]{%
Uhl.2021}
\APACinsertmetastar {%
Uhl.2021}%
\begin{APACrefauthors}%
Uhl, M.%
, Lindenmann, A.%
\BCBL {}\ \BBA {} Matthiesen, S.%
\end{APACrefauthors}%
\unskip\
\newblock
\APACrefYearMonthDay{2021}{}{}.
\newblock
{\BBOQ}\APACrefatitle {{Analysis of factors influencing the productivity of
  hammer drilling - user forces, human fatigue, drilling direction, and drill
  bit}} {{Analysis of factors influencing the productivity of hammer drilling -
  user forces, human fatigue, drilling direction, and drill bit}}.{\BBCQ}
\newblock
\APACjournalVolNumPages{{Applied Ergonomics}}{92}{}{103338}.
\newblock
\begin{APACrefDOI} \doi{10.1016/j.apergo.2020.103338} \end{APACrefDOI}
\PrintBackRefs{\CurrentBib}

\bibitem [\protect \citeauthoryear {%
Unhelkar%
\ \protect \BOthers {.}}{%
Unhelkar%
\ \protect \BOthers {.}}{%
{\protect \APACyear {2018}}%
}]{%
Unhelkar.2018}
\APACinsertmetastar {%
Unhelkar.2018}%
\begin{APACrefauthors}%
Unhelkar, V\BPBI V.%
, Lasota, P\BPBI A.%
, Tyroller, Q.%
, Buhai, R\BHBI D.%
, Marceau, L.%
, Deml, B.%
\BCBL {}\ \BBA {} Shah, J\BPBI A.%
\end{APACrefauthors}%
\unskip\
\newblock
\APACrefYearMonthDay{2018}{}{}.
\newblock
{\BBOQ}\APACrefatitle {{Human-Aware Robotic Assistant for Collaborative
  Assembly: Integrating Human Motion Prediction With Planning in Time}}
  {{Human-Aware Robotic Assistant for Collaborative Assembly: Integrating Human
  Motion Prediction With Planning in Time}}.{\BBCQ}
\newblock
\APACjournalVolNumPages{{IEEE Robotics and Automation
  Letters}}{3}{3}{2394--2401}.
\newblock
\begin{APACrefDOI} \doi{10.1109/lra.2018.2812906} \end{APACrefDOI}
\PrintBackRefs{\CurrentBib}

\bibitem [\protect \citeauthoryear {%
Uno%
, Kawato%
\BCBL {}\ \BBA {} Suzuki%
}{%
Uno%
\ \protect \BOthers {.}}{%
{\protect \APACyear {1989}}%
}]{%
Uno.1989}
\APACinsertmetastar {%
Uno.1989}%
\begin{APACrefauthors}%
Uno, Y.%
, Kawato, M.%
\BCBL {}\ \BBA {} Suzuki, R.%
\end{APACrefauthors}%
\unskip\
\newblock
\APACrefYearMonthDay{1989}{}{}.
\newblock
{\BBOQ}\APACrefatitle {{Formation and control of optimal trajectory in human
  multijoint arm movement. Minimum torque-change model}} {{Formation and
  control of optimal trajectory in human multijoint arm movement. Minimum
  torque-change model}}.{\BBCQ}
\newblock
\APACjournalVolNumPages{{Biological Cybernetics}}{61}{2}{89--101}.
\newblock
\begin{APACrefDOI} \doi{10.1007/BF00204593} \end{APACrefDOI}
\PrintBackRefs{\CurrentBib}

\bibitem [\protect \citeauthoryear {%
{van der Laan}%
, Heino%
\BCBL {}\ \BBA {} de Waard%
}{%
{van der Laan}%
\ \protect \BOthers {.}}{%
{\protect \APACyear {1997}}%
}]{%
vanderLaan.1997}
\APACinsertmetastar {%
vanderLaan.1997}%
\begin{APACrefauthors}%
{van der Laan}, J\BPBI D.%
, Heino, A.%
\BCBL {}\ \BBA {} de Waard, D.%
\end{APACrefauthors}%
\unskip\
\newblock
\APACrefYearMonthDay{1997}{}{}.
\newblock
{\BBOQ}\APACrefatitle {{A simple procedure for the assessment of acceptance of
  advanced transport telematics}} {{A simple procedure for the assessment of
  acceptance of advanced transport telematics}}.{\BBCQ}
\newblock
\APACjournalVolNumPages{{Transportation Research Part C: Emerging
  Technologies}}{5}{1}{}.
\PrintBackRefs{\CurrentBib}

\bibitem [\protect \citeauthoryear {%
{van Erp}%
, Veltman%
\BCBL {}\ \BBA {} Grootjen%
}{%
{van Erp}%
\ \protect \BOthers {.}}{%
{\protect \APACyear {2010}}%
}]{%
vanErp.2010}
\APACinsertmetastar {%
vanErp.2010}%
\begin{APACrefauthors}%
{van Erp}, J\BPBI B\BPBI F.%
, Veltman, H.%
\BCBL {}\ \BBA {} Grootjen, M.%
\end{APACrefauthors}%
\unskip\
\newblock
\APACrefYearMonthDay{2010}{}{}.
\newblock
{\BBOQ}\APACrefatitle {{Brain-based indices for user system symbiosis}}
  {{Brain-based indices for user system symbiosis}}.{\BBCQ}
\newblock
\BIn{} \APACrefbtitle {{Brain-Computer Interfaces}} {{Brain-Computer
  Interfaces}}\ (\BPGS\ 201--219).
\newblock
\APACaddressPublisher{}{{Springer, London}}.
\newblock
\begin{APACrefDOI} \doi{10.1007/978-1-84996-272-8_12} \end{APACrefDOI}
\PrintBackRefs{\CurrentBib}

\bibitem [\protect \citeauthoryear {%
Vantrepotte%
, Berberian%
, Pagliari%
\BCBL {}\ \BBA {} Chambon%
}{%
Vantrepotte%
\ \protect \BOthers {.}}{%
{\protect \APACyear {2021}}%
}]{%
Vantrepotte.2021}
\APACinsertmetastar {%
Vantrepotte.2021}%
\begin{APACrefauthors}%
Vantrepotte, Q.%
, Berberian, B.%
, Pagliari, M.%
\BCBL {}\ \BBA {} Chambon, V.%
\end{APACrefauthors}%
\unskip\
\newblock
\APACrefYearMonthDay{2021}{}{}.
\newblock
\APACrefbtitle {{Leveraging human agency to improve confidence and
  acceptability in human-machine interactions}.} {{Leveraging human agency to
  improve confidence and acceptability in human-machine interactions}.}
\newblock
\begin{APACrefDOI} \doi{10.31234/osf.io/6pvnh} \end{APACrefDOI}
\PrintBackRefs{\CurrentBib}

\bibitem [\protect \citeauthoryear {%
Vedder%
\ \BBA {} Carey%
}{%
Vedder%
\ \BBA {} Carey%
}{%
{\protect \APACyear {2005}}%
}]{%
Vedder.2005}
\APACinsertmetastar {%
Vedder.2005}%
\begin{APACrefauthors}%
Vedder, J.%
\BCBT {}\ \BBA {} Carey, E.%
\end{APACrefauthors}%
\unskip\
\newblock
\APACrefYearMonthDay{2005}{}{}.
\newblock
{\BBOQ}\APACrefatitle {{A multi-level systems approach for the development of
  tools, equipment and work processes for the construction industry}} {{A
  multi-level systems approach for the development of tools, equipment and work
  processes for the construction industry}}.{\BBCQ}
\newblock
\APACjournalVolNumPages{{Applied Ergonomics}}{36}{4}{471--480}.
\newblock
\begin{APACrefDOI} \doi{10.1016/j.apergo.2005.01.004} \end{APACrefDOI}
\PrintBackRefs{\CurrentBib}

\bibitem [\protect \citeauthoryear {%
Venkatesh%
\ \BBA {} Bala%
}{%
Venkatesh%
\ \BBA {} Bala%
}{%
{\protect \APACyear {2008}}%
}]{%
Venkatesh.2008}
\APACinsertmetastar {%
Venkatesh.2008}%
\begin{APACrefauthors}%
Venkatesh, V.%
\BCBT {}\ \BBA {} Bala, H.%
\end{APACrefauthors}%
\unskip\
\newblock
\APACrefYearMonthDay{2008}{}{}.
\newblock
{\BBOQ}\APACrefatitle {{Technology Acceptance Model 3 and a research agenda on
  interventions}} {{Technology Acceptance Model 3 and a research agenda on
  interventions}}.{\BBCQ}
\newblock
\APACjournalVolNumPages{{Decision Sciences}}{39}{2}{273--315}.
\newblock
\begin{APACrefDOI} \doi{10.1111/j.1540-5915.2008.00192.x} \end{APACrefDOI}
\PrintBackRefs{\CurrentBib}

\bibitem [\protect \citeauthoryear {%
Volpe%
\ \protect \BOthers {.}}{%
Volpe%
\ \protect \BOthers {.}}{%
{\protect \APACyear {2001}}%
}]{%
Volpe.2001}
\APACinsertmetastar {%
Volpe.2001}%
\begin{APACrefauthors}%
Volpe, R.%
, Nesnas, I.%
, Estlin, T.%
, Mutz, D.%
, Petras, R.%
\BCBL {}\ \BBA {} Das, H.%
\end{APACrefauthors}%
\unskip\
\newblock
\APACrefYearMonthDay{2001}{}{}.
\newblock
{\BBOQ}\APACrefatitle {{The CLARAty architecture for robotic autonomy}} {{The
  CLARAty architecture for robotic autonomy}}.{\BBCQ}
\newblock
\BIn{} \APACrefbtitle {{2001 IEEE Aerospace Conference Proceedings (Cat.
  No.01TH8542)}.} {{2001 IEEE Aerospace Conference Proceedings (Cat.
  No.01TH8542)}.}
\newblock
\APACaddressPublisher{}{IEEE}.
\newblock
\begin{APACrefDOI} \doi{10.1109/aero.2001.931701} \end{APACrefDOI}
\PrintBackRefs{\CurrentBib}

\bibitem [\protect \citeauthoryear {%
K\BHBI J.~Wang%
, Sun%
, Xia%
\BCBL {}\ \BBA {} Mao%
}{%
K\BHBI J.~Wang%
\ \protect \BOthers {.}}{%
{\protect \APACyear {2016}}%
}]{%
Wang.2016}
\APACinsertmetastar {%
Wang.2016}%
\begin{APACrefauthors}%
Wang, K\BHBI J.%
, Sun, M.%
, Xia, R.%
\BCBL {}\ \BBA {} Mao, Z\BHBI H.%
\end{APACrefauthors}%
\unskip\
\newblock
\APACrefYearMonthDay{2016}{}{}.
\newblock
{\BBOQ}\APACrefatitle {{Human-robot symbiosis framework on exoskeleton
  devices}} {{Human-robot symbiosis framework on exoskeleton devices}}.{\BBCQ}
\newblock
\BIn{} \APACrefbtitle {{2016 IEEE International Conference on Industrial
  Technology (ICIT)}.} {{2016 IEEE International Conference on Industrial
  Technology (ICIT)}.}
\newblock
\begin{APACrefDOI} \doi{10.1109/icit.2016.7474982} \end{APACrefDOI}
\PrintBackRefs{\CurrentBib}

\bibitem [\protect \citeauthoryear {%
L.~Wang%
\ \protect \BOthers {.}}{%
L.~Wang%
\ \protect \BOthers {.}}{%
{\protect \APACyear {2019}}%
}]{%
Wang.2019}
\APACinsertmetastar {%
Wang.2019}%
\begin{APACrefauthors}%
Wang, L.%
, Gao, R.%
, V{\'a}ncza, J.%
, Kr{\"u}ger, J.%
, Wang, X\BPBI V.%
, Makris, S.%
\BCBL {}\ \BBA {} Chryssolouris, G.%
\end{APACrefauthors}%
\unskip\
\newblock
\APACrefYearMonthDay{2019}{}{}.
\newblock
{\BBOQ}\APACrefatitle {{Symbiotic human-robot collaborative assembly}}
  {{Symbiotic human-robot collaborative assembly}}.{\BBCQ}
\newblock
\APACjournalVolNumPages{{CIRP Annals}}{68}{2}{701--726}.
\newblock
\begin{APACrefDOI} \doi{10.1016/j.cirp.2019.05.002} \end{APACrefDOI}
\PrintBackRefs{\CurrentBib}

\bibitem [\protect \citeauthoryear {%
Wen%
, Kuroki%
\BCBL {}\ \BBA {} Asama%
}{%
Wen%
\ \protect \BOthers {.}}{%
{\protect \APACyear {2019}}%
}]{%
Wen.2019}
\APACinsertmetastar {%
Wen.2019}%
\begin{APACrefauthors}%
Wen, W.%
, Kuroki, Y.%
\BCBL {}\ \BBA {} Asama, H.%
\end{APACrefauthors}%
\unskip\
\newblock
\APACrefYearMonthDay{2019}{}{}.
\newblock
{\BBOQ}\APACrefatitle {{The sense of agency in driving automation}} {{The sense
  of agency in driving automation}}.{\BBCQ}
\newblock
\APACjournalVolNumPages{{Frontiers in Psychology}}{10}{}{}.
\newblock
\begin{APACrefDOI} \doi{10.3389/fpsyg.2019.02691} \end{APACrefDOI}
\PrintBackRefs{\CurrentBib}

\end{thebibliography}
\endgroup

\end{document}